\documentclass[aps,prd,amsfonts,groupedaddress,showpacs]{revtex4}

\usepackage{graphicx}
\usepackage{axodraw}

\begin{document}

\def\intdk{\int\frac{d^4k}{(2\pi)^4}}
\def\sla{\hspace{-0.17cm}\slash}

\title{\Large{\bf One-Loop Renormalization of Non-Abelian Gauge Theory \\
and $\beta$ Function Based on Loop Regularization Method}}
\author{Jian-Wei Cui and Yue-Liang Wu}
\affiliation{ Kavli Institute for Theoretical Physics China,
Institute of Theoretical Physics \\ Chinese Academy of Science
(KITPC/ITP-CAS), Beijing,100080, P.R.China}

\date{January 15, 2008}

\begin{abstract}
All one-loop renormalization constants for Non-Abelian gauge theory
are computed in details by using the symmetry-preserving Loop
Regularization method proposed in\cite{LR1,LR2}. The resulting
renormalization constants are manifestly shown to satisfy
Ward-Takahaski-Slavnov-Taylor identities, and lead to the well-known
one loop $\beta$ function for Non-Abelian gauge theory of
QCD\cite{GWP}. The loop regularization method is realized in the
dimension of original field theories, it maintains not only
symmetries but also divergent behaviors of original field theories
with the introduction of two energy scales. Such two scales play the
roles of characterizing and sliding energy scales as well as
ultraviolet and infrared cutoff energy scales. An explicit Check of
those identities provides a clear demonstration how the
symmetry-preserving Loop Regularization method can consistently be
applied to non-Abelian gauge theories.
\end{abstract}

\maketitle

\section{Introduction}

It is known that quantum field theories (QFTs) can not be defined by
the straightforward perturbative expansion because of the
ultraviolet (UV) divergences. In order to make meaningful for QFTs,
it is necessary to remove infinities from perturbative calculations
by renormalizing the fields, masses, and coupling constants. A
successful renormalization of QFT was firstly realized in 1940s by
Tomonaga\cite{tomonaga}, Schwinger\cite{schwinger},
Feynman\cite{fenyman} and Dyson\cite{dyson} for the case of QED,
while it took until the early of 1970s when Wilson\cite{wilson} gave
it full physical meaning on QFTs.

The first step before renormalization is to modify the behavior of
field theory at very large momentum so that all Feynman diagrams
become well-defined finite quantities. This procedure is usually
called regularization. The most important properties needed for a
good regularization method are that it must preserve all symmetries
of original field theories and meantime maintain the divergent
behavior of original Feynman integrals. In fact, many regularization
and renormalization methods have been proposed in the last several
decades such as: cut-off regularization\cite{cutoff}, Pauli-Villars
regularization\cite{PV}, Schwinger's proper time
regularization\cite{propertime}, dimensional
regularization\cite{DR}, lattice regularization\cite{lattice},
constrained differential renormalization\cite{CDR} and so on. As
discussed in\cite{LR1,LR2}, each of them has its advantage in
applying to different situations. Up to now, there exists no single
regularization which is suitable to all purposes in QFTs. In
refs.\cite{LR1,LR2}, a new symmetry-preserving loop
regularization(LR) was introduced to meet the request mentioned
above. The key concept in such a new regularization method is the
introduction of irreducible loop integrals(ILIs) \cite{LR1,LR2}
which are evaluated from Feynman integrals.
The gauge symmetry requires a set of necessary and sufficient
conditions called consistency conditions\cite{LR1} which are held
between the regularized tensor type ILIs and scalar type ILIs. The
loop regularization method was realized to satisfy those consistency
conditions\cite{LR1,LR2} in the existence of two energy scales. We
shall give a brief introduction for the loop regularization below.
For more details on the loop regularization including motivations
and concrete computation methods as well as general properties, we
refer the original papers\cite{LR1,LR2} to readers. Some interesting
applications of this new method have been investigated in
\cite{DW,MW1,MW2}.

This paper is devoted to explicitly demonstrate how the loop
regularization preserves non-Abelian gauge symmetry by evaluating
all the renormalization constants at one loop level and verifying
the Ward-Takahaski-Slavnov-Taylor identities among the
renormalization constants. The paper is organized as follows: in
section II, we shape the gauge symmetry into the well-known
Ward-Takahaski-Slavnov-Taylor identities, and give the conditions
that the renormalization constants must satisfy. In section III, we
briefly outline the LR method. In section IV, we explicitly evaluate
all the one-loop divergent Feynman diagrams to yield all the
renormalization constants of non-Abelian gauge theory by using the
loop regularization method, and derive the well-known $\beta$
function\cite{GWP} once checking manifestly the
Ward-Takahaski-Slavnov-Taylor identities among the obtained
renormalization constants. The results are found to be consistent
with those obtained via the dimensional regularization as the
quadratic divergent parts cancel each other due to gauge symmetry.
The conclusions and remarks are presented in the last section.

\section{Renormalization of Gauge Theory and Ward-Takahaski-Slavnov-Taylor identities}

The lagrangian of gauge theory with Dirac spinor fields $\psi_n\
(n=1,...,N_f)$ interacting with gauge field $A^a_\mu\
(a=1,...,d_G)$ is:
\begin{eqnarray}
\mathcal{L}=\bar{\psi}_n(i\gamma^{\mu}D_{\mu}-m)\psi_n-\frac{1}{4}F^a_{\mu\nu}F^{a\mu\nu}
\end{eqnarray}
where:
\begin{eqnarray}
& &F_{\mu\nu}^a=\partial_{\mu}A_{\nu}^a-\partial_{\nu}A_{\mu}^a+gf^{abc}A_{\mu}^bA_{\nu}^c\\
& &D_{\mu}\psi_n=(\partial_{\mu}-igT^aA_{\mu}^a)\psi_n
\end{eqnarray}
According to the Faddeev-Popov\cite{faddeevpopov} quantization
method, some ghost fields are necessary to be introduced when fixing
a gauge. In the covariant gauge, the lagrangian has the following
form:
\begin{eqnarray}
\mathcal{L}_{eff}&=&\bar{\psi}_n(i\gamma^{\mu}D_{\mu}-m)\psi_n-\frac{1}{4}F^a_{\mu\nu}F^{a\mu\nu}-\frac{1}{2\xi}(\partial^{\mu}A_{\mu}^a)^2+\partial^{\mu}\bar{c}^a(\partial_{\mu}\delta^{ac}+gf^{abc}A_{\mu}^b)c^c\nonumber\\
&=&[\bar{\psi}_n(i\gamma^{\mu}\partial_{\mu}-m)\psi_n]+[-\frac{1}{4}(\partial_{\mu}A^a_{\nu}-\partial_{\nu}A^a_{\mu})^2-\frac{1}{2\xi}(\partial^{\mu}A_{\mu}^a)^2]+[\partial^{\mu}\bar{c}^a\delta^{ac}\partial_{\mu}c^c]\nonumber\\
& &+g\bar{\psi}_n\gamma_{\mu}A^{a\mu}T^a\psi_{n}-\frac{1}{2}gf^{abc}(\partial_{\mu}A^a_{\nu}-\partial{\nu}A^a_{\mu})A^{b\mu}A^{c\nu}+\frac{1}{4}g^2f^{abc}f^{ade}A^b_{\mu}A^c_{\nu}A^{d\mu}A^{e\nu}\nonumber\\
& &+gf^{abc}\partial^{\mu}\bar{c}^aA_{\mu}^bc^c
\end{eqnarray}
The corresponding Feynman Rules for this lagrangian are presented in
App.B. All one loop Feynman diagrams are shown below (for
simplicity, the permutation graphs are omitted):

\includegraphics[scale=0.7]{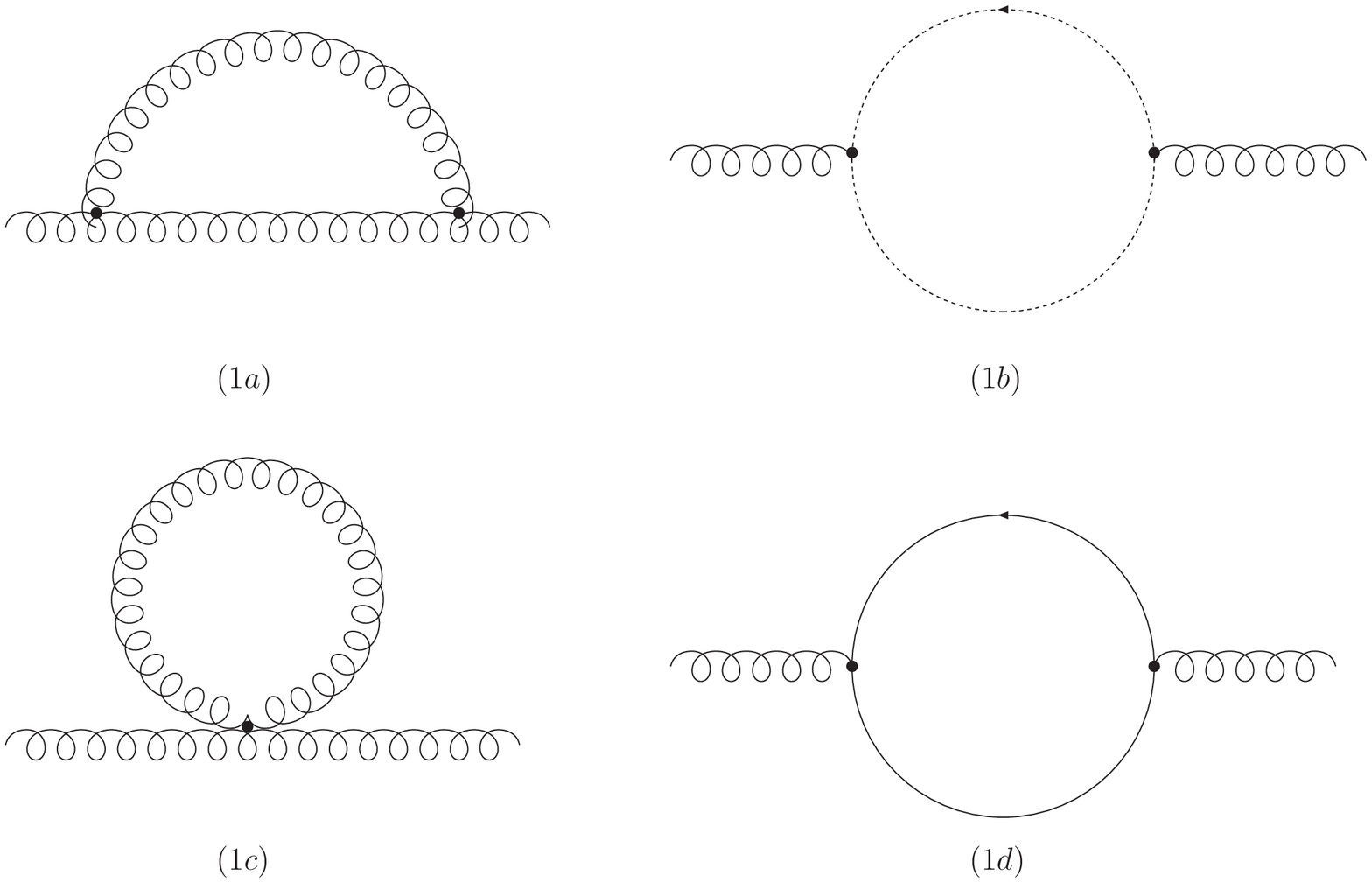}
\includegraphics[scale=0.7]{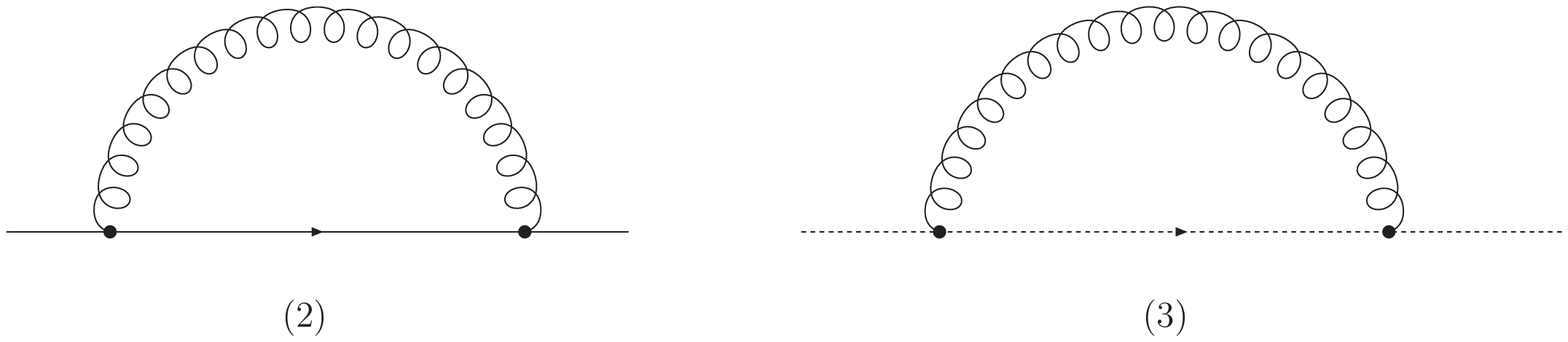}
\includegraphics[scale=0.7]{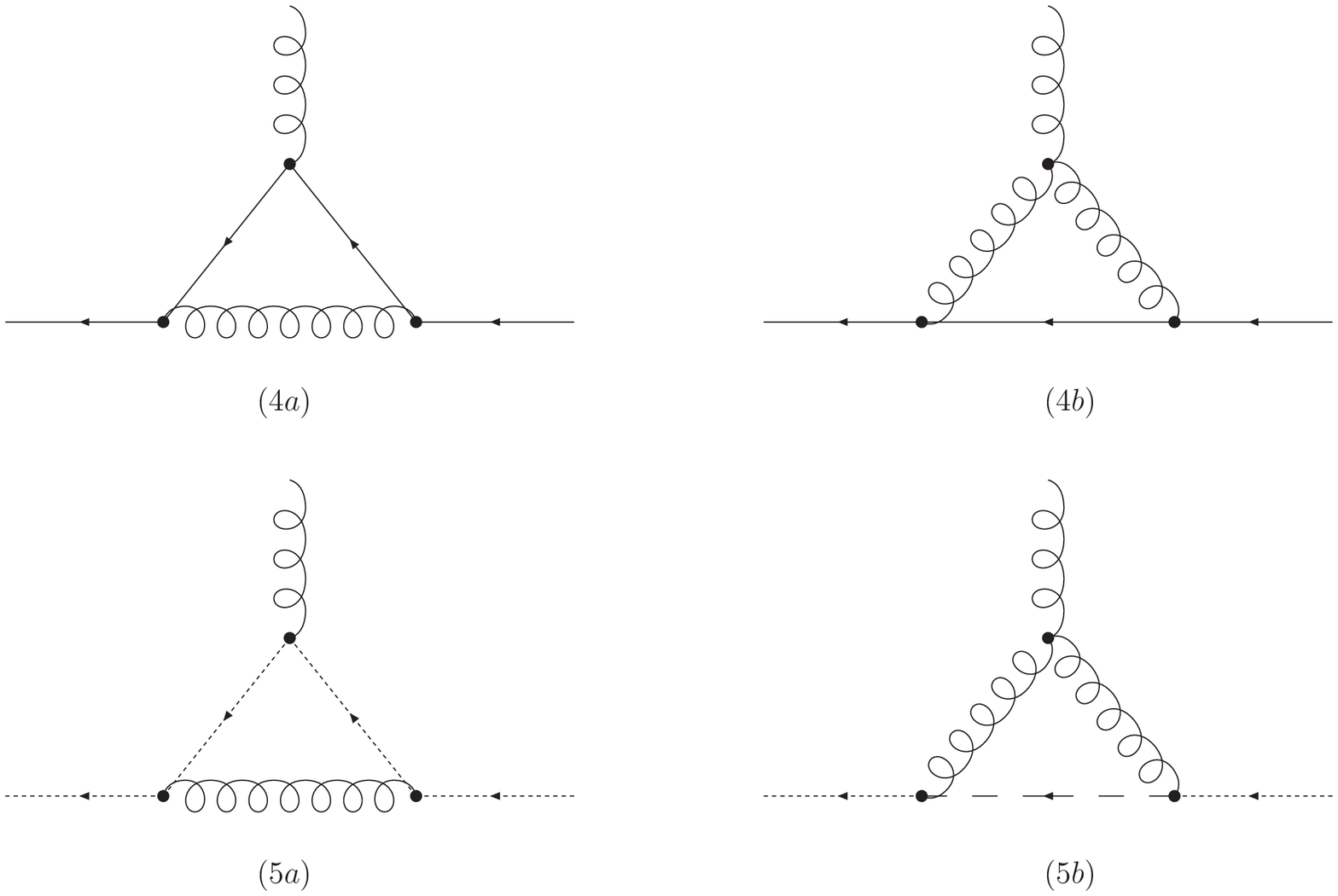}
\includegraphics[scale=0.7]{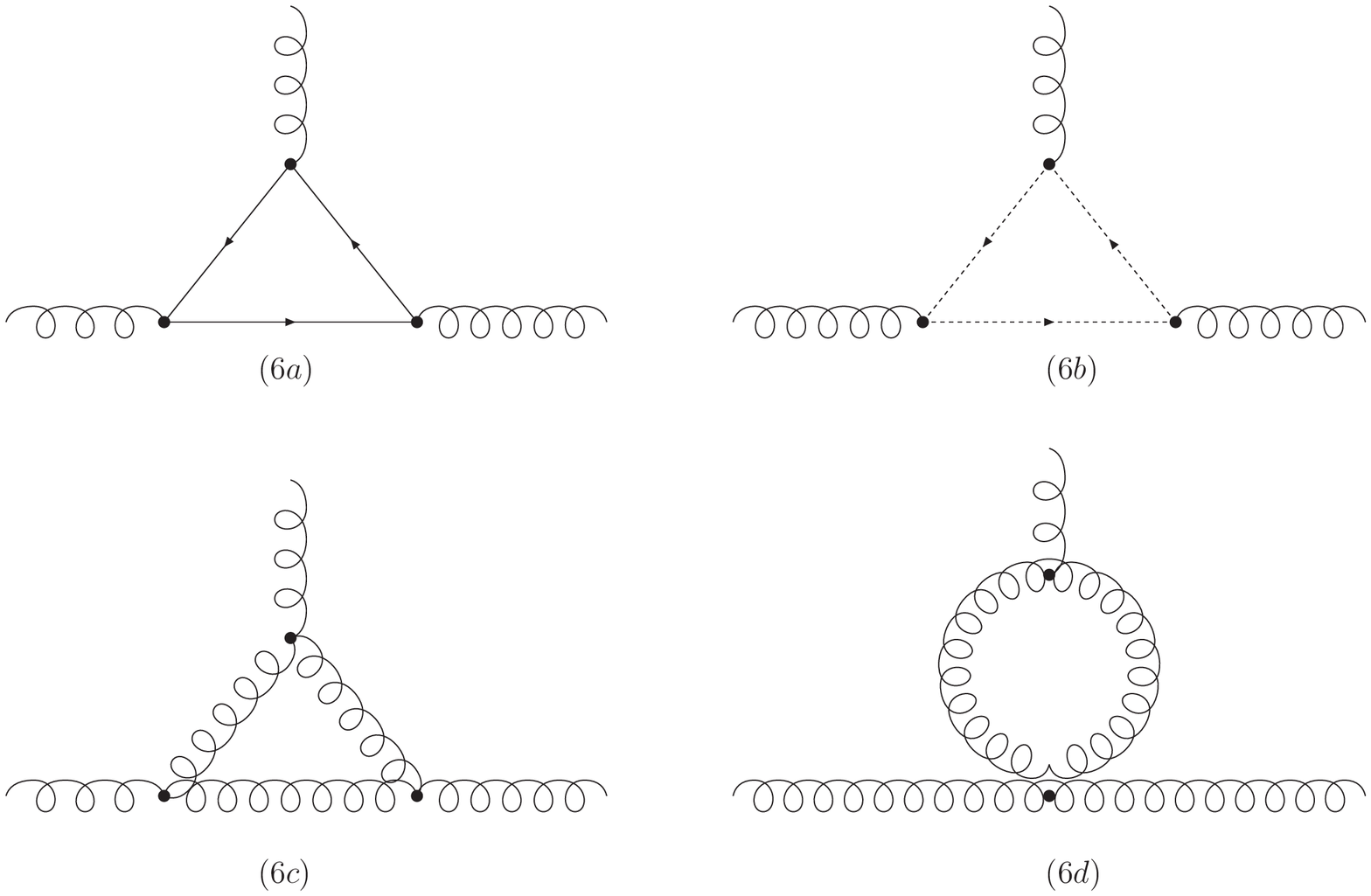}
\includegraphics[scale=0.7]{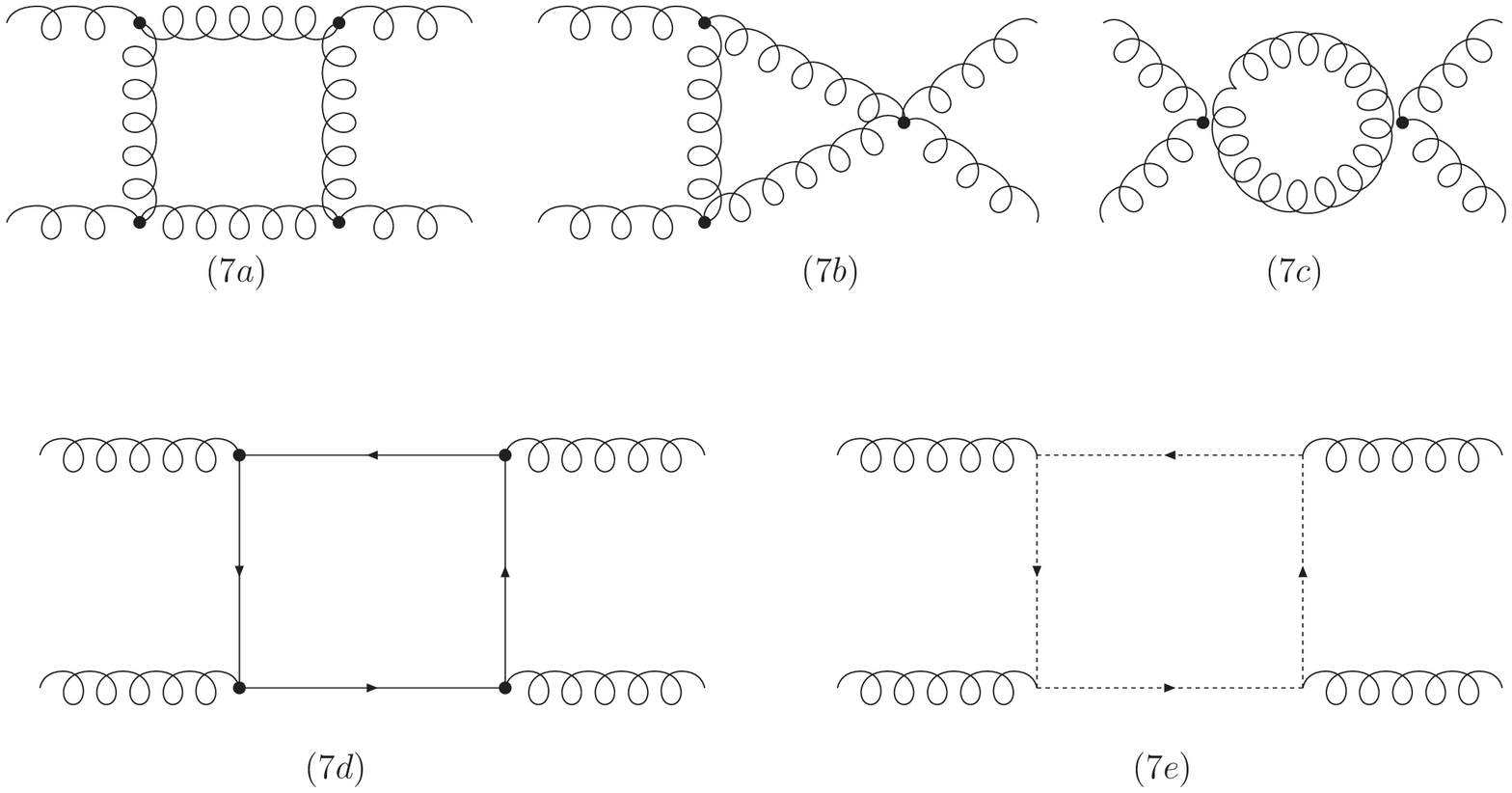}
\begin{center}{\sl Fig.1.}\end{center}

Though all loop diagrams contain divergent integrals, it was proved
that gauge theories are
renormalizable\cite{renormalizability1,renormalizability2,renormalizability3,renormalizability4,renormalizability5}.
To remove the divergence, it is necessary to renormalize the theory
by rescaling the fields and redefining the masses and coupling
constant. This procedure is equivalent to the introduction of some
counterterms to the Lagrangian
\begin{eqnarray}
\delta\mathcal{L}&=&[(z_2-1)\bar{\psi}_ni\gamma^{\mu}\partial_{\mu}\psi_n-(z_2z_m-1)m\bar{\psi}_n\psi_n]
+(z_3-1)[-\frac{1}{4}(\partial_{\mu}A^a_{\nu}-\partial_{\nu}A^a_{\mu})^2]\nonumber\\
& &+(\tilde{z}_3-1)[\partial^{\mu}\bar{c}^a\delta^{ac}\partial_{\mu}c^c]
+(z_{1F}-1)g\bar{\psi}_n\gamma_{\mu}A^{a\mu}T^a\psi_{n}\nonumber\\
& &-(z_1-1)\frac{1}{2}gf^{abc}(\partial_{\mu}A^a_{\nu}-\partial{\nu}A^a_{\mu})A^{b\mu}A^{c\nu}
+(z_4-1)\frac{1}{4}g^2f^{abc}f^{ade}A^b_{\mu}A^c_{\nu}A^{d\mu}A^{e\nu}\nonumber\\
& &+(\tilde{z_1}-1)gf^{abc}\partial^{\mu}\bar{c}^aA_{\mu}^bc^c
\end{eqnarray}
where $z_1, \cdots, z_4$ are the so-called renormalization
constants. They are not independent and must satisfy the relations
called Slavnov-Taylor identities\cite{st} which are the
generalization of the usual Ward-Takahaski identities. Those
identities are actually consequence of the gauge symmetry. To obtain
the relations, one can make the BRST transformation\cite{BRST} which
leads to some identities for the generating functional. Then
performing a Lengendre transformation we obtained the identities for
the 1PI generating functional. Taking the functional derivatives of
the 1PI generating functional, one arrives at the relations between
the 1PI Green functions. Those relations are the strict restriction
of the solution of the gauge symmetry. As a consequence, the
renormalization constants should satisfy the following
identities\cite{relationofz}:
\begin{eqnarray}
\frac{z_{1F}}{z_3^{1/2}z_2}=\frac{\tilde{z_1}}{z_3^{1/2}\tilde{z}_3}=\frac{z_1}{z_3^{3/2}}=\frac{z_4^{1/2}}{z_3}
\end{eqnarray}
There is a more intuitive method to yield the relations among the
renormalization constants. In fact, the gauge independence and the
unitarity of the renormalized S matrix require that the gauge
symmetry must be maintained after renormalization\cite{smatrix},
which means that the renormalization constants of $g$ obtained from
each vertex renormalization  must be the same. From such a
requirement, one can arrive at above identities. The two-, three-
and four-point renormalization constants were evaluated in
refs.\cite{CG,PS} by using the dimensional regularization. For
completeness, we shall perform in this note a detailed calculation
for all two-, three- and four-point renormalization constants by
using the loop regularization method. As our calculations for the
renormalization constants are carried out only at one loop level,
which does not involve the renormalization scheme dependence, so we
shall not discuss in this note the relevant issues. A detailed
discussion on the renormalization scheme prescription in loop
regularization will be considered elsewhere.


\section{brief introduction to loop regularization method}

In this section we shall briefly introduce the loop regularization
method. For our current consideration, we demonstrate only the one
loop case. The key concept of the loop regularization is the
introduction of irreducible loop integrals (ILIs). It has been shown
in\cite{LR1,LR2} that by adopting the Feynman parameterization
method with appropriately shifting the integration variables, all
one loop Feynman integrals can be expressed in terms of the
following 1-fold ILIs:
\begin{eqnarray}
I_{-2\alpha}&=&\intdk\frac{1}{(k^2-M^2)^{2+\alpha}},\nonumber\\
I_{-2\alpha\ \mu\nu}&=&\intdk\frac{k_{\mu}k_\nu}{(k^2-M^2)^{3+\alpha}},\hspace{8mm}\alpha=-1,0,1,2,...\\
I_{-2\alpha\
\mu\nu\rho\sigma}&=&\intdk\frac{k_{\mu}k_{\nu}k_{\rho}k_{\sigma}}{(k^2-M^2)^{4+\alpha}}\nonumber
\end{eqnarray}
Here $M^2$ is in general a function of the external momenta $p_i$,
the masses of particles $m_i$ and the Feynman parameters. Where
$I_2$ and $I_0$ are corresponding to the quadratic and logarithmic
divergent integrals.

To maintain the gauge invariance, the regularized 1-fold ILIs should
satisfy a set of consistency conditions\cite{LR1,LR2}:
\begin{eqnarray}
& & I_{2\mu\nu}^R = \frac{1}{2} g_{\mu\nu}\ I_2^R, \quad
I_{2\mu\nu\rho\sigma }^R = \frac{1}{8} (g_{\mu\nu}g_{\rho\sigma} +
g_{\mu\rho}g_{\nu\sigma} +
g_{\mu\sigma}g_{\rho\nu})\ I_2^R  , \nonumber \\
& & I_{0\mu\nu}^R = \frac{1}{4} g_{\mu\nu} \ I_0^R, \quad
I_{0\mu\nu\rho\sigma }^R = \frac{1}{24} (g_{\mu\nu}g_{\rho\sigma} +
g_{\mu\rho}g_{\nu\sigma} + g_{\mu\sigma}g_{\rho\nu})\ I_0^R .
\end{eqnarray}
where the superscript "R" denotes the regularized ILIs.

A simple prescription of loop regularization \cite{LR1,LR2} was
realized to ensure the above consistency conditions. The procedure
is: Rotating to the four dimensional Euclidean space of momentum,
replacing in the ILIs the loop integrating variable $k^2$ and the
loop integrating measure $\int{d^4k}$ by the corresponding
regularized ones $[k^2]_l$ and $\int[d^4k]_l$:
\begin{eqnarray}
& & \quad k^2 \rightarrow [k^2]_l \equiv k^2+M^2_l\ ,
\nonumber \\
& & \int{d^4k} \rightarrow \int[d^4k]_l \equiv \lim_{N,
M_l^2}\sum_{l=0}^{N}c_l^N\int{d^4k}
\end{eqnarray}
where $M_l^2$ ($ l= 0,1,\ \cdots $) may be regarded as the mass
factors of loop regulators. If there is no IR divergence in the
integrals, one can take the initial conditions $M_0^2 = 0$ and
$c_0^N = 1$ to recover the original integrals in the limit $M_l^2
\to \infty$ ($l=1,2,\cdots$ ). For IR divergent integrals, one may
set $M_0^2=\mu_s^2$ to regularize it. The regularized ILIs in the
Euclidean space-time are then given by:
\begin{eqnarray}
I_{-2\alpha}^R&=& i (-1)^{\alpha} \lim_{N,
M_l^2}\sum_{l=0}^{N}c_l^N\intdk\frac{1}{(k^2 + M^2 + M_l^2)^{2+\alpha}},\nonumber\\
I_{-2\alpha\ \mu\nu}^R&=& -i (-1)^{\alpha} \lim_{N,
M_l^2}\sum_{l=0}^{N}c_l^N\intdk\frac{k_{\mu}k_\nu}{(k^2+M^2+M_l^2)^{3+\alpha}},\hspace{8mm}\alpha=-1,0,1,2,...\\
I_{-2\alpha\ \mu\nu\rho\sigma}^R&=& i (-1)^{\alpha} \lim_{N,
M_l^2}\sum_{l=0}^{N}c_l^N\intdk\frac{k_{\mu}k_{\nu}k_{\rho}k_{\sigma}}{(k^2+M^2+M_l^2)^{4+\alpha}}\nonumber
\end{eqnarray}

The coefficients $c_l^N$ are chosen to satisfy the following
conditions:
\begin{eqnarray}
\lim_{N, M_l^2}\sum_{l=0}^{N}c_l^N(M_l^2)^n = 0 \quad
       (n= 0, 1, \cdots)\label{cl conditions}
\end{eqnarray}
One can easily verify that the following set is the simplest
solution of the above conditions:
\begin{equation}
M_l^2=\mu_s^2+lM_R^2, \quad
c_l^N=(-1)^l\frac{N!}{(N-l)!l!}\label{mus}
\end{equation}
Here $M_R$ may be regarded as a basic mass scale of loop regulator
and the notation $\lim_{N, M_l^2}$ stands for the limit $\lim_{N,
M_R^2\rightarrow \infty}$. It has been shown in \cite{LR2} that the
above regularization prescription can be understood in terms of
Schwinger proper time formulation with an appropriate regulating
distribution function. Note that the loop regularization is
different from the Pauli-Villars regularization in which the
regularization prescription is realized through introducing super
heavy particles, so that the Pauli-Villars regularization cannot
directly be applied to non-Abelian gauge theories. Unlike the
Pauli-Villars regularization, the loop regularization is applicable
to non-Abelian gauge theories via above regularization prescription
on the ILIs.

With the simple solution for $M_l^2$ and $c_l^N$ in above equation,
the regularized ILIs $I_0^R$ and $I_2^R$ can be evaluated explicitly
as \cite{LR1,LR2}:
\begin{eqnarray}
I_2^R&=&\frac{-i}{16\pi^2}\{M_c^2-\mu^2[ln\frac{M_c^2}{\mu^2}-\gamma_w+1+y_2(\frac{\mu^2}{M_c^2})]\}  \nonumber \\
I_0^R&=&\frac{i}{16\pi^2}[ln\frac{M_c^2}{\mu^2}-\gamma_w+y_0(\frac{\mu^2}{M_c^2})]
\end{eqnarray}
with $\mu^2=\mu_s^2+M^2$, and
\begin{eqnarray}
& & \gamma_w \equiv \lim_{N}\{ \ \sum_{l=1}^{N} c_l^N \ln l +
     \ln [\ \sum_{l=1}^{N} c_l^N\ l \ln l \ ] \} = \gamma_E=0.5772\cdots, \nonumber \\
& & y_0(x)=\int_0^x d\sigma \frac{1-e^{-\sigma}}{\sigma},
 \quad  y_1(x)=\frac{e^{-x}-1+x}{x}\nonumber \\
& & y_2(x)=y_0(x)-y_1(x),\quad \lim_{x\rightarrow0}y_i(x)\rightarrow
0,\ i=0,1,2 \\
& & M_c^2\equiv \lim_{N,M_R} M_R^2 \sum_{l=1}^{N}c_l^N(l \ln l) =\lim_{N,M_R}M_R^2/\ln N \nonumber\\
\end{eqnarray}
By comparing the above results with the ones obtained by naive
cutoff regularizaton, it is easily seen that the $\mu_s $ sets an IR
`cutoff' at $M^2 =0$ and $M_c$ provides an UV `cutoff'. For
renormalizable quantum field theories, $M_c$ can be taken to be
infinity $(M_c\rightarrow\infty)$. In a theory without infrared
divergence, $\mu_s$ can safely run to $\mu_s=0$. Actually, in the
case that $M_c\to\infty$ and $\mu_s=0$, one recovers the initial
integral. Also once $M_R$ and $N$ are taken to be infinity, the
regularized theory becomes independent of the regularization
prescription. These are main properties needed for a proper
regularization. For a detailed description and an explicit treatment
for higher loop Feynman integrals, it is referred to the original
paper on loop regularization \cite{LR1,LR2}. Note that to evaluate
the ILIs, the algebraic computing for multi $\gamma$ matrices
involving loop momentum $k\sla$ such as $k\sla\gamma_{\mu}k\sla$
should be carried out to be expressed in terms of the independent
components: $\gamma_\mu$, $\sigma_{\mu\nu}$, $\gamma_5\gamma_{\mu}$,
$\gamma_5$.

It is known that in all the regularization schemes, there is an
important issue that for a divergent integral it is in general not
appropriate to shift the integration variable. In the loop
regularization method we have actually shifted the integration
variables before taking the regularization prescription, one may
doubt wether such a treatment is well justified. The answer is yes.
In fact, we can take the loop regularization prescription before
shifting the integration variables, and the results are the same as
what we get when shifting the integration variables first. For an
illustration, let us examine a simple logarithmic divergent Feynman
integral:
\begin{eqnarray}
L={\intdk}\frac{1}{k^2-m_1^2}\frac{1}{(k-p)^2-m_2^2}
\end{eqnarray}

Following the standard process of the loop regularization method,
the first step is to apply the general Feynman parameter formula
\begin{eqnarray}
\frac{1}{a_1^{\alpha_1}a_2^{\alpha_2}{\cdots}a_n^{\alpha_n}}& = &
\frac{\Gamma(\alpha_1+\cdots+\alpha_n)}{\Gamma(\alpha_1)\cdots\Gamma(\alpha_n)}
\int_0^1dx_1\int_0^{x_1}dx_2\cdots\int_0^{x_{n-2}}dx_{n-1} \nonumber
\\
& &
\frac{(1-x_1)^{\alpha_1-1}(x_1-x_2)^{\alpha_2-1}{\cdots}x_{n-1}^{\alpha_n-1}}
{[a_1(1-x_1)+a_2(x_1-x_2)+\cdots+a_nx_{n-1}]^{\alpha_1+\cdots+\alpha_n}}
\end{eqnarray}
to the Feyman integral. For the above Feynman integral, we then
obtain the following integral
\begin{eqnarray}
L&=&{\intdk}\int_0^1dx\frac{1}{\{(1-x)(k^2-m_1^2)+x[(k-p)^2-m_2^2]\}^2}\nonumber\\
&=&{\intdk}\int_0^1dx\frac{1}{\{(k-xp)^2-[(1-x)m_1^2+xm_2^2-x(1-x)p^2]\}^2}\nonumber\\
&=&\int_0^1dx{\intdk}\frac{1}{( (k-xp)^2 -M^2)^2}
\end{eqnarray}
with $M^2=(1-x)m_1^2+xm_2^2-x(1-x)p^2$. When shifting the
integration variable, we arrive at the standard scalar type ILI
\begin{eqnarray}
L&=& \int_0^1dx{\intdk}\frac{1}{( k^2 -M^2)^2} = \int_0^1dx\  I_0
\end{eqnarray}
By making Wick rotation and applying the loop regularization
prescription to such an integral, we then obtain the regularized
Feynman integral
\begin{eqnarray}
L^R= i \int_0^1dx\lim_{N,
M_l^2}\sum_{l=0}^{N}c_l^N\intdk\frac{1}{(k^2+M^2+M_l^2)^2}
\end{eqnarray}

Alternatively, one can also apply for the regularization
prescription before shifting the integration variable, i.e.,
$(k-xp)^2 \to (k-xp)^2 + M_l^2$, we then have
\begin{eqnarray}
L^{\prime R}= i \lim_{N,
M_l^2}\sum_{l=0}^{N}c_l^N\intdk\frac{1}{[(k-xp)^2+M^2+M_l^2]^2}
\end{eqnarray}
which becomes a well defined integral, so that we can safely shift
the integration variable:
\begin{eqnarray}
L^{\prime R}=\int_0^1dx\lim_{N,
M_l^2}\sum_{l=0}^{N}c_l^N\intdk\frac{1}{(k^2+M^2+M_l^2)^2} \equiv
L^R
\end{eqnarray}
which explicitly shown that in loop regularization method, one can
safely shift the integration variables and express all the Feynman
integrals in terms of ILIs before applying for the regularization
prescription. In fact, it was found from the calculation of triangle
anomaly that even for the linear divergent integral, only when
firstly making a shift of integral variable, which then allows one
to eliminate the ambiguities and leads to a consistent result. The
reason is simple that loop regularization is translational
invariant.

\section{Checking Ward-Takahaski-Slavnov-Taylor identities with explicit calculations of
Renormalization Constants and $\beta$ function}

With the above analyzes, we are in the position to calculate the
renormalization constants of Non-Abelian gauge theory at one loop
level by using the loop regularization method. More details can be
found in Appendix C where we evaluate all the one-loop divergent
diagrams in terms of the explicit forms of ILIs.

\subsection{ Renormalization constant for fermion fields strength}

As there is only one diagram which contributes the one-loop
renormalization for the fermion fields strength, the divergent part
of this diagram has been evaluated in detail in the Appendix C and
explicitly given in terms of the ILIs. Here we only write down the
regularized divergent part for the purpose of defining the relevant
renormalization constant
\begin{eqnarray}
L(2)_{div}&=&(-g^2C_2)\int_0^1dx_1\{[x_1(3x_1-4)(\xi-1)-2x_1]p{\sla}+[2x_1(\xi-1)+4]m\}I_0^R
\end{eqnarray}
The explicit form of $I_0^R$ is given in loop regularization by the
following form
\begin{eqnarray}
I_0^R&=&\frac{i}{16\pi^2}[ln\frac{M_c^2}{\mu^2}-\gamma_\omega+y_0(\frac{\mu^2}{M_c^2})]
\end{eqnarray}

The next step is to introduce appropriate renormalization conditions
to make a suitable subtraction. Namely we shall find a prescription
to divide the Feynman integral into divergent part and finite part,
and cancel the divergent part by the counterterms. Such a
prescription will fix the renormalization constants uniquely. Many
different ways to introduce the renormalization conditions have been
put forward in literature, they are referred as various
renormalization schemes, such as: On-Shell renormalization scheme,
Momentum Subtraction scheme, Minimal Subtraction scheme, and so on.
Different renormalization schemes will lead to different definitions
of the renormalized parameters. Nevertheless, the physics content of
the theory, i.e. the renormalized S matrix elements, should not
depend on the choices of renormalization schemes\cite{Collins}.

As is well-known, no matter under which renormalization schemes, it
is inevitable to involve a mass dimensional parameter into the
original theory, even though the original theory contains only
dimensionless parameters. For example, in Momentum Subtraction
scheme, one needs set the reference momentum point for subtraction,
and in Minimal Subtraction scheme one has to introduce a mass
dimensional parameter $\mu$. In fact, this is the essential reason
of the dimension transmutation\cite{dimtrans}. Any choice for the
involved parameter is as good as any other, the physics should be
invariant under the transformations which merely change this
parameter. This is actually the consequence of renormalization
group. Such a mass dimensional parameter plays the role of
physically interesting sliding energy scale.

To remove the infinities, it needs to specify the subtraction
scheme. In the loop regularization method, we may adopt, for
simplicity, a subtraction scheme similar to the Modified Minimal
Subtraction scheme in dimensional regularization. Notice that the
arbitrary mass parameter $\mu_s$ plays the role of the sliding
energy scale, one may rewrite $I_0^R$ as follows
\begin{eqnarray}
I_0^R=\frac{i}{16\pi^2}[ ln\frac{M_c^2}{\mu_s^2}
 -\gamma_\omega ] +\frac{i}{16\pi^2}[ln\frac{\mu_s^2}{\mu^2}
+y_0(\frac{\mu^2}{M_c^2})]\label{I0R}
\end{eqnarray}
Since the term $y_0$ approaches to zero $y_0\to 0$ in the limit
$M_c\to \infty$. For the massless case with on mass shell condition,
we have $\mu^2 = \mu_s^2$ and $\ln \mu_s^2/\mu^2 = 0$. Thus the
substraction scheme is chosen so that the terms proportional to
$\frac{i}{16\pi^2}(\ln\frac{M_c^2}{\mu_s^2}-\gamma_\omega)$ in the
Feynman integral are canceled by the introduction of counterterms.
As such a term doesn't depend on the Feynman parameter $x_1$, one
can integrate $x_1$ easily. Final results are given by:
\begin{eqnarray}
L(2)_{div}&=&(-g^2C_2)\int_0^1dx_1\{[x_1(3x_1-4)(\xi-1)-2x_1]p{\sla}
+[2x_1(\xi-1)+4]m\}\frac{i}{16\pi^2}(\ln\frac{M_c^2}{\mu_s^2}-\gamma_\omega)\nonumber\\
&=&\frac{-ig^2}{8\pi^2}C_2[(-\xi)p{\sla}+(\xi+3)m]\frac{1}{2}(\ln\frac{M_c^2}{\mu_s^2}-\gamma_\omega)
\end{eqnarray}
From the condition $i(z_2-1)p\sla+L(2)_{div}=0$, we then obtain the
renormalization constant $z_2$:
\begin{eqnarray}
z_2=1-\frac{g^2}{8\pi^2}C_2\xi\frac{1}{2}(ln\frac{M_c^2}{\mu_s^2}-\gamma_\omega)
\end{eqnarray}


\subsection{Renormalization constant for gluon fields strength }

Four diagrams can contribute to the $A_\mu^a$'s renormalization as
shown in Fig.1. These four diagrams have explicitly been evaluated
in \cite{LR1,LR2} with the result:
\begin{eqnarray}
L^{ab}_{R\mu\nu}&=&g^2\delta^{ab}(p^2g_{\mu\nu}-p_{\mu}p_{\nu})
{\int_0^1dx}\{C_1[1+4x(1-x)+\frac{1}{2}(1-\xi)]I_0^R\nonumber\\
&
&-N_fT_28x(1-x)I_0^R(m)-4C_1(1-\xi)[1-\frac{1}{8}(1-\xi)]x(1-x)p^2I_{-2}^R\}
\end{eqnarray}
where $I_0^R$ is the renormalized divergent ILIs and given by
Eq.\ref{I0R} in the loop regularization. Thus the purely
renormalized divergent term turns out to have the following form:
\begin{eqnarray}
L^{ab}_{\mu\nu;div}&=&g^2\delta^{ab}(p^2g_{\mu\nu}-p_{\mu}p_{\nu}){\int_0^1dx}\{C_1[1+4x(1-x)+\frac{1}{2}(1-\xi)]
\frac{i}{16\pi^2}(ln\frac{M_c^2}{\mu_s^2}-\gamma_\omega)\nonumber\\
&
&-N_fT_28x(1-x)\frac{i}{16\pi^2}(ln\frac{M_c^2}{\mu_s^2}-\gamma_\omega)\}\nonumber\\
&=&i\{\frac{g^2}{16\pi^2}(\frac{13}{3}-\xi)C_1\frac{1}{2}(ln\frac{M_c^2}{\mu_s^2}-\gamma_\omega)-\frac{g^2}{6\pi^2}N_fT_2\frac{1}{2}(ln\frac{M_c^2}{\mu_s^2}-\gamma_\omega)\}\delta^{ab}(p^2g_{\mu\nu}-p_{\mu}p_{\nu})
\end{eqnarray}
The above divergent term can be canceled by introducing the
counterterm
\begin{eqnarray}
i(z_3-1)\delta^{ab}(p^2g_{\mu\nu}-p_{\mu}p_{\nu})=L^{ab}_{\mu\nu;div}\nonumber
\end{eqnarray}
with the renormalization constant $z_3$
\begin{eqnarray}
z_3=1+\frac{g^2}{16\pi^2}\left[(\frac{13}{3}-\xi)C_1-\frac{g^2}{6\pi^2}N_fT_2
\right] \frac{1}{2}(ln\frac{M_c^2}{\mu_s^2}-\gamma_\omega)
\end{eqnarray}


\subsection{Ghost self-energy diagram and renormalization of ghost fields}

There is only one diagram (fig.3) which contributes to the one-loop
renormalization for the ghost fields strength. The divergent part of
this diagram is evaluated in the Appendix C and reads in terms of
the renormalized divergent ILIs as follows
\begin{eqnarray}
L(3)^{cd}_{div}&=&-C_1g^2\delta^{cd}\int_0^1dx{\intdk}(x-(1-\frac{3}{2}x)(\xi-1))p^2I_0^R
\end{eqnarray}
Using Eq.(\ref{I0R}) and noticing that the subtracting divergent
term $\frac{i}{16\pi^2}ln\frac{M_c^2}{\mu_s^2}$ is independent of
the Feynman parameter $x$, we have
\begin{eqnarray}
L(3)^{cd}_{div}=\frac{ig^2}{16\pi^2}(\frac{1}{2}\xi-\frac{3}{2})C_1\delta^{cd}p^2\frac{1}{2}(ln\frac{M_c^2}{\mu_s^2}-\gamma_\omega)
\end{eqnarray}
The counterterm should satisfy the condition
\begin{eqnarray}
i(\tilde{z}_3-1)p^2\delta^{cd}+L(3)^{cd}_{div}=0
\end{eqnarray}
which leads the renormaliztion constant $\tilde{z_3}$ to be
\begin{eqnarray}
\tilde{z_3}&=&1+\frac{g^2}{16\pi^2}C_1(\frac{3}{2}-\frac{\xi}{2})\frac{1}{2}(ln\frac{M_c^2}{\mu_s^2}-\gamma_\omega)
\end{eqnarray}


\subsection{Fermion-gluon vertex renormalization}

Two kind of diagrams including their permutation (fig.4) contribute
to the one-loop renormalization for the fermion-gluon vertex. They
are explicitly evaluated in the Appendix C, the divergent parts are
given in terms of the renormalized divergent ILIs as follows
\begin{eqnarray}
L(4a)^{aR}_{\mu;div}&=&g^3(C_2-\frac{1}{2}C_1)T^a\int_0^1dx_1\int_0^{x_1}dx_2[2+6(1-x_1)(\xi-1)]\gamma_{\mu}I_0^R(M_{4a})\\
L(4b)^{aR}_{\mu;div}&=&g^3C_1T^a\gamma_{\mu}\int_0^1dx_1\int_0^{x_1}dx_2[3+\frac{9}{4}x_1(\xi-1)]I_0^R(M_{4b})
\end{eqnarray}
The corresponding subtracting divergent terms are found to be
\begin{eqnarray}
L(4a)^{aR}_{\mu;div}&=&\frac{ig^3}{8\pi^2}(C_2-\frac{1}{2}C_1)\xi\gamma_{\mu}T^a\frac{1}{2}(ln\frac{M_c^2}{\mu_s^2}-\gamma_\omega)\\
L(4b)^{aR}_{\mu;div}&=&\frac{ig^3}{8\pi^2}\frac{3}{4}(\xi+1)C_1T^a\gamma_{\mu}\frac{1}{2}(ln\frac{M_c^2}{\mu_s^2}-\gamma_\omega)
\end{eqnarray}

The total contribution is given
\begin{eqnarray}
L(4)^{aR}_{\mu;div}&=&L(4a)^{aR}_{\mu;div}+L(4b)^{aR}_{\mu;div}\nonumber\\
&=&\frac{ig^3}{8\pi^2}[(\frac{3}{4}+\frac{1}{4}\xi)C_1+{\xi}C_2]T^a\gamma_{\mu}\frac{1}{2}(ln\frac{M_c^2}{\mu_s^2}-\gamma_\omega)
\end{eqnarray}
From the renormaliztion condition
$(z_{1F}-1)igT^a\gamma_\mu+L(4)^{aR}_{\mu;div}=0$, the
renormalization constant $z_{1F}$ reads:
\begin{eqnarray}
z_{1F}&=&1-\frac{g^2}{8\pi^2}\left[(\frac{3}{4}+\frac{\xi}{4})C_1+{\xi}C_2
\right] \frac{1}{2}(ln\frac{M_c^2}{\mu_s^2}-\gamma_\omega)
\end{eqnarray}


\subsection{Ghost-gluon vertex renormalization}

For the one-loop renormalization of three-gluon vertex, there are
two diagrams including their permutation (fig.5). Their explicit
evaluation is presented in the Appendix C. The divergent parts are
given in terms of the renormalized divergent ILIs as follows
\begin{eqnarray}
L(5a)^{acb}_{\mu;div}&=&-\frac{i}{2}g^3C_1f^{acb}\int_0^1dx_1\int_0^{x_1}dx_2\frac{1}{2}{\xi}p_{2\mu}I_0^R(M_{5a})\\
L(5b)^{acb}_{\mu;div}&=&-\frac{3i}{4}g^3C_1f^{acb}\int_0^1dx_1\int_0^{x_1}dx_2(3(x_1-x_2)(\xi-1)+1)p_{2\mu}I_0^R(M_{5b})
\end{eqnarray}
The corresponding subtracting divergent terms are given by
integrating over Feynman parameters $x_1$, $x_2$
\begin{eqnarray}
L(5a)^{acb}_{\mu;div}&=&\frac{g^3}{16\pi^2}\frac{1}{4} \xi C_1f^{acb}p_{2\mu}\frac{1}{2}(ln\frac{M_c^2}{\mu_s^2}-\gamma_\omega)\\
L(5b)^{acb}_{\mu;div}&=&\frac{g^3}{16\pi^2}\frac{3}{4}{\xi}C_1f^{acb}p_{2\mu}\frac{1}{2}(ln\frac{M_c^2}{\mu_s^2}-\gamma_\omega)
\end{eqnarray}
with the final result
\begin{eqnarray}
L(5)^{acb}_{\mu;div}&=&L(5a)^{acb}_{\mu;div}+L(5b)^{acb}_{\mu;div}\nonumber\\
&=&\frac{g^3}{16\pi^2}{\xi}C_1f^{acb}p_{2\mu}\frac{1}{2}(ln\frac{M_c^2}{\mu_s^2}-\gamma_\omega)
\end{eqnarray}
From the renormalization condition
$(\tilde{z_1}-1)gf^{acb}p_{2\mu}+L(5)^{acb}_{\mu;div}=0$, the
renormalization constant $\tilde{z_1}$ is given by:
\begin{eqnarray}
\tilde{z_1}=1-\frac{g^2}{16\pi^2}{\xi}C_1\frac{1}{2}(ln\frac{M_c^2}{\mu_s^2}-\gamma_\omega)
\end{eqnarray}


\subsection{Three-gluon vertex renormalization}

Four loop diagrams including their permutation graphs will
contribute to the one-loop renormalization of three-gluon vertex.
More detailed evaluation is presented in the Appendix C, the
divergent parts in terms of the renormalized divergent ILIs read
\begin{eqnarray}
L(6a)_{\mu\nu\lambda;div}^{abc}&=&2ig^3f^{abc}T_2{\int_0^1dx_1}{\int_0^{x_1}dx_2}I_0^R(M_{6a})[4(-x_1+x_2+1)(\frac{1}{2}k_{2\mu}g_{\nu\lambda}+\nonumber\\
&
&+\frac{1}{2}k_{2\nu}g_{\mu\lambda}+k_{2\lambda}g_{\mu\nu}-k_{2\mu}g_{\nu\lambda}-k_{2\nu}g_{\mu\lambda}-\frac{1}{2}k_{2\lambda}g_{\mu\nu})+x_2(\frac{1}{2}k_{3\mu}g_{\nu\lambda}+\nonumber\\
&
&+\frac{1}{2}k_{3\nu}g_{\mu\lambda}+k_{3\lambda}g_{\mu\nu}-k_{3\mu}g_{\nu\lambda}-k_{3\nu}g_{\mu\lambda}-\frac{1}{2}k_{3\lambda}g_{\mu\nu})+\nonumber\\
&
&+(x_2-x_1)(\frac{1}{2}k_{2\nu}g_{\lambda\mu}+\frac{1}{2}k_{2\lambda}g_{\nu\mu}+k_{2\mu}g_{\nu\lambda}-k_{2\nu}g_{\lambda\mu}-k_{2\lambda}g_{\nu\mu}-\frac{1}{2}k_{2\mu}g_{\nu\lambda})+\nonumber\\
&
&+x_2(\frac{1}{2}k_{3\nu}g_{\lambda\mu}+\frac{1}{2}k_{3\lambda}g_{\nu\mu}+k_{3\mu}g_{\nu\lambda}-k_{3\nu}g_{\lambda\mu}-k_{3\lambda}g_{\nu\mu}-\frac{1}{2}k_{3\mu}g_{\nu\lambda})+\nonumber\\
&
&+(x_2-x_1)(\frac{1}{2}k_{2\lambda}g_{\mu\nu}+\frac{1}{2}k_{2\mu}g_{\lambda\nu}+k_{2\nu}g_{\lambda\mu}-k_{2\lambda}g_{\mu\nu}-k_{2\mu}g_{\lambda\nu}-\frac{1}{2}k_{\nu}g_{\lambda\mu})+\nonumber\\
&
&+(x_2-1)(\frac{1}{2}k_{3\lambda}g_{\mu\nu}+\frac{1}{2}k_{3\mu}g_{\lambda\nu}+k_{3\nu}g_{\lambda\mu}-k_{3\lambda}g_{\mu\nu}-k_{3\mu}g_{\lambda\nu}-\frac{1}{2}k_{3\nu}g_{\lambda\mu})]\\
L(6b)_{\mu\nu\lambda;div}^{abc}&=&2ig^3f^{anm}f^{bmp}f^{cpn}{\int_0^1dx_1}{\int_0^{x_1}dx_2}I_0^R(M_{6b})\nonumber\\
&
&\{[-(1-x_1)k_2+x_2k_3]_\lambda{\frac{1}{4}g_{\mu\nu}}+(x_1k_2+x_2k_3)_\nu{\frac{1}{4}g_{\mu\lambda}}+[-(1-x_1)k_2-(1-x_2)k_3]_\mu{\frac{1}{4}g_{\nu\lambda}}\}\nonumber\\
& &+(k_2{\rightarrow}k_3,\nu\rightarrow\lambda,b{\rightarrow}c)\\
L(6c)_{\mu\nu\lambda;div}^{abc}&=&-iC_1g^3f^{abc}\int_0^1dx_1\int_0^{x_1}dx_2I_0^R(M_{6c})\nonumber\\
& & \times
\{(\frac{1}{4}g^\alpha\gamma{g}_{\lambda\nu}-\frac{2}{4}g^\alpha_\lambda{g}_{\nu\gamma}
-\frac{1}{4}g_{\gamma\lambda}g^\alpha_\nu-\frac{1}{4}g^\alpha_\nu{g}_{\lambda\gamma}
-\frac{2}{4}g_{\nu\gamma}g^\alpha_\lambda+\frac{4}{4}g_{\nu\lambda}g^\alpha_\gamma
+ g^\alpha_\nu{g}_{\gamma\lambda})\nonumber\\
& &\times
[((-1-x_1)k_2+(-1-x_2)k_3)^\gamma{g_{\mu\alpha}}+((-1+2x_1)k_2+(-1+2x_2)k_3)_\mu{g_\alpha^\gamma}
+((2-x_1)k_2 \nonumber\\
& & +(2-x_2)k_3)_\alpha{g^\gamma_\mu}]+
(-\frac{2}{4}g_{\alpha\lambda}g^\rho_\mu-\frac{1}{4}g_{\alpha\mu}g^\rho_\lambda
+\frac{4}{4}g_{\mu\lambda}g^\rho_\alpha+\frac{1}{4}g^\rho_\alpha{g}_{\mu\lambda}
-\frac{1}{4}g^\rho_\lambda{g}_{\mu\alpha}-\frac{2}{4}g^\rho_\mu{g}_{\alpha\lambda}
+g^\rho_\lambda{g}_{\alpha\mu})\nonumber\\
& &\times [((2-x_1)k_2-x_2k_3)^\alpha{g_{\nu\rho}}+((-1+2x_1)k_2+2x_2k_3)_\nu{g_\rho^\alpha}
+((-1-x_1)k_2-x_2k_3)_\rho{g^\alpha_\nu}]\nonumber\\
& & + (-\frac{1}{4}g^\gamma_\mu{g}_{\nu\rho}+g^\gamma_\mu{g}_{\nu\rho}-\frac{2}{4}g^\gamma_\nu{g}{\mu\rho}
+\frac{4}{4}g_{\mu\nu}g^\gamma_\rho+\frac{1}{4}g^\gamma_\rho{g}_{\mu\nu}-\frac{2}{4}g_{\mu\rho}g^\gamma_\nu
-\frac{1}{4}g_{\nu\rho}g^\gamma_\mu)\nonumber\\
& & \times [((1-x_1)k_2+(2-x_2)k_3)^\rho{g_{\lambda\gamma}}
+((-2+2x_1)k_2+(-1+2x_2)k_3)_\lambda{g_\gamma^\rho} \nonumber \\
& & +((1-x_1)k_2+(-1-x_2)k_3)_\gamma{g^\rho_\lambda}]\}\\
L(6d)_{\mu\nu\lambda;div}^{abc}&=&\frac{3i}{4}g^3C_1f^{abc}(g_{\nu\rho}g_{\lambda\sigma}-g_{\rho\lambda}
g_{\sigma\nu})\int_0^1dx_1I_0^R(M_{6d})[(1+x_1)k_1^\sigma{g}^\rho_\mu+(1-2x_1)k_{1\mu}g^{\rho\sigma}+(-2+x_1)k_1^\rho{g}^\sigma_\mu]\nonumber\\
& &+permutation\ graphs
\end{eqnarray}
The corresponding subtracting divergent terms are simply obtained by
integrating over the Feynman parameters $x_1$, $x_2$
\begin{eqnarray}
L(6a)_{\mu\nu\lambda;div}^{abc}&=&-\frac{4}{3}ig^2T_2\frac{i}{16\pi^2}(ln\frac{M_c^2}{\mu_s^2}-\gamma_\omega)
gf^{abc}[g_{\mu\nu}(k_{1\lambda}-k_{2\lambda})+g{\nu\lambda}(k_{2\mu}-k_{3\mu})+g{\lambda\mu}(k_{3\nu}-k_{1\nu})]\\
L(6b)_{\mu\nu\lambda;div}^{abc}&=&\frac{i}{24}g^2C_1\frac{i}{16\pi^2}(ln\frac{M_c^2}{\mu_s^2}-\gamma_\omega)
gf^{abc}[g_{\mu\nu}(k_1-k_2)_\lambda+g_{\nu\lambda}(k_2-k_3)_\mu+g_{\lambda\mu}(k_3-k_1)_\nu]\\
L(6c)_{\mu\nu\lambda;div}^{abc}&=&-\frac{13i}{8}C_1g^2\frac{i}{16\pi^2}(ln\frac{M_c^2}{\mu_s^2}-\gamma_\omega)
gf^{abc}[g_{\mu\nu}(k_1-k_2)_\lambda+g_{\nu\lambda}(k_2-k_3)_\mu+g_{\lambda\mu}(k_3-k_1)_\nu]\\
L(6d)_{\mu\nu\lambda;div}^{abc}&=&\frac{9i}{4}C_1g^2\frac{i}{16\pi^2}(ln\frac{M_c^2}{\mu_s^2}-\gamma_\omega)
gf^{abc}[g_{\mu\nu}(k_1-k_2)_\lambda+g_{\nu\lambda}(k_2-k_3)_\mu+g_{\lambda\mu}(k_3-k_1)_\nu]
\end{eqnarray}
with the total result being given by summing over the four diagrams
including their permutation graphs
\begin{eqnarray}
L(6)_{\mu\nu\lambda;div}^{abc}&=&N_fL(6a)_{\mu\nu\lambda;div}^{abc}+L(6b)_{\mu\nu\lambda;div}^{abc}+L(6c)_{\mu\nu\lambda;div}^{abc}+L(6d)_{\mu\nu\lambda;div}^{abc}\nonumber\\
&=&[(\frac{2}{3}ig^2C_1-\frac{4}{3}ig^2N_fT_2)\frac{i}{16\pi^2}(ln\frac{M_c^2}{\mu_s^2}-\gamma_\omega)]gf^{abc}[g_{\mu\nu}(k_{1\lambda}-k_{2\lambda})+g_{\nu\lambda}(k_{2\mu}-k_{3\mu})+g_{\lambda\mu}(k_{3\nu}-k_{1\nu})]\nonumber\\
\end{eqnarray}

Using the renormalization condition
\begin{eqnarray}
(z_1-1)gf^{abc}[g_{\mu\nu}(k_{1\lambda}-k_{2\lambda})+g_{\nu\lambda}(k_{2\mu}-k_{3\mu})+g_{\lambda\mu}(k_{3\nu}-k_{1\nu})]+L(6)_{\mu\nu\lambda;div}^{abc}=0
\end{eqnarray}
we obtain the renormalization constant $z_1$ in Feynman gauge $\xi
=1$ to be
\begin{eqnarray}
z_1=1+(\frac{g^2}{12\pi^2}C_1-\frac{g^2}{6\pi^2}N_fT_2)\frac{1}{2}(ln\frac{M_c^2}{\mu_s^2}-\gamma_\omega)
\end{eqnarray}
The evaluation in the $\xi$ gauge is rather length, the result is
\begin{eqnarray}
z_1=1+\left[\frac{g^2}{12\pi^2}[1 + \frac{9}{8} (1 -
\xi)]C_1-\frac{g^2}{6\pi^2}N_fT_2 \right]
\frac{1}{2}(ln\frac{M_c^2}{\mu_s^2}-\gamma_\omega)
\end{eqnarray}


\subsection{Four-gluon vertex renormalization}

We finally consider the four-gluon vertex renormalization, there are
five loop diagrams which contribute to its renormalizaion. The
detailed evaluation can be found in the Appendix C, we present here
only the divergent parts in terms of the renormalized divergent ILIs
\begin{eqnarray}
L(7a)_{\mu\nu\lambda\rho;div}^{abcd}&=&6g^4f^{aef}f^{bfj}f^{cjn}f^{dne}\int_0^1dx_1\int_0^{x_1}dx_2\int_0^{x_2}dx_3[\frac{5}{2}(g_{\mu\nu}g_{\lambda\rho}+g_{\mu\rho}g_{\nu\lambda})+\nonumber\\
&
&\frac{34}{24}(g_{\mu\nu}g_{\lambda\rho}+g_{\mu\lambda}g_{\nu\rho}+g_{\mu\rho}g_{\nu\lambda})]I_{0}^R(M_{7a})+2\
permutations\\
L(7b)_{\mu\nu\lambda\rho;div}^{abcd}&=&2g^4f^{aef}f^{dme}
[f^{lfb}f^{lcm}(g^{\beta}_{\lambda}g_{\nu}^{\chi}-g^{\beta\chi}g_{\nu\lambda})+f^{lfc}f^{lmb}(g^{\beta\chi}g_{\lambda\nu}-g^{\beta}_{\nu}g_{\lambda}^{\chi})+f^{lfm}f^{lbc}(g^{\beta}_{\nu}g^{\chi}_{\lambda}-g^{\beta}_{\lambda}g^{\chi}_{\nu})]\nonumber\\
&
&\int_0^1\int_0^{x_1}dx_1dx_2(g_{\beta\mu}g_{\rho\chi}-\frac{1}{4}g_{\beta\mu}g_{\rho\chi}-\frac{2}{4}g_{\beta\rho}
g_{\mu\chi}+\frac{4}{4}g_{\mu\rho}g_{\beta\chi}+\frac{1}{4}g_{\beta\chi}g_{\mu\rho}-\frac{2}{4}g_{\mu\chi}
g_{\beta\rho}-\frac{1}{4}g_{\rho\chi}g_{\mu\beta})I_0^R(M_{7b})\nonumber\\
& &+5\ permutations\\
L(7c)_{\mu\nu\lambda\rho;div}^{abcd}&=&\frac{1}{2}g^4[f^{eai}f^{ejd}(g_{\mu\beta}g_{\alpha\rho}-g_{\mu\rho}g_{\alpha\beta})+f^{eaj}f^{edi}(g_{\mu\rho}g_{\beta\alpha}-g_{\mu\alpha}g_{\beta\rho})+f^{ead}f^{eij}(g_{\mu\alpha}g_{\rho\beta}-g_{\mu\beta}g_{\rho\alpha})]\times\nonumber\\
& &[f^{fib}f^{fcj}(g^{\alpha}_{\lambda}g_{\nu}^{\beta}-g^{\alpha\beta}g_{\nu\lambda})+f^{fic}f^{fjb}(g^{\alpha\beta}g_{\lambda\nu}-g^{\alpha}_{\nu}g_{\lambda}^{\beta})+f^{fij}f^{fbc}(g^{\alpha}_{\nu}g^{\beta}_{\lambda}-g^{\alpha}_{\lambda}g^{\beta}_{\nu})]\nonumber\\
& &\int_0^1dx_1I_0^R(M_{7c})+2\ permutations\\
L(7d)_{\mu\nu\lambda\rho;div}^{abcd}&=&-\frac{1}{4}g^4f^{aie}f^{bmi}f^{cpm}f^{dep}(g_{\mu\nu}g_{\lambda\rho}
+g_{\mu\lambda}g_{\nu\rho}+g_{\mu\rho}g_{\nu\lambda})\int_0^1dx_1\int_0^{x_1}dx_2\int_0^{x_2}dx_3I_{0}^R(M_{7d})+\nonumber\\
& &+5\ permutations\\
L(7e)_{\mu\nu\lambda\rho;div}^{abcd}&=&-8g^4Tr(T^{a}T^{d}T^{c}T^{b})(g_{\mu\nu}g_{\lambda\rho}-2g_{\mu\lambda}
g_{\nu\rho}+g_{\mu\rho}g_{\nu\lambda})\int_0^1dx_1\int_0^{x_1}dx_2\int_0^{x_2}dx_3I_0^R(M_{7e})+\nonumber\\
& & 5\ permutations
\end{eqnarray}
The corresponding subtracting divergent terms are yielded by
integrating over the Feynman parameters $x_1$, $x_2$, $x_3$
\begin{eqnarray}
L(7a)_{\mu\nu\lambda\rho;div}^{abcd}&=&g^4\frac{i}{16\pi^2}(ln\frac{M_c^2}{\mu_s^2}-\gamma_\omega)[g_{\mu\nu}g_{\lambda\rho}(\frac{47}{12}F^{abcd}+\frac{17}{12}F^{acbd}+\frac{47}{12}F^{abdc})+\nonumber\\
& &g_{\mu\lambda}g_{\nu\rho}(\frac{17}{12}F^{abcd}+\frac{47}{12}F^{acbd}+\frac{47}{12}F^{abdc})+g_{\mu\rho}g_{\nu\lambda}(\frac{47}{12}F^{abcd}+\frac{47}{12}F^{acbd}+\frac{17}{12}F^{abdc})]\\
L(7b)_{\mu\nu\lambda\rho;div}^{abcd}&=&g^4\frac{i}{16\pi^2}(ln\frac{M_c^2}{\mu_s^2}-\gamma_\omega)[g_{\mu\nu}g_{\lambda\rho}(-\frac{17}{2}F^{abcd}+2F^{acbd}-\frac{17}{2}F^{abdc}-\frac{3}{2}C_1f^{lad}f^{lbc}-\frac{3}{2}C_1f^{lac}f^{lbd})+\nonumber\\
&
&g_{\mu\lambda}g_{\nu\rho}(2F^{abcd}-\frac{17}{2}F^{acbd}-\frac{17}{2}F^{abdc}+\frac{3}{2}C_1f^{lad}f^{lbc}-\frac{3}{2}C_1f^{lab}f^{lcd})+\nonumber\\
&
&g_{\mu\rho}g_{\nu\lambda}(-\frac{17}{2}F^{abcd}-\frac{17}{2}F^{acbd}+2F^{abdc}+\frac{3}{2}C_1f^{lac}f^{lbd}+\frac{3}{2}C_1f^{lab}f^{lcd})]\\
L(7c)_{\mu\nu\lambda\rho;div}^{abcd}&=&g^4\frac{i}{16\pi^2}(ln\frac{M_c^2}{\mu_s^2}-\gamma_\omega)[g_{\mu\nu}g_{\lambda\rho}(2C_1f^{ead}f^{ebc}+2C_1f^{eac}f^{ebd}+3F^{abcd}+3F^{abdc})+\nonumber\\
& &g_{\mu\lambda}g_{\nu\rho}(2C_1f^{eab}f^{ecd}-2C_1f^{ead}f^{ebc}+3F^{abdc}+3F^{acbd})+\nonumber\\
& &g_{\mu\rho}g_{\nu\lambda}(-2C_1f^{eab}f^{ecd}-2C_1f^{eac}f^{ebd}+3F^{abcd}+3F^{acbd})\nonumber\\
L(7d)_{\mu\nu\lambda\rho;div}^{abcd}&=&g^4\frac{i}{16\pi^2}(ln\frac{M_c^2}{\mu_s^2}-\gamma_\omega)[g_{\mu\nu}g_{\lambda\rho}(-\frac{1}{12}F^{abcd}-\frac{1}{12}F^{acbd}-\frac{1}{12}F^{abdc})+\nonumber\\
& &g_{\mu\lambda}g_{\nu\rho}(-\frac{1}{12}F^{abcd}-\frac{1}{12}F^{acbd}-\frac{1}{12}F^{abdc})+g_{\mu\rho}g_{\nu\lambda}(-\frac{1}{12}F^{abcd}-\frac{1}{12}F^{acbd}-\frac{1}{12}F^{abdc})]\\
L(7e)_{\mu\nu\lambda\rho;div}^{abcd}&=&-\frac{4}{3}T_2g^4\frac{i}{16\pi^2}(ln\frac{M_c^2}{\mu_s^2}-\gamma_\omega)\nonumber\\
&
&\{g_{\mu\nu}g_{\lambda\rho}(f^{adl}f^{bcl}+f^{acl}f^{bdl})+g_{\mu\lambda}g_{\nu\rho}(f^{abl}f^{cdl}-f^{adl}f^{bcl})+g_{\mu\rho}g_{\nu\lambda}(-f^{abl}f^{cdl}-f^{acl}f^{bdl})\}
\end{eqnarray}
with $F^{abcd}{\equiv}f^{aef}f^{bfg}f^{cgh}f^{dhe}$. By adding those
five diagrams together, we have
\begin{eqnarray}
L(7)_{\mu\nu\lambda\rho;div}^{abcd}&=&L(7a)_{\mu\nu\lambda\rho;div}^{abcd}+L(7b)_{\mu\nu\lambda\rho;div}^{abcd}+L(7c)_{\mu\nu\lambda\rho;div}^{abcd}+L(7d)_{\mu\nu\lambda\rho;div}^{abcd}+N_fL(7e)_{\mu\nu\lambda\rho;div}^{abcd}\nonumber\\
&=&g^4\frac{i}{16\pi^2}(ln\frac{M_c^2}{\mu_s^2}-\gamma_\omega)[g_{\mu\nu}g_{\lambda\rho}(\frac{47}{12}F^{abcd}+\frac{17}{12}F^{acbd}+\frac{47}{12}F^{abdc})+\nonumber\\
&&g_{\mu\lambda}g_{\nu\rho}(\frac{17}{12}F^{abcd}+\frac{47}{12}F^{acbd}+\frac{47}{12}F^{abdc})+g_{\mu\rho}g_{\nu\lambda}(\frac{47}{12}F^{abcd}+\frac{47}{12}F^{acbd}+\frac{17}{12}F^{abdc})]\nonumber\\
&&+g^4\frac{i}{16\pi^2}(ln\frac{M_c^2}{\mu_s^2}-\gamma_\omega)[g_{\mu\nu}g_{\lambda\rho}(-\frac{17}{2}F^{abcd}+2F^{acbd}-\frac{17}{2}F^{abdc}-\frac{3}{2}C_1f^{lad}f^{lbc}-\frac{3}{2}C_1f^{lac}f^{lbd})+\nonumber\\
&&g_{\mu\lambda}g_{\nu\rho}(2F^{abcd}-\frac{17}{2}F^{acbd}-\frac{17}{2}F^{abdc}+\frac{3}{2}C_1f^{lad}f^{lbc}-\frac{3}{2}C_1f^{lab}f^{lcd})+\nonumber\\
&&g_{\mu\rho}g_{\nu\lambda}(-\frac{17}{2}F^{abcd}-\frac{17}{2}F^{acbd}+2F^{abdc}+\frac{3}{2}C_1f^{lac}f^{lbd}+\frac{3}{2}C_1f^{lab}f^{lcd})]\nonumber\\
&&+g^4\frac{i}{16\pi^2}(ln\frac{M_c^2}{\mu_s^2}-\gamma_\omega)[g_{\mu\nu}g_{\lambda\rho}(2C_1f^{ead}f^{ebc}+2C_1f^{eac}f^{ebd}+3F^{abcd}+3F^{abdc})+\nonumber\\
&&g_{\mu\lambda}g_{\nu\rho}(2C_1f^{eab}f^{ecd}-2C_1f^{ead}f^{ebc}+3F^{abdc}+3F^{acbd})+\nonumber\\
&&g_{\mu\rho}g_{\nu\lambda}(-2C_1f^{eab}f^{ecd}-2C_1f^{eac}f^{ebd}+3F^{abcd}+3F^{acbd})\nonumber\\
&&+g^4\frac{i}{16\pi^2}(ln\frac{M_c^2}{\mu_s^2}-\gamma_\omega)[g_{\mu\nu}g_{\lambda\rho}(-\frac{1}{12}F^{abcd}-\frac{1}{12}F^{acbd}-\frac{1}{12}F^{abdc})+\nonumber\\
& &g_{\mu\lambda}g_{\nu\rho}(-\frac{1}{12}F^{abcd}-\frac{1}{12}F^{acbd}-\frac{1}{12}F^{abdc})+g_{\mu\rho}g_{\nu\lambda}(-\frac{1}{12}F^{abcd}-\frac{1}{12}F^{acbd}-\frac{1}{12}F^{abdc})]\nonumber\\
&&-\frac{4}{3}N_fT_2g^4\frac{i}{16\pi^2}(ln\frac{M_c^2}{\mu_s^2}-\gamma_\omega)\nonumber\\
&&\{g_{\mu\nu}g_{\lambda\rho}(f^{adl}f^{bcl}+f^{acl}f^{bdl})+g_{\mu\lambda}g_{\nu\rho}(f^{abl}f^{cdl}-f^{adl}f^{bcl})+g_{\mu\rho}g_{\nu\lambda}(-f^{abl}f^{cdl}-f^{acl}f^{bdl})\}\nonumber\\
&=&g^4\frac{i}{16\pi^2}(ln\frac{M_c^2}{\mu_s^2}-\gamma_\omega)[g_{\mu\nu}g_{\lambda\rho}(\frac{1{}}2C_1f^{ead}f^{ebc}+\frac{1}{2}C_1f^{eac}f^{ebd}+\frac{-5}{3}F^{abcd}+\frac{10}{3}F^{acbd}+\frac{-5}{3}F^{abdc})+\nonumber\\
&&g_{\mu\lambda}g_{\nu\rho}(\frac{1}{2}C_1f^{eab}f^{ecd}-\frac{1}{2}C_1f^{ead}f^{ebc}+\frac{-5}{3}F^{abcd}+\frac{10}{3}F^{acbd}+\frac{-5}{3}F^{abdc})+\nonumber\\
&&g_{\mu\rho}g_{\nu\lambda}(-\frac{1}{2}C_1f^{eab}f^{ecd}-\frac{1}{2}C_1f^{eac}f^{ebd}+\frac{-5}{3}F^{abcd}+\frac{10}{3}F^{acbd}+\frac{-5}{3}F^{abdc})]\nonumber\\
&&-\frac{4}{3}N_fT_2g^4\frac{i}{16\pi^2}(ln\frac{M_c^2}{\mu_s^2}-\gamma_\omega)\nonumber\\
&&\{g_{\mu\nu}g_{\lambda\rho}(f^{adl}f^{bcl}+f^{acl}f^{bdl})+g_{\mu\lambda}g_{\nu\rho}(f^{abl}f^{cdl}-f^{adl}f^{bcl})+g_{\mu\rho}g_{\nu\lambda}(-f^{abl}f^{cdl}-f^{acl}f^{bdl})\}\nonumber\\
&=&[-\frac{1}{3}C_1-\frac{4}{3}N_fT_2]g^4\frac{i}{16\pi^2}(ln\frac{M_c^2}{\mu_s^2}-\gamma_\omega)\nonumber\\
&&\{g_{\mu\nu}g_{\lambda\rho}(f^{adl}f^{bcl}+f^{acl}f^{bdl})+g_{\mu\lambda}g_{\nu\rho}(f^{abl}f^{cdl}-f^{adl}f^{bcl})+g_{\mu\rho}g_{\nu\lambda}(-f^{abl}f^{cdl}-f^{acl}f^{bdl})\}
\end{eqnarray}
By applying for the renormalization condition
\begin{eqnarray}
& &
(z_4-1)(-ig^2)\{g_{\mu\nu}g_{\lambda\rho}(f^{adl}f^{bcl}+f^{acl}f^{bdl})
+g_{\mu\lambda}g_{\nu\rho}(f^{abl}f^{cdl}-f^{adl}f^{bcl})+g_{\mu\rho}g_{\nu\lambda}
(-f^{abl}f^{cdl}-f^{acl}f^{bdl})\} \nonumber
\\
& & +L(7)_{\mu\nu\lambda;div}^{abc}=0
\end{eqnarray}
we then obtain in the Feyman gauge $\xi = 1$ the renormalization
constant $z_4$
\begin{eqnarray}
z_4=1-(\frac{g^2}{24\pi^2}C_1+\frac{g^2}{6\pi^2}N_fT_2)\frac{1}{2}(ln\frac{M_c^2}{\mu_s^2}-\gamma_\omega)
\end{eqnarray}
A similar but length evaluation in the $\xi$ gauge leads to the
result
\begin{eqnarray}
z_4=1-\left[\frac{g^2}{24\pi^2}(1 + 3(\xi -1)
)C_1+\frac{g^2}{6\pi^2}N_fT_2\right]\frac{1}{2}(ln\frac{M_c^2}{\mu_s^2}-\gamma_\omega)
\end{eqnarray}


\subsection{Ward-Takahaski-Slavnov-Taylor identities and $\beta$ function}

   We shall summarize all the renormalization constants in this section
to check Ward-Takahaski-Slavnov-Taylor identities and calculate
$\beta$ function. All the results are listed as below:
\begin{eqnarray}
z_2&=&1-\frac{g^2}{8\pi^2}C_2\xi\frac{1}{2}(ln\frac{M_c^2}{\mu_s^2}-\gamma_\omega)\nonumber\\
z_3&=&1+[\frac{g^2}{16\pi^2}(\frac{13}{3}-\xi)C_1-\frac{g^2}{6\pi^2}N_fT_2]\frac{1}{2}
(ln\frac{M_c^2}{\mu_s^2}-\gamma_\omega)\nonumber\\
\tilde{z_3}&=&1+\frac{g^2}{16\pi^2}C_1(\frac{3}{2}-\frac{\xi}{2})\frac{1}{2}
(ln\frac{M_c^2}{\mu_s^2}-\gamma_\omega)\nonumber\\
z_{1F}&=&1-\frac{g^2}{8\pi^2}[(\frac{3}{4}+\frac{\xi}{4})C_1+{\xi}C_2]\frac{1}{2}
(ln\frac{M_c^2}{\mu_s^2}-\gamma_\omega)\nonumber\\
\tilde{z_1}&=&1-\frac{g^2}{16\pi^2}{\xi}C_1\frac{1}{2}(ln\frac{M_c^2}{\mu_s^2}-\gamma_\omega)\nonumber\\
z_1&=&1+[\frac{g^2}{12\pi^2}(\frac{17}{8} - \frac{9}{8}\xi) C_1-\frac{g^2}{6\pi^2}N_fT_2]
\frac{1}{2}(ln\frac{M_c^2}{\mu_s^2}-\gamma_\omega)\nonumber\\
z_4&=&1-[\frac{g^2}{24\pi^2}(-2+ 3\xi)C_1 +
\frac{g^2}{6\pi^2}N_fT_2]\frac{1}{2}(ln\frac{M_c^2}{\mu_s^2}-\gamma_\omega)\nonumber
\end{eqnarray}

It is straight forward to verify explicitly the
Ward-Takahaski-Slavnov-Taylor identities:
\begin{eqnarray}
z_g =
\frac{z_{1F}}{z_3^{1/2}z_2}=\frac{\tilde{z_1}}{z_3^{1/2}\tilde{z}_3}=\frac{z_1}{z_3^{3/2}}=\frac{z_4^{1/2}}{z_3}
\end{eqnarray}
which leads to the gauge independent renormalization constant for
the gauge coupling constant $g = z_g^{-1} g_0$
\begin{eqnarray}
z_g&=&1-(\frac{11}{48\pi^2}C_1-\frac{1}{12\pi^2}N_fT_2)g^2\frac{1}{2}(ln\frac{M_c^2}{\mu_s^2}-\gamma_\omega)
\end{eqnarray}

In the loop regularization method, the energy scale $\mu_s$ plays
the role of the sliding energy scale. According to the definition of
$\beta$ function, we obtain the one-loop $\beta$ function:
\begin{eqnarray}
\beta(g)&{\triangleq}&\lim_{M_c\to \infty} \mu_s\frac{\partial}{\partial\mu_s}g\mid_{g_0,m_0}\nonumber\\
&=& \lim_{M_c\to \infty} g\mu_s\frac{\partial}{\partial\mu_s}ln{z_g}\mid_{g_0,m_0}\nonumber\\
&{\simeq}&g\mu_s\frac{\partial}{\partial\mu_s}[(\frac{11}{48\pi^2}C_1-\frac{1}{12\pi^2}N_fT_2)g^2
\frac{1}{2}(ln\frac{M_c^2}{\mu_s^2}-\gamma_\omega)]\nonumber\\
&{\simeq}&g^3\mu_s(\frac{11}{48\pi^2}C_1-\frac{1}{12\pi^2}N_fT_2)\frac{-1}{\mu_s}\nonumber\\
&=&-\frac{g^3}{(4\pi)^2}(\frac{11}{3\pi^2}C_1-\frac{4}{3\pi^2}N_fT_2)
\end{eqnarray}
which agrees with the well-known result obtained by using dimesional
regularization. It is noticed that a simple corresponding for the
logarithmic divergences between the loop regularization method and
dimensional regularization scheme is
\begin{eqnarray}
\frac{2}{\varepsilon}\longleftrightarrow ln\frac{M_c^2}{\mu_s^2}
\end{eqnarray}
with $\varepsilon \to 0$ and $M_c\to \infty$.


\section{conclusion}

We have performed a complete calculation for all one loop diagrams
of non-Abelian gauge theory by using the loop regularization
method\cite{LR1,LR2} and provided an explicit check for the
consistency of loop regularization method from the
Ward-Takahaski-Slavnov-Taylor identities satisfied among the
renormalization constants. It has been shown that the loop
regularization method can lead to a consistent $\beta$ function.

From above explicit calculations, the conclusions stated in
\cite{LR1,LR2} become manifest that the loop regularization method
preserves not only non-Abelian gauge symmetry, but also Lorentz and
translational symmetries though the existence of two energy scales
$M_c$ and $\mu_s$ introduced intrinsically in this method. As the
scales $M_c$ and $\mu_s$ play the role of ultraviolet divergent
cutoff and infrared divergent cutoff respectively, the loop
regularization method can deal with both the ultraviolet and
infrared divergences. The existence of two energy scales also makes
the loop regularization to maintain the divergent behavior of
original theories, while the quadratic divergences in gauge theories
are found to cancel each other as the loop regularization preserves
gauge symmetry. Thus both loop regularization and dimensional
regularization lead to the same renormalization constants for gauge
theories with making a simple replacement between $\ln M_c/\mu_s$
and $1/\varepsilon$. The possible distinguishable properties between
loop regularization and dimensional regularization may occur for
treating chiral field theories with anomaly action concerning the
$\gamma_5$ matrix \cite{MW1,MW2}, and for deriving effective field
theories with dynamically generated spontaneous symmetry
breaking\cite{DW} as well as for applying to supersymmetric theories
involving the exact dimension\cite{CTW}. Finally, we would like to
point out that the renormalization scheme dependence is not involved
in our present consideration as our computation for the
renormalization constants is only at the one loop level and our
focus in this note is mainly on the check of
Ward-Takahaski-Slavnov-Taylor identities among the renormalization
constants. It is interesting to see that the loop regularization
method generally allows one to make on-shell renormalization
prescription due to the existence of the energy scale $\mu_s$ which
plays the role of infrared cutoff and sliding energy scale, such a
feature may provide a practical way for reducing the renormalization
scheme dependence, which is worthwhile to be further investigated
elsewhere.

\acknowledgments \label{ACK}

The authors would like to thank Einhorn Marty for valuable
discussions during the KITPC program. This work was supported in
part by the National Science Foundation of China (NSFC) under the
grant 10475105, 10491306 and the Project of Knowledge Innovation
Program (PKIP) of Chinese Academy of Sciences.

\appendix

\section{some useful formulae of compact simple Lie group}

In this appendix, we shall present some useful formulae about the
structure constants and the traces of the representation matrices
for Lie group.

For compact simple Lie group, one can choose the killing form of
group to be in proportion to a unit matrix, then the Lie algebras
satisfy the following identities:
\begin{eqnarray}
& &Tr(T^aT^b)=T_2\delta^{ab};\ \ [T^a,T^b]=if^{abc}T^c;\\
& &f^{abd}f^{dce}+f^{bcd}f^{dae}+f^{cad}f^{dbe}=0;\\
& &f^{abc}f^{dbc}=C_1\delta^{ad};\ \ T^aT^a=C_2I;\label{g const}
\end{eqnarray}
where $f^{abc}=-iT^{-1}_2tr(T^aT^bT^c-T^bT^aT^c)$ is completely
antisymmetric, and $T^a$ are group generators in fundamental
representation. Using the above relations, one can easily prove the
following relations: \
\[ f^{anm}f^{bmp}f^{cpn}=\frac{1}{2}C_1f^{abc}, \quad
T^aT^bT^a=(C_2-\frac{1}{2}C_1)T^b \]
\[
T^{abcd}+T^{abdc}+T^{acdb}+T^{adcb}-2T^{acbd}-2T^{adbc}=T_2(f^{adl}f^{bcl}+f^{acl}f^{bdl})\]
\[
F^{abcd}-2F^{acbd}+F^{abdc}=\frac{1}{2}C_1(f^{adl}f^{bcl}+f^{acl}f^{bdl})\]
where $T^{abcd}{\equiv}Tr(T^{a}T^{b}T^{c}T^{d})$ with
$T^{abcd}=T^{cycle}$, and
$F^{abcd}{\equiv}f^{aef}f^{bfg}f^{cgh}f^{dhe}$ with
$F^{abcd}=F^{cycle}=F^{inverse}$

\section{Feynman Rules for Gauge Theory}

\begin{eqnarray}
\mathcal{L}=\bar{\psi}_n(i\gamma^{\mu}D_{\mu}-m)\psi_n-\frac{1}{4}F^a_{\mu\nu}F^{a\mu\nu}-\frac{1}{2\xi}(\partial^{\mu}A_{\mu}^a)^2+\partial^{\mu}\bar{c}^a(\partial_{\mu}\delta^{ac}+gf^{abc}A_{\mu}^b)c^c\nonumber\\
\end{eqnarray}

\includegraphics[scale=0.6]{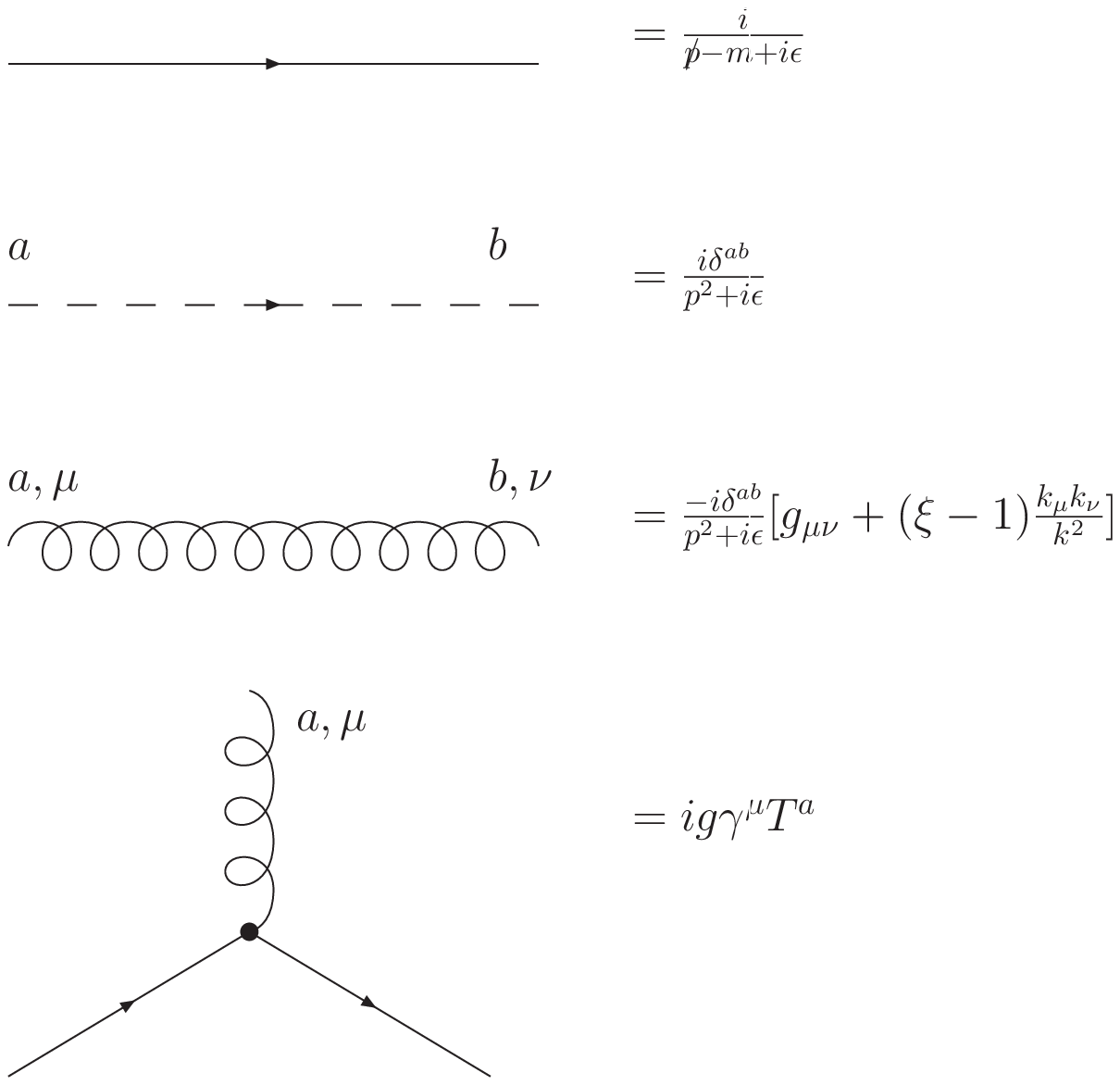}
\includegraphics[scale=0.5]{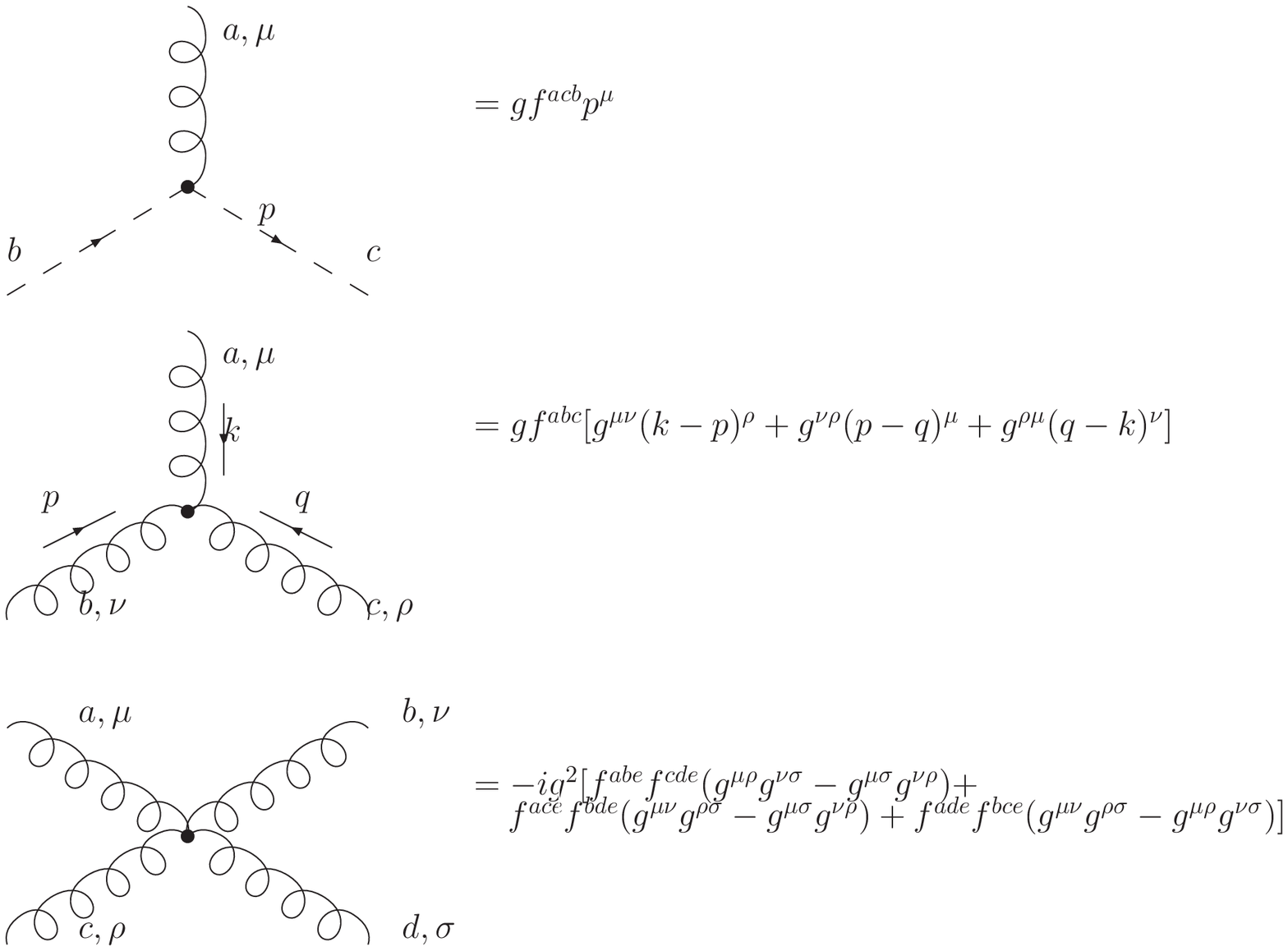}

\section{calculation of renormalization constants in loop regularization}

\subsection{fermion self-energy diagram}

There is only one diagram which contribute to the one-loop
renormalization for the fermion fields strength as shown in Fig.2.
\begin{center}
\includegraphics[scale=1]{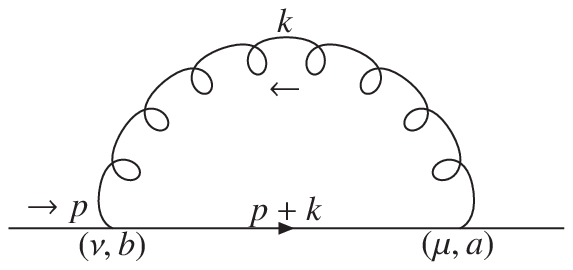}
\end{center}
\begin{center}{\sl Fig.2.}\end{center}
Following the Feynman rules given in Appendix B, we write down the
Feynman integral corresponding to this diagram (for simplification
we will ignore the $i\epsilon$ prescription in the propagators
throughout this paper, this dose not make any confusion if we keep
the prescription in mind):
\begin{eqnarray}
L(2)&=&\intdk(ig\gamma_{\mu}T_a)\frac{i}{(p\sla+k\sla)-m}\frac{-i\delta^{ab}}{k^2}
[g^{\mu\nu}+(\xi-1)\frac{k^{\mu}k^\nu}{k^2}](ig\gamma_{\nu}T^a)\nonumber\\
&=&\intdk(-g^2T^aT^b\delta^{ab})\gamma_{\mu}\frac{p\sla+k\sla+m}{(p+k)^2-m^2}\gamma_{\nu}\frac{1}{k^2}
[g^{\mu\nu}+(\xi-1)\frac{k^{\mu}k^\nu}{k^2}]\nonumber\\
&=&(-g^2C_2)\intdk\gamma_{\mu}\frac{p\sla+k\sla+m}{(p+m)^2-m^2}\gamma_{\nu}
(\frac{g^{\mu\nu}}{k^2}+\frac{(\xi-1)k^{\mu}k^{\nu}}{k^4})
\end{eqnarray}
we now apply the Feynman parameter method to the denominators in the
integral in order to squeeze those denominator factors into a single
quadratic polynomial in $k$. Then we can get:
\begin{eqnarray}
L(2)&=&(-g^2C_2)\int_0^1dx_1\intdk\gamma_{\mu}(p\sla+k\sla+m)\gamma_{\nu}
(\frac{\Gamma(2)g^{\mu\nu}}{\Gamma(1)\Gamma(1)[(1-x_1)((p+k)^2-m^2)+x_1k^2]^2}+\nonumber\\
& &\frac{\Gamma(3)x_1(\xi-1)k^{\mu}k^{\nu}}{\Gamma(1)\Gamma(2)[(1-x_1)((p+k)^2-m^2)+x_1k^2]^3})\nonumber\\
&=&(-g^2C_2)\int_0^1dx_1\intdk\gamma_{\mu}(p\sla+k\sla+m)\gamma_{\nu}(\frac{g^{\mu\nu}}{(k^2-M_2^2)^2}+\nonumber\\
&
&\frac{2x_1(\xi-1)(k-(1-x_1)p)^{\mu}(k-(1-x_1)p)^{\nu}}{(k^2-M_2^2)^3})
\end{eqnarray}
where we have shifted the integrating variable by a constant to
complete the square in the denominators and have introduced the
notation $M_2^2\equiv(1-x_1)m^2-x_1(1-x_1)p^2$. Then the divergent
part can be extracted to be:
\begin{eqnarray}
L(2)_{div}&=&(-g^2C_2)\int_0^1dx_1\intdk(\frac{g^{\mu\nu}\gamma_{\mu}(x_1p\sla+m)\gamma_{\nu}}{(k^2-M_2^2)^2}
+\frac{2x_1(\xi-1)k^{\mu}k^{\nu}\gamma_{\mu}(x_1p\sla+m)\gamma_{\nu}}{(k^2-M_2^2)^3}+\nonumber\\
& &\frac{2x_1(\xi-1)(1-x_1)(-k^{\mu}p^{\nu}-k^{\mu}p^{\mu})\gamma_{\mu}k\sla\gamma_{\nu}}{(k^2-M_2^2)^3})\nonumber\\
&=&(-g^2C_2)\int_0^1dx_1[(-2x_1p\sla+4m)I_0+2\gamma_{\mu}(x_1p\sla+m)\gamma_{\nu}x_1(\xi-1)I_0^{\mu\nu}+\nonumber\\
&
&2x_1(x_1-1)(\xi-1)\gamma_{\mu}\gamma_{\alpha}\gamma_{\nu}(p^{\mu}I_0^{\alpha\nu}+p^{\nu}I_0^{\alpha\mu})]
\end{eqnarray}
It is seen that the Feynman integral can be expressed in terms of
1-fold ILIs $I_0$ and $I_0^{\mu\nu}$. To regularize this Feynman
integral we only need apply the loop regularization prescription to
the relevant ILIs. We mention that the explicit forms of all the
regularized ILIs have been worked out in \cite{LR1,LR2}, what we
need here is to use the relation for the regularized ILIs:
$I_{0\mu\nu}^R=\frac{1}{4}g_{\mu\nu}I_0^R$. As a consequence, the
regularized divergent parts of the Feynman diagram can be expressed
only in term of the regularized scalar divergent ILIs $I_0^R$
\begin{eqnarray}
L(2)_{div}^R&=&(-g^2C_2)\int_0^1dx_1\{[x_1(3x_1-4)(\xi-1)-2x_1]p{\sla}+[2x_1(\xi-1)+4]m\}I_0^R
\end{eqnarray}

\subsection{gluon self-energy diagram}

There are four gluon self-energy diagrams as shown in Fig.1, which
have been evaluated in \cite{LR1,LR2}, the results read:
\begin{eqnarray}
L^{ab}_{R\mu\nu}&=&g^2\delta^{ab}(p^2g_{\mu\nu}-p_{\mu}p_{\nu}){\int_0^1dx}\{C_1[1+4x(1-x)+\frac{1}{2}(1-\xi)]I_0^R\nonumber\\
&
&-N_fT_28x(1-x)I_0^R(m)-4C_1(1-\xi)[1-\frac{1}{8}(1-\xi)]x(1-x)p^2I_{-2}^R\}
\end{eqnarray}

\subsection{ghost self-energy diagram}

There is only one diagram which contributes to the one-loop
renormalization of the ghost fields strength as shown in Fig.3.
\begin{center}
\includegraphics[scale=1]{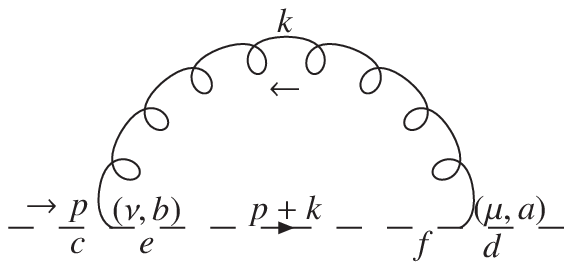}
\end{center}
\begin{center}{\sl Fig.3.}\end{center}
\begin{eqnarray}
L(3)^{cd}&=&\intdk(gf^{adf}p_{\mu})\frac{i\delta^{ef}}{(p+k)^2}\frac{-i\delta^{ab}}{k^2}[g^{\mu\nu}+(\xi-1)\frac{k^{\mu}k^{\nu}}{k^2}](gf^{bec}(p+k)_{\nu})\nonumber\\
&=&-C_1g^2\delta^{cd}{\intdk}p_{\mu}\frac{1}{(p+k)^2}[\frac{g^{\mu\nu}}{k^2}+(\xi-1)\frac{k^{\mu}k^{\nu}}{k^4}](p_{\nu}+k_{\nu})\nonumber\\
&=&-C_1g^2\delta^{cd}{\intdk}[\frac{p^2+{\xi}p_{\mu}k^{\mu}}{(p+k)^2k^2}+(\xi-1)\frac{p_{\mu}p_{\nu}k^{\mu}k^{\nu}}{(p+k)^2k^4}]\nonumber\\
&=&-C_1g^2\delta^{cd}\int_0^1dx{\intdk}[\frac{\Gamma(2)}{\Gamma(1)\Gamma(1)}\frac{p^2+{\xi}p_{\mu}k^{\mu}}{((1-x)(p+k)^2+xk^2)^2}+(\xi-1)\frac{\Gamma(3)}{\Gamma(1)\Gamma(2)}\frac{xp_{\mu}p_{\nu}k^{\mu}k^{\nu}}{((1-x)(p+k)^2+xk^2)^3}]\nonumber\\
&=&-C_1g^2\delta^{cd}\int_0^1dx{\intdk}[\frac{p^2+{\xi}p_{\mu}(k-(1-x)p)^{\mu}}{(k^2-M_3^2)^2}+\frac{2x(\xi-1)p_{\mu}p_{\nu}(k-(1-x)p)^{\mu}(k-(1-x)p)^{\nu}}{(k^2-M_3^2)^3}]
\end{eqnarray}
where we have introduced the notation $M_3^2=-x(1-x)p^2$. Thus the
divergent part can be found to be:
\begin{eqnarray}
L(3)^{cd}_{div}&=&-C_1g^2\delta^{cd}\int_0^1dx{\intdk}[\frac{(1-\xi(1-x))p^2}{(k^2-M_3^2)^2}+\frac{2x(\xi-1)p_{\mu}p_{\nu}k^{\mu}k^{\nu}}{(k^2-M_3^2)^3}]\nonumber\\
&=&-C_1g^2\delta^{cd}\int_0^1dx{\intdk}[(1-\xi(1-x))p^2I_0+2x(\xi-1)p_{\mu}p_{\nu}I_0^{\mu\nu}]
\end{eqnarray}
Applying for the relation
$I_{0\mu\nu}^R=\frac{1}{4}g_{\mu\nu}I_0^R$, all the divergent parts
can be expressed in term of $I_0^R$ and given by
\begin{eqnarray}
L(3)^{cdR}_{div}&=&-C_1g^2\delta^{cd}\int_0^1dx{\intdk}(x-(1-\frac{3}{2}x)(\xi-1))p^2I_0^R
\end{eqnarray}

\subsection{fermion-gluon vertex renormalization}

Two diagrams including their permutation can contribute to the
one-loop renormalization of fermion-gluon vertex. Let's begin with
the calculation for Fig.4a.
\begin{center}
\includegraphics[scale=1]{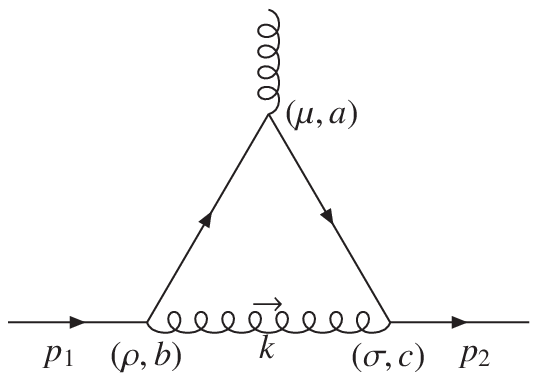}
\end{center}
\begin{center}{\sl Fig.4a.}\end{center}
\begin{eqnarray}
L(4a)^a_{\mu}&=&\intdk(ig\gamma_{\sigma}T^c)\frac{i}{p\sla_2-k\sla-m}(ig\gamma_{\mu}T^a)\frac{i}{p\sla_1-k\sla-m}(ig\gamma_{\rho}T^b)\frac{-i\delta^{bc}}{k^2}[g^{\rho\sigma}+(\xi-1)\frac{k^{\rho}k^{\sigma}}{k^2}]\nonumber\\
&=&g^3T^bT^aT^b\intdk\gamma_{\sigma}\frac{p\sla_2-k\sla+m}{(p_2-k)^2-m^2}\gamma_{\mu}\frac{p\sla_1-k\sla+m}{(p_1-k)^2-m^2}\gamma_{\rho}\frac{1}{k^2}[g^{\rho\sigma}+(\xi-1)\frac{k^{\rho}k^{\sigma}}{k^2}]\nonumber\\
&=&g^3(C_2-\frac{1}{2}C_1)T^a\intdk\frac{\gamma_{\sigma}(p\sla_2-k\sla+m)\gamma_{\mu}(p\sla_1-k\sla+m)\gamma_{\rho}}{k^2[(p_2-k)^2-m^2][(p_1-k)^2-m^2]}[g^{\rho\sigma}+(\xi-1)\frac{k^{\rho}k^{\sigma}}{k^2}]\nonumber\\
&=&g^3(C_2-\frac{1}{2}C_1)T^a\int_0^1dx_1\int_0^{x_1}dx_2\intdk\nonumber\\
& &\{\frac{\Gamma(3)}{\Gamma(1)^3}\frac{\gamma^{\rho}(p\sla_2-k\sla+m)\gamma_{\mu}(p\sla_1-k\sla+m)\gamma_{\rho}}{\{(1-x_1)k^2+(x_1-x_2)[(p_2-k)^2-m^2]+x_2[(p_1-k)^2-m^2]\}^3}\nonumber\\
& &+\frac{\Gamma(4)}{\Gamma(2)\Gamma(1)^2}\frac{(1-x_1)(\xi-1)k^{\rho}k^{\sigma}\gamma_{\sigma}(p\sla_2-k\sla+m)\gamma_{\mu}(p\sla_1-k\sla+m)\gamma_{\rho}}{\{(1-x_1)k^2+(x_1-x_2)[(p_2-k)^2-m^2]+x_2[(p_1-k)^2-m^2]\}^4}\}\nonumber\\
&=&g^3(C_2-\frac{1}{2}C_1)T^a\int_0^1dx_1\int_0^{x_1}dx_2\intdk\nonumber\\
& &\{\frac{1}{(k^2-M_{4a}^2)^3}[2\gamma^{\rho}(k\sla+(x_1-x_2-1)p\sla_2+x_2p\sla_1-m)\gamma_{\mu}(k\sla+(x_1-x_2)p\sla_2+(x_2-1)p\sla_1-m)\gamma_{\rho}]\nonumber\\
& &+\frac{1}{(k^2-M_{4a}^2)^4}[6(1-x_1)(\xi-1)(k+(x_1-x_2)p_2+x_2p_1)^{\rho}(k+(x_1-x_2)p_2+x_2p_1)^{\sigma}\gamma_{\sigma}\nonumber\\
&
&(k\sla+(x_1-x_2-1)p\sla_2+x_2p\sla_1-m)\gamma_{\mu}(k\sla+(x_1-x_2)p\sla_2+(x_2-1)p\sla_1-m)\gamma_{\rho}]\}
\end{eqnarray}
with definition:
$M_{4a}^2=-(x_1-x_2)(1-x_1+x_2)p_2^2+x_2(1-x_2)p_1^2-2x_2(x_1-x_2)p_1{\cdot}p_2-x_1m^2$,
and the divergent part is found to be:
\begin{eqnarray}
L(4a)^a_{\mu;div}&=&g^3(C_2-\frac{1}{2}C_1)T^a\int_0^1dx_1\int_0^{x_1}dx_2\intdk[2\frac{\gamma^{\rho}k\sla\gamma_{\mu}k\sla\gamma_{\rho}}{(k^2-M_{4a}^2)^3}\nonumber\\
& &+6(1-x_1)(\xi-1)\frac{k\sla{k}\sla\gamma_\mu{k}\sla{k}\sla}{(k^2-M_{4a}^2)^4}]\nonumber\\
&=&g^3(C_2-\frac{1}{2}C_1)T^a\int_0^1dx_1\int_0^{x_1}dx_2\intdk[2\frac{-4k_{\mu}k_\alpha\gamma^\alpha+2k^2\gamma_\mu}{(k^2-M_{4a}^2)^3}\nonumber\\
& &+6(1-x_1)(\xi-1)\frac{k^4\gamma_\mu}{(k^2-M_{4a}^2)^4}]\nonumber\\
&\sim&g^3(C_2-\frac{1}{2}C_1)T^a\int_0^1dx_1\int_0^{x_1}dx_2[-8\gamma^{\alpha}I_{0\mu\alpha}+4\gamma_{\mu}I_0+6(1-x_1)(\xi-1)\gamma_{\mu}I_0]
\end{eqnarray}
By adopting the relation $I_{0\mu\nu}^R=\frac{1}{4}g_{\mu\nu}I_0^R$
the divergent parts can be expressed in term of $I_0^R$
\begin{eqnarray}
L(4a)^{aR}_{\mu;div}&=&g^3(C_2-\frac{1}{2}C_1)T^a\int_0^1dx_1\int_0^{x_1}dx_2[2+6(1-x_1)(\xi-1)]\gamma_{\mu}I_0^R
\end{eqnarray}

\begin{center}
\includegraphics[scale=1]{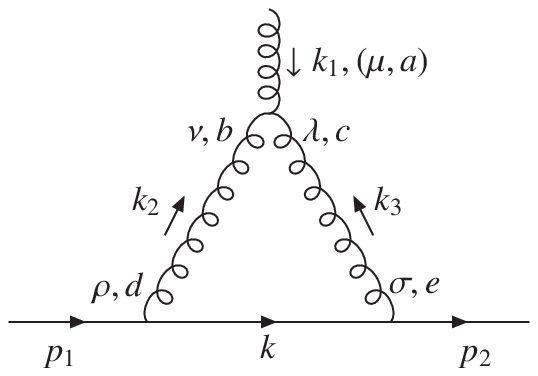}
\end{center}
\begin{center}{\sl Fig.4b.}\end{center}

The second diagram (Fig.4b) that contributes to the fermion-gluon
vertex has the following form
\begin{eqnarray}
L(4b)^a_{\mu}&=&\intdk(ig\gamma_{\sigma}T^e)\frac{i}{k\sla-m}(ig\gamma_{\rho}T^d)\frac{-i\delta^{ec}}{k_3^2}[g^{\sigma\lambda}+(\xi-1)\frac{k_3^{\sigma}k_3^{\lambda}}{k_3^2}]\times\nonumber\\
& &gf^{abc}[(k_1-k_2)_{\lambda}g_{\mu\nu}+(k_2-k_3)_{\mu}g_{\nu\lambda}+(k_3-k_1)_{\nu}g_{\mu\lambda}]\frac{-i\delta^{bd}}{k_2^2}[g^{\nu\rho}+(\xi-1)\frac{k_2^{\nu}k_2^{\rho}}{k_2^2}]\nonumber\\
&=&\frac{1}{2}g^3C_1T^a\intdk\gamma_{\sigma}\frac{k\sla+m}{k^2-m^2}\gamma_{\rho}[\frac{g^{\sigma\lambda}}{k_3^2}+(\xi-1)\frac{k_3^{\sigma}k_3^{\lambda}}{k_3^4}]\times\nonumber\\
& &[(k_1-k_2)_{\lambda}g_{\mu\nu}+(k_2-k_3)_{\mu}g_{\nu\lambda}+(k_3-k_1)_{\nu}g_{\mu\lambda}][\frac{g^{\nu\rho}}{k_2^2}+(\xi-1)\frac{k_2^{\nu}k_2^{\rho}}{k_2^4}]\nonumber\\
&=&\frac{1}{2}g^3C_1T^a\intdk\gamma_{\sigma}\frac{k\sla+m}{k^2-m^2}\gamma_{\rho}[(k_1-k_2)_{\lambda}g_{\mu\nu}+(k_2-k_3)_{\mu}g_{\nu\lambda}+(k_3-k_1)_{\nu}g_{\mu\lambda}]\times\nonumber\\
& &[\frac{g^{\sigma\lambda}g^{\nu\rho}}{k_3^2k_2^2}+(\xi-1)\frac{g^{\sigma\lambda}k_2^{\nu}k_2^{\rho}}{k_3^2k_2^4}+(\xi-1)\frac{g^{\nu\rho}k_3^{\sigma}k_3^{\lambda}}{k_3^4k_2^2}+(\xi-1)^2\frac{k_3^{\sigma}k_3^{\lambda}k_2^{\nu}k_2^{\rho}}{k_3^4k_2^4}]\nonumber\\
&=&\frac{1}{2}g^3C_1T^a\int_0^1dx_1\int_0^{x_1}dx_2\intdk\gamma_{\sigma}(k\sla+m)\gamma_{\rho}[(k_1-k_2)_{\lambda}g_{\mu\nu}+(k_2-k_3)_{\mu}g_{\nu\lambda}+(k_3-k_1)_{\nu}g_{\mu\lambda}]\times\nonumber\\
& &[\frac{\Gamma(3)}{\Gamma(1)^3}\frac{g^{\sigma\lambda}g^{\nu\rho}}{[(1-x_1)(k^2-m^2)+(x_1-x_2)k_3^2+x_2k_2^2]^3}+\nonumber\\
& &+(\xi-1)\frac{\Gamma(4)}{\Gamma(1)^2\Gamma(2)}\frac{x_2g^{\sigma\lambda}k_2^{\nu}k_2^{\rho}}{[(1-x_1)(k^2-m^2)+(x_1-x_2)k_3^2+x_2k_2^2]^4}\nonumber\\
& &+(\xi-1)\frac{\Gamma(4)}{\Gamma(1)^2\Gamma(2)}\frac{(x_1-x_2)g^{\nu\rho}k_3^{\sigma}k_3^{\lambda}}{[(1-x_1)(k^2-m^2)+(x_1-x_2)k_3^2+x_2k_2^2]^4}+\nonumber\\
& &+(\xi-1)^2\frac{\Gamma(5)}{\Gamma(1)\Gamma(2)^2}\frac{(x_1-x_2)x_2k_3^{\sigma}k_3^{\lambda}k_2^{\nu}k_2^{\rho}}{[(1-x_1)(k^2-m^2)+(x_1-x_2)k_3^2+x_2k_2^2]^5}]\nonumber\\
\end{eqnarray}
Introducing the notation
$M_{4b}^2=-(x_1-x_2)(1-x_1+x_2)p_2^2+x_2(1-x_2)p_1^2-2x_2(x_1-x_2)p_1{\cdot}p_2-x_1m^2$,
the divergent part can be written down as follows:
\begin{eqnarray}
L(4b)^a_{\mu;div}&=&\frac{1}{2}g^3C_1T^a\int_0^1dx_1\int_0^{x_1}dx_2\intdk\gamma_{\sigma}k\sla\gamma_{\rho}(k_{\lambda}g_{\mu\nu}-2k_{\mu}g_{\nu\lambda}+k_{\nu}g_{\mu\lambda})\times\nonumber\\
& &[\frac{2g^{\sigma\lambda}g^{\nu\rho}}{(k^2-M_{4b}^2)^3}+(\xi-1)\frac{6x_2g^{\sigma\lambda}k^{\nu}k^{\rho}}{(k^2-M_{4b}^2)^4}+(\xi-1)\frac{6(x_1-x_2)g^{\nu\rho}k^{\sigma}k^{\lambda}}{(k^2-M_{4b}^2)^4}+(\xi-1)^2\frac{24(x_1-x_2)x_2k^{\sigma}k^{\lambda}k^{\nu}k^{\rho}}{(k^2-M_{4b}^2)^5}]\nonumber\\
&=&\frac{1}{2}g^3C_1T^a\int_0^1dx_1\int_0^{x_1}dx_2[2\gamma^{\lambda}\gamma^{\alpha}\gamma_{\mu}I_{0\alpha\lambda}+2\gamma_{\mu}\gamma^{\alpha}\gamma^{\nu}I_{0\alpha\nu}-4\gamma^{\lambda}\gamma^{\alpha}\gamma_{\lambda}I_{0\alpha\mu}+6x_2(\xi-1)\gamma^{\lambda}\gamma^{\alpha}\gamma^{\rho}I_{0\alpha\rho\lambda\mu}\nonumber\\
& &-12x_2(\xi-1)\gamma^{\lambda}\gamma^{\alpha}\gamma^{\rho}I_{0\alpha\rho\lambda\mu}+6x_2(\xi-1)\gamma_{\mu}\gamma^{\alpha}\gamma^{\rho}I_{0\alpha\rho}+6(x_1-x_2)(\xi-1)\gamma^{\sigma}\gamma^{\alpha}\gamma_{\mu}I_{0\alpha\sigma}\nonumber\\
& &-12(x_1-x_2)(\xi-1)\gamma^{\sigma}\gamma^{\alpha}\gamma^{\nu}I_{0\alpha\sigma\mu\nu}+6(x_1-x_2)(\xi-1)\gamma^{\sigma}\gamma^{\alpha}\gamma^{\nu}I_{0\alpha\sigma\mu\nu}+0]\nonumber\\
&=&g^3C_1T^a\int_0^1dx_1\int_0^{x_1}dx_2[4\gamma^{\alpha}I_{0\alpha\mu}+(3(x_1-x_2)(\xi-1)+1)\gamma^{\lambda}\gamma^{\sigma}\gamma_{\mu}I_{0\lambda\alpha}+\nonumber\\
&
&+(3x_2(\xi-1)+1)\gamma_{\mu}\gamma^{\lambda}\gamma^{\alpha}I_{0\lambda\alpha}-3x_1(\xi-1)\gamma^{\alpha}\gamma^{\sigma}\gamma^{\rho}I_{0\alpha\sigma\rho\mu}]
\end{eqnarray}
Using the relations $I_{0\mu\nu}^R=\frac{1}{4}g_{\mu\nu}I_0^R$ and
$I_{0\rho\sigma\alpha\beta}^R=\frac{1}{24}(g_{\rho\sigma}g_{\alpha\beta}
+g_{\rho\alpha}g_{\sigma\beta}+g_{\rho\beta}g_{\sigma\alpha})I_0^R$,
the divergent parts can be simplified to be
\begin{eqnarray}
L(4b)^{aR}_{\mu;div}=g^3C_1T^a\gamma_{\mu}\int_0^1dx_1\int_0^{x_1}dx_2[3+\frac{9}{4}x_1(\xi-1)]I_0^R
\end{eqnarray}

\subsection{ghost-gluon vertex renormalization}

There are two diagrams which contribute to the one-loop
renormalization of the ghost-gluon vertex. Let's consider the first
diagram (Fig.5a)
\begin{center}
\includegraphics[scale=1]{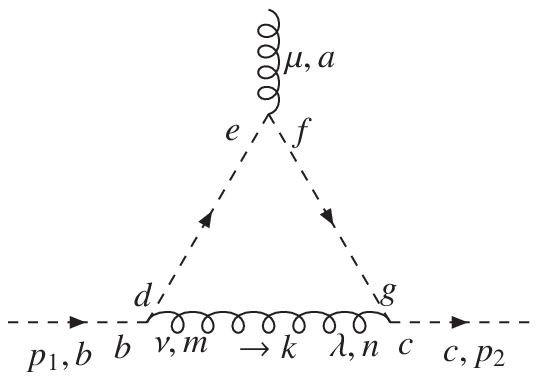}
\end{center}
\begin{center}{\sl Fig.5a.}\end{center}
\begin{eqnarray}
L(5a)^{acb}_{\mu}&=&\intdk(gf^{ncg}p_{2\lambda})\frac{i\delta^{fg}}{(p_2-k)^2}(gf^{afe}(p_2-k)_\mu)\frac{i\delta^{ed}}{(p_1-k)^2}(gf^{mdb}(p_1-k)_\nu)\nonumber\\
& &\frac{-i\delta^{mn}}{k^2}[g^{\nu\lambda}+(\xi-1)\frac{k^{\nu}k^{\lambda}}{k^2}]\nonumber\\
&=&-\frac{i}{2}g^3C_1f^{acb}{\intdk}p_{2\lambda}\frac{(p_2-k)_\mu}{(p_2-k)^2}\frac{(p_1-k)_\nu}{(p_1-k)^2}[\frac{g^{\nu\lambda}}{k^2}+(\xi-1)\frac{k^{\nu}k^{\lambda}}{k^4}]\nonumber\\
&=&-\frac{i}{2}g^3C_1f^{acb}{\intdk}[\frac{(p_1{\cdot}p_2)p_{2\mu}-(p_1{\cdot}p_2)k_{\mu}-{\xi}p_2^{\nu}p_{2\mu}k_\nu+{\xi}p_2^{\nu}k_{\mu}k_{\nu}}{(p_2-k)^2(p_1-k)^2k^2}+\nonumber\\
& &+(\xi-1)\frac{p_{2\lambda}p_{2\mu}p_{1\nu}k^{\nu}k^{\lambda}-p_{2\lambda}p_{1\nu}k_{\mu}k^{\nu}k^{\lambda}}{(p_2-k)^2(p_1-k)^2k^4}]\nonumber\\
&=&-\frac{i}{2}g^3C_1f^{acb}\int_0^1dx_1\int_0^{x_1}dx_2{\intdk}[\frac{\Gamma(3)}{\Gamma(1)^3}\frac{(p_1{\cdot}p_2)p_{2\mu}-(p_1{\cdot}p_2)k_{\mu}-{\xi}p_2^{\nu}p_{2\mu}k_\nu+{\xi}p_2^{\nu}k_{\mu}k_{\nu}}{[(1-x_1)(p_2-k)^2+(x_1-x_2)(p_1-k)^2+x_2k^2]^3}+\nonumber\\
&
&+(\xi-1)\frac{\Gamma(4)}{\Gamma(1)\Gamma(1)\Gamma(2)}\frac{x_2(p_{2\lambda}p_{2\mu}p_{1\nu}k^{\nu}k^{\lambda}-p_{2\lambda}p_{1\nu}k_{\mu}k^{\nu}k^{\lambda})}{[(1-x_1)(p_2-k)^2+(x_1-x_2)(p_1-k)^2+x_2k^2]^4}]
\end{eqnarray}
Taking the notation
$M_{5a}^2=-(1-x_1)x_1p_2^2+(x_1-x_2)(1-x_1+x_2)p_1^2-2(1-x_1)(x_1-x_2)p_1{\cdot}p_2$,
the divergent part is given by
\begin{eqnarray}
L(5a)^{acb}_{\mu;div}&=&-\frac{i}{2}g^3C_1f^{acb}\int_0^1dx_1\int_0^{x_1}dx_2{\intdk}
\frac{2{\xi}p_2^{\nu}k_{\mu}k_{\nu}}{(k^2-M_{5a}^2)^3}\nonumber\\
&=&-\frac{i}{2}g^3C_1f^{acb}\int_0^1dx_1\int_0^{x_1}dx_22{\xi}p_2^{\nu}I_{0\mu\nu}
\end{eqnarray}
with the relation $I_{0\mu\nu}^R=\frac{1}{4}g_{\mu\nu}I_0^R$, the
divergent part is simply given by
\begin{eqnarray}
L(5a)^{acbR}_{\mu;div}&=&-\frac{i}{2}g^3C_1f^{acb}\int_0^1dx_1\int_0^{x_1}dx_2\frac{1}{2}{\xi}p_{2\mu}I_0^R
\end{eqnarray}

We now evaluate the second diagram (fig.5b) which contributes to the
ghost-gluon vertex as:
\begin{center}
\includegraphics[scale=1]{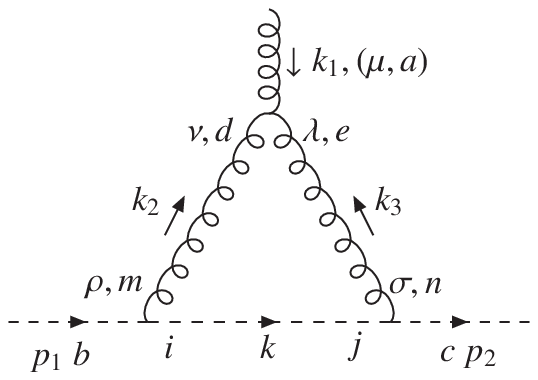}
\end{center}
\begin{center}{\sl Fig.5b.}\end{center}
\begin{eqnarray}
L(5b)^{acb}_{\mu}&=&\intdk(gf^{ncj}p_{2\sigma})\frac{-i\delta^{en}}{k_3^2}[g^{\lambda\sigma}+(\xi-1)\frac{k_3^{\lambda}k_3^{\sigma}}{k_3^2}](gf^{ade})[(k_1-k_2)_{\lambda}g_{\mu\nu}+(k_2-k_3)_{\mu}g_{\nu\lambda}+(k_3-k_1)_{\nu}g_{\lambda\mu}]\nonumber\\
& &\frac{-i\delta^{dm}}{k_2^2}[g^{\rho\nu}+(\xi-1)\frac{k_2^{\rho}k_2^{\nu}}{k_2^2}]\frac{i\delta^{ij}}{k^2}(gf^{mib}k_\rho)\nonumber\\
&=&-\frac{i}{2}g^3C_1f^{acb}\int dk\frac{p_{2\sigma}k_\rho}{k^2}
[\frac{g^{\lambda\sigma}}{k_3^2}+(\xi-1)\frac{k_3^{\lambda}k_3^{\sigma}}{k_3^4}]
[\frac{g^{\rho\nu}}{k_2^2}+(\xi-1)\frac{k_2^{\rho}k_2^{\nu}}{k_2^4}]\times\nonumber\\
& &[(k_1-k_2)_{\lambda}g_{\mu\nu}+(k_2-k_3)_{\mu}g_{\nu\lambda}+(k_3-k_1)_{\nu}g_{\lambda\mu}]\nonumber\\
&=&-\frac{i}{2}g^3C_1f^{acb}\int
dk[(k_1-k_2)_{\lambda}g_{\mu\nu}+(k_2-k_3)_{\mu}g_{\nu\lambda}
+(k_3-k_1)_{\nu}g_{\lambda\mu}]\times \int_0^1 dx_1 \int_0^{x_1} dx_2 \nonumber\\
& &[\frac{\Gamma(3)}{\Gamma(1)^3}\frac{p_{2\sigma}k_{\rho}g^{\lambda\sigma}g^{\rho\nu}}{[(1-x_1)k^2
+(x_1-x_2)k_2^2+x_2k_3^2]^3}+\frac{\Gamma(4)}{\Gamma(1)^2\Gamma(2)}
\frac{x_2(\xi-1)p_{2\sigma}k_{\rho}k_3^{\lambda}k_3^{\sigma}g^{\rho\nu}}{[(1-x_1)k^2
+(x_1-x_2)k_2^2+x_2k_3^2]^4}\nonumber\\
& &+\frac{\Gamma(4)}{\Gamma(1)^2\Gamma(2)}\frac{(x_1-x_2)(\xi-1)p_{2\sigma}k_{\rho}k_2^{\rho}k_2^{\nu}
g^{\lambda\sigma}}{[(1-x_1)k^2+(x_1-x_2)k_2^2+x_2k_3^2]^4}+\frac{\Gamma(5)}{\Gamma(1)\Gamma(2)^2}
\frac{x_2(x_1-x_2)(\xi-1)^2p_{2\sigma}k_{\rho}k_3^{\lambda}k_3^{\sigma}k_2^{\rho}k_2^{\nu}}{[(1-x_1)k^2
+(x_1-x_2)k_2^2+x_2k_3^2]^5}]\nonumber\\
\end{eqnarray}
Introducing the notation
$M_{5b}^2=-(x_1-x_2)(1-x_1+x_2)p_1^2+x_2(1-x_2)p_2^2-2x_2(x_1-x_2)p_1{\cdot}p_2$,
the divergent part can be expressed as:
\begin{eqnarray}
L(5b)^{acb}_{\mu;div}&=&-\frac{i}{2}g^3C_1f^{acb}\int_0^1dx_1\int_0^{x_1}dx_2\intdk[k_{\lambda}g_{\mu\nu}-2k_{\mu}g_{\nu\lambda}+k_{\nu}g_{\lambda\mu}][\frac{2p_{2\sigma}k_{\rho}g^{\lambda\sigma}g^{\rho\nu}}{(k^2-M_{5b}^2)^3}+\nonumber\\
& &\frac{6x_2(\xi-1)p_{2\sigma}k_{\rho}k^{\lambda}k^{\sigma}g^{\rho\nu}}{(k^2-M_{5b}^2)^4}+\frac{6(x_1-x_2)(\xi-1)p_{2\sigma}k_{\rho}k^{\rho}k^{\nu}g^{\lambda\sigma}}{(k^2-M_{5b}^2)^4}+\frac{24x_2(x_1-x_2)(\xi-1)^2p_{2\sigma}k_{\rho}k^{\rho}k^{\nu}k^{\lambda}k^{\sigma}}{(k^2-M_{5b}^2)^5}]\nonumber\\
&=&-\frac{i}{2}g^3C_1f^{acb}\int_0^1dx_1\int_0^{x_1}dx_2\intdk[\frac{-2p_2^{\sigma}k_{\sigma}k_{\mu}}{(k^2-M_{5b}^2)^3}+\frac{2p_{2\mu}}{(k^2-M_{5b}^2)^3}+0+\frac{-6(x_1-x_2)(\xi-1)p_2^{\sigma}k_{\sigma}k_{\mu}}{(k^2-M_{5b}^2)^3}\nonumber\\
& &+\frac{6(x_1-x_2)(\xi-1)p_{2\mu}}{(k^2-M_{5b}^2)^3}+0]\nonumber\\
&=&-ig^3C_1f^{acb}\int_0^1dx_1\int_0^{x_1}dx_2[(3(x_2-x_1)(\xi-1)-1)p_2^{\sigma}I_{0\sigma\mu}+(-3(x_2-x_1)(\xi-1)+1)p_{2\mu}I_0]
\end{eqnarray}
with the relation  $I_{0\mu\nu}^R=\frac{1}{4}g_{\mu\nu}I_0^R$, all
the divergent parts can be expressed in term of $I_0^R$
\begin{eqnarray}
L(5b)^{acb R}_{\mu;div}
&=&-\frac{3i}{4}g^3C_1f^{acb}\int_0^1dx_1\int_0^{x_1}dx_2(3(x_1-x_2)(\xi-1)+1)p_{2\mu}I_0^R
\end{eqnarray}

\subsection{three-gluon vertex renormalization}

Four diagrams including their permutation graphs can contribute to
the one-loop renormalization of three-gluon vertex, let's evaluate
each of them. We begin with the first graph (Fig.6a)
\begin{center}
\includegraphics[scale=1]{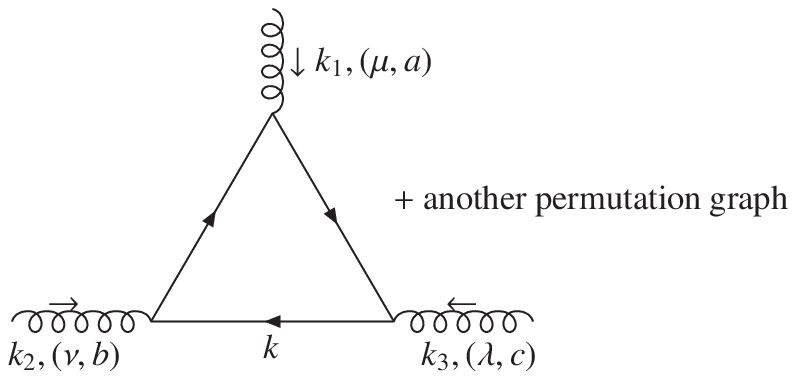}
\end{center}
\begin{center}{\sl Fig.6a.}
\end{center}

which is calculated via the following form
\begin{eqnarray}
L(6a)_{\mu\nu\lambda}^{abc}&=&-Tr{\intdk}ig\gamma_{\mu}T^a\frac{1}{k\sla+k\sla_2-m}ig\gamma_{\nu}
T_b\frac{i}{k\sla-m}ig\gamma_{\lambda}T^c\frac{i}{k\sla-k\sla_3-m}\nonumber\\
&
&--Tr{\intdk}ig\gamma_{\mu}T^a\frac{1}{-k\sla+k\sla_3-m}ig\gamma_{\nu}
T_b\frac{i}{1k\sla-m}ig\gamma_{\lambda}T^c\frac{i}{-k\sla-k\sla_2-m}\nonumber\\
&=&-[Tr(T^aT^bT^c)+Tr(T^aT^cT^b)]{\intdk}tr(ig\gamma_{\mu}\frac{1}{k\sla+k\sla_2-m}ig\gamma_{\nu}
\frac{i}{k\sla-m}ig\gamma_{\lambda}\frac{i}{k\sla-k\sla_3-m})\nonumber\\
&=&-Tr(if_{abd}T^dT^c){\intdk}(-g^3)\frac{tr[\gamma_{mu}(k\sla+k\sla_2+m)\gamma_{\nu}(k\sla+m)
\gamma_{\lambda}(k\sla-k\sla_3-m)]}{[(k+k+_2)^2-m^2][k^2-m^2][(k-k_3)^2-m^2]}\nonumber\\
&=&ig^3f^{gbc}T_2{\intdk}{\int_0^1dx_1}{\int_0^{x_1}dx_2}\frac{\Gamma(3)}{\Gamma(1)^3}\nonumber\\
& &\frac{tr[\gamma_{mu}(k\sla+k\sla_2+m)\gamma_{\nu}(k\sla+m)\gamma_{\lambda}(k\sla-k\sla_3-m)]}{\{(k^2-m^2)(1-x_1)
+[(k+k+_2)^2-m^2](x_1-x_2)+[(k-k_3)^2-m^2]x_2\}^3}\nonumber\\
&=&2ig^3f^{abc}T_2{\intdk}{\int_0^1dx_1}{\int_0^{x_1}dx_2}\frac{\Gamma(3)}{\Gamma(1)^3}\nonumber\\
& &\frac{tr[\gamma_{mu}(k\sla+k\sla_2+m)\gamma_{\nu}(k\sla+m)
\gamma_{\lambda}(k\sla-k\sla_3-m)]}{[(k+(x_1-x_2)k_2-x_2k_3)^2-m^2+(x_1-x_2)(1-x_1+x_2)k_2^2
+x_2(1-x_2)k_3^2]^3}\nonumber\\
&=&2ig^3f^{abc}T_2{\intdk}{\int_0^1dx_1}{\int_0^{x_1}dx_2}\frac{1}{(k^2-M_{6a}^2)^3}\times\nonumber\\
& &tr[\gamma_{\mu}(k\sla-(x_1-x_2)k\sla_2+x_2k\sla_3+k\sla_2+m)
\gamma_{\nu}(k\sla-(x_1-x_2)k\sla_2+x_2k\sla_3+m)\nonumber\\
& &\gamma_{\lambda}(k\sla-(x_1-x_2)k\sla_2+x_2k\sla_3-k\sla_3-m)]
\end{eqnarray}
with $M_{6a}^2=m^2-(x_1-x_2)(1-x_1+x_2)k_2^2-x_2(1-x_2)k_3^2$. The
divergent part is given by
\begin{eqnarray}
L(6a)_{\mu\nu\lambda;div}^{abc}&=&2ig^3f^{abc}T_2{\intdk}{\int_0^1dx_1}{\int_0^{x_1}dx_2}
\frac{1}{(k^2-M_{6a}^2)^3}\times\nonumber\\
& &tr[\gamma_{\mu}k\sla\gamma_{\nu}k\sla\gamma_{\lambda}(-(x_1-x_2)k\sla_2+x_2k\sla_3-k\sla_3+m)+\nonumber\\
& &+\gamma_{\mu}k\sla\gamma_{\nu}(-(x_1-x_2)k\sla_2+x_2k\sla_3+m)\gamma_{\lambda}k\sla+\nonumber\\
& &+\gamma_{\mu}(-(x_1-x_2)k\sla_2+x_2k\sla_3+k\sla_2+m)\gamma_{\nu}k\sla\gamma_{\lambda}k\sla]\nonumber\\
&=&2ig^3f^{abc}T_2{\intdk}{\int_0^1dx_1}{\int_0^{x_1}dx_2}\frac{1}{(k^2-M_{6a}^2)^3}
[(x_2-x_1)tr(\gamma_{\mu}k\sla\gamma_{\nu}k\sla\gamma_{\gamma}k\sla_2)+\nonumber\\
& &+(x_2-1)tr(\gamma_{\mu}k\sla\gamma_{\nu}k\sla\gamma_{\gamma}k\sla_3)+(x_2-x_1)
tr(\gamma_{\mu}k\sla\gamma_{\nu}k\sla_2\gamma_{\gamma}k\sla)
+x_2tr(\gamma_{\mu}k\sla\gamma_{\nu}k\sla_3\gamma_{\gamma}k\sla)+\nonumber\\
& &+(x_2-x_1+1)tr(\gamma_{\mu}k\sla_2\gamma_{\nu}k\sla\gamma_{\gamma}k\sla)
+x_2tr(\gamma_{\mu}k\sla_3\gamma_{\nu}k\sla\gamma_{\gamma}k\sla)]\nonumber\\
&=&2ig^3f^{abc}T_2{\intdk}{\int_0^1dx_1}{\int_0^{x_1}dx_2}\frac{1}{(k^2-M_{6a}^2)^3}
[4(-x_1+x_2+1)(2k_{2\mu}k_{\nu}k_{\lambda}+\nonumber\\
&
&+2k_{\mu}k_{2\nu}k_{\lambda}+k_{2\lambda}g_{\mu\nu}k^2-k_{2\mu}g_{\nu\lambda}k^2-k_{2\nu}
g_{\mu\lambda}k^2-2k_{\lambda}g_{\mu\nu}k{\cdot}k_2)+x_2(2k_{3\mu}k_{\nu}k_{\lambda}+\nonumber\\
&
&+2k_{\mu}k_{3\nu}k_{\lambda}+k_{3\lambda}g_{\mu\nu}k^2-k_{3\mu}g_{\nu\lambda}k^2-k_{3\nu}
g_{\mu\lambda}k^2-2k_{\lambda}g_{\mu\nu}k{\cdot}k_3)+\nonumber\\
&
&+(x_2-x_1)(2k_{2\nu}k_{\lambda}k_{\mu}+2k_{\nu}k_{2\lambda}k_{\mu}+k_{2\mu}
g_{\nu\lambda}k^2-k_{2\nu}g_{\lambda\mu}k^2-k_{2\lambda}g_{\nu\mu}k^2-2k_{\mu}g_{\nu\lambda}k{\cdot}k_2)+\nonumber\\
&
&+x_2(2k_{3\nu}k_{\lambda}k_{\mu}+2k_{\nu}k_{3\lambda}k_{\mu}+k_{3\mu}
g_{\nu\lambda}k^2-k_{3\nu}g_{\lambda\mu}k^2-k_{3\lambda}g_{\nu\mu}k^2-2k_{\mu}
g_{\nu\lambda}k{\cdot}k_3)+\nonumber\\
&
&+(x_2-x_1)(2k_{2\lambda}k_{\mu}k_{\nu}+2k_{\lambda}k_{2\mu}k_{\nu}+k_{2\nu}
g_{\lambda\mu}k^2-k_{2\lambda}g_{\mu\nu}k^2-k_{2\mu}g_{\lambda\nu}k^2-2k_{\nu}
g_{\lambda\mu}k{\cdot}k_2)+\nonumber\\
&
&+(x_2-1)(2k_{3\lambda}k_{\mu}k_{\nu}+2k_{\lambda}k_{3\mu}k_{\nu}+k_{3\nu}
g_{\lambda\mu}k^2-k_{3\lambda}g_{\mu\nu}k^2-k_{3\mu}g_{\lambda\nu}k^2-2k_{\nu}g_{\lambda\mu}k{\cdot}k_3)]\nonumber
\end{eqnarray}
with the relation $I_{0\mu\nu}^R=\frac{1}{4}g_{\mu\nu}I_0^R$, all
the divergent parts can be expressed in term of $I_0^R$. The result
is given by
\begin{eqnarray}
L(6a)_{\mu\nu\lambda;div}^{abcR}&=&2ig^3f^{abc}T_2{\int_0^1dx_1}{\int_0^{x_1}dx_2}I_0^R
[4(-x_1+x_2+1)(\frac{1}{2}k_{2\mu}g_{\nu\lambda}+\nonumber\\
&
&+\frac{1}{2}k_{2\nu}g_{\mu\lambda}+k_{2\lambda}g_{\mu\nu}-k_{2\mu}g_{\nu\lambda}-k_{2\nu}g_{\mu\lambda}
-\frac{1}{2}k_{2\lambda}g_{\mu\nu})+x_2(\frac{1}{2}k_{3\mu}g_{\nu\lambda}+\nonumber\\
&
&+\frac{1}{2}k_{3\nu}g_{\mu\lambda}+k_{3\lambda}g_{\mu\nu}-k_{3\mu}g_{\nu\lambda}-k_{3\nu}g_{\mu\lambda}
-\frac{1}{2}k_{3\lambda}g_{\mu\nu})+\nonumber\\
&
&+(x_2-x_1)(\frac{1}{2}k_{2\nu}g_{\lambda\mu}+\frac{1}{2}k_{2\lambda}g_{\nu\mu}+k_{2\mu}g_{\nu\lambda}
-k_{2\nu}g_{\lambda\mu}-k_{2\lambda}g_{\nu\mu}-\frac{1}{2}k_{2\mu}g_{\nu\lambda})+\nonumber\\
&
&+x_2(\frac{1}{2}k_{3\nu}g_{\lambda\mu}+\frac{1}{2}k_{3\lambda}g_{\nu\mu}+k_{3\mu}g_{\nu\lambda}
-k_{3\nu}g_{\lambda\mu}-k_{3\lambda}g_{\nu\mu}-\frac{1}{2}k_{3\mu}g_{\nu\lambda})+\nonumber\\
&
&+(x_2-x_1)(\frac{1}{2}k_{2\lambda}g_{\mu\nu}+\frac{1}{2}k_{2\mu}g_{\lambda\nu}+k_{2\nu}g_{\lambda\mu}
-k_{2\lambda}g_{\mu\nu}-k_{2\mu}g_{\lambda\nu}-\frac{1}{2}k_{\nu}g_{\lambda\mu})+\nonumber\\
&
&+(x_2-1)(\frac{1}{2}k_{3\lambda}g_{\mu\nu}+\frac{1}{2}k_{3\mu}g_{\lambda\nu}+k_{3\nu}g_{\lambda\mu}
-k_{3\lambda}g_{\mu\nu}-k_{3\mu}g_{\lambda\nu}-\frac{1}{2}k_{3\nu}g_{\lambda\mu})]\nonumber\\
\end{eqnarray}

The second diagram (Fig.6b) has the following form
\begin{center}
\includegraphics[scale=1]{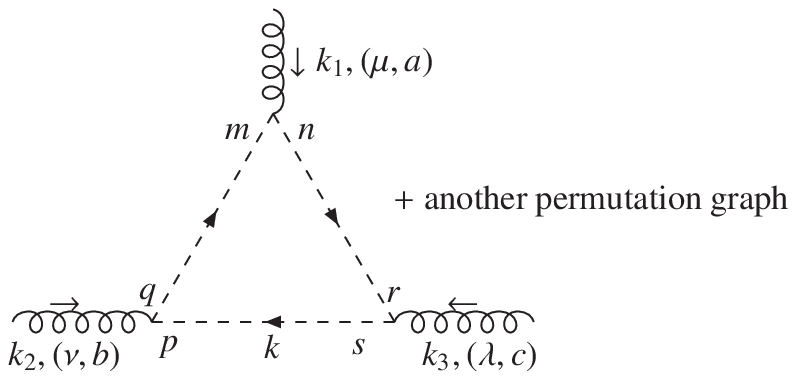}
\end{center}
\begin{center}{\sl Fig.6b.}\end{center}

\begin{eqnarray}
L(6b)_{\mu\nu\lambda}^{abc}&=&-{\intdk}gf^{anm}(k-k_3)_\mu\frac{i\delta^{mq}}{(k+k_2)^2}gf^{bqp}(k+k_2)_\nu
\frac{i\delta^{ps}}{k^2}gf^{csr}k_\lambda\frac{i\delta^{rn}}{(k-k_3)^2}\nonumber\\
&
&-{\intdk}gf^{anm}(k-k_2)_\mu\frac{i\delta^{mq}}{(k+k_3)^2}gf^{cqp}(k+k_3)_\lambda
\frac{i\delta^{ps}}{k^2}gf^{bsr}k_\nu\frac{i\delta^{rn}}{(k-k_2)^2}\nonumber\\
&=&ig^3f^{anm}f^{bmp}f^{cpn}{\intdk}\frac{(k-k_3)_\mu(k+k_2)_\nu
k_\lambda}{(k+k_2)^2k^2(k-k_3)^2}+(k_2{\rightarrow}k_3,\nu\rightarrow\lambda,b{\rightarrow}c)\nonumber\\
&=&ig^3f^{anm}f^{bmp}f^{cpn}{\int_0^1dx_1}{\int_0^{x_1}dx_2}{\intdk}\frac{\Gamma(3)}{\Gamma(1)^3}
\frac{(k-k_3)_\mu(k+k_2)_\nu
k_\lambda}{[(1-x_1)(k+k_2)^2+(x_1-x_2)k^2+x_2(k-k_3)^2]^3}\nonumber\\
& &+(k_2{\rightarrow}k_3,\nu\rightarrow\lambda,b{\rightarrow}c)\nonumber\\
&=&2ig^3f^{anm}f^{bmp}f^{cpn}{\int_0^1dx_1}{\int_0^{x_1}dx_2}{\intdk}\frac{(k-k_3)_\mu(k+k_2)_\nu
k_\lambda}{\{[k+(1-x_1)k_2-x_2k_3]^2+x_1(1-x_1)k_2^2+x_2(1-x_2)k_3^2\}^3}\nonumber\\
& &+(k_2{\rightarrow}k_3,\nu\rightarrow\lambda,b{\rightarrow}c)\nonumber\\
&=&2ig^3f^{anm}f^{bmp}f^{cpn}{\int_0^1dx_1}{\int_0^{x_1}dx_2}\nonumber\\
& &{\intdk}\frac{[k-(1-x_1)k_2-(1-x_2)k_3]_\mu(k+x_1k_2+x_2k_3)_\nu
[k-(1-x_1)k_2+x_2k_3]_\lambda}{(k^2-M_{6b}^2)^3}\nonumber\\
& &+(k_2{\rightarrow}k_3,\nu\rightarrow\lambda,b{\rightarrow}c)\nonumber\\
\end{eqnarray}
with $M_{6b}^2=-x_1(1-x_1)k_2^2-x_2(1-x_2)k_3^2$. The divergent part
is found to be
\begin{eqnarray}
L(6b)_{\mu\nu\lambda;div}^{abc}&=&2ig^3f^{anm}f^{bmp}f^{cpn}{\int_0^1dx_1}{\int_0^{x_1}dx_2}\nonumber\\
& &{\intdk}\frac{{k\mu}{k_\nu}[-(1-x_1)k_2+x_2k_3]_\lambda+{k_\mu}{k_\lambda}(x_1k_2+x_2k_3)_\nu+{k_\nu}{k_\lambda}[-(1-x_1)k_2-(1-x_2)k_3]_\mu}{(k^2-M_{6b}^2)^3}\nonumber\\
& &+(k_2{\rightarrow}k_3,\nu\rightarrow\lambda,b{\rightarrow}c)\nonumber\\
&=&2ig^3f^{anm}f^{bmp}f^{cpn}{\int_0^1dx_1}{\int_0^{x_1}dx_2}\{[-(1-x_1)k_2+x_2k_3]_\lambda{I_{0\mu\nu}(M_{6b}^2)}+(x_1k_2+x_2k_3)_\nu{I_{0\mu\lambda}(M_{6b}^2)}\nonumber\\
&
&+[-(1-x_1)k_2-(1-x_2)k_3]_\mu{I_{0\nu\lambda}(M_{6b}^2)}\}+(k_2{\rightarrow}k_3,\nu\rightarrow\lambda,b{\rightarrow}c)
\end{eqnarray}
Taking the relation $I_{0\mu\nu}^R=\frac{1}{4}g_{\mu\nu}I_0^R$, we
arrive at the result
\begin{eqnarray}
L(6b)_{\mu\nu\lambda;div}^{abcR}&=&2ig^3f^{anm}f^{bmp}f^{cpn}{\int_0^1dx_1}{\int_0^{x_1}dx_2}\nonumber\\
&
&\{[-(1-x_1)k_2+x_2k_3]_\lambda{\frac{1}{4}g_{\mu\nu}}+(x_1k_2+x_2k_3)_\nu{\frac{1}{4}g_{\mu\lambda}}+[-(1-x_1)k_2-(1-x_2)k_3]_\mu{\frac{1}{4}g_{\nu\lambda}}\}I_0^R(M_{6b}^2)\nonumber\\
& &+(k_2{\rightarrow}k_3,\nu\rightarrow\lambda,b{\rightarrow}c)
\end{eqnarray}

The third diagram (Fig.6c) is given by
\begin{center}
\includegraphics[scale=1]{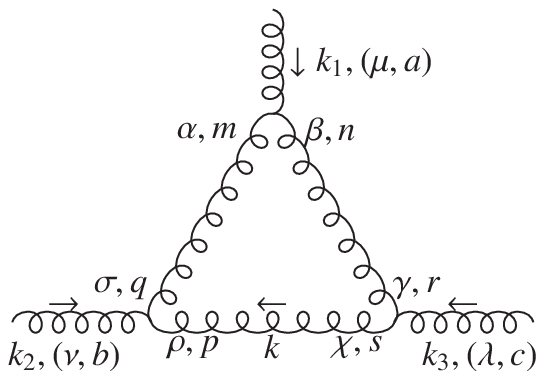}
\end{center}
\begin{center}{\sl Fig.6c.}\end{center}

\begin{eqnarray}
L(6c)_{\mu\nu\lambda}^{abc}&=&{\intdk}gf^{amn}[(k_1-k-k_2)_\beta{g_{\mu\alpha}}
+(k+k_2-k_3+k)_\mu{g_{\alpha\beta}}+(k_3-k-k_1)_\alpha{g_{\beta\mu}}]
\frac{-i\delta^{qm}g^{\sigma\alpha}}{(k+k_2)^2}\nonumber\\
& &\times
gf^{bpq}[(k_2-k)_\sigma{g_{\nu\rho}}+(k+k+k_2)_\nu{g_{\rho\sigma}}
+(-k-k_2-k_2)_\rho{g_{\sigma\nu}}]\frac{-i\delta^{ps}g^{\rho\chi}}{k^2}\nonumber\\
& &\times
gf^{crs}[(k_3-k+k_3)_\chi{g_{\lambda\gamma}}+(k-k_3+k)_\lambda{g_{\gamma\chi}}
+(-k-k_3)_\gamma{g_{\chi\lambda}}]\frac{-i\delta^{rn}g^{\gamma\beta}}{(k-k_3)^2}\nonumber\\
&=&(-ig)^3f^{amn}f^{bpm}f^{cnp}{\intdk}[(k_1-k-k_2)^\gamma{g_{\mu\alpha}}+(k+k_2-k_3+k)_\mu{g_\alpha^\gamma}
+(k_3-k-k_1)_\alpha{g^\gamma_\mu}]\nonumber\\
& &\times [(k_2-k)^\alpha{g_{\nu\rho}}+(k+k+k_2)_\nu{g_\rho^\alpha}+(-k-k_2-k_2)_\rho{g^\alpha_\nu}] \nonumber\\
& &\times [(k_3-k+k_3)^\rho{g_{\lambda\gamma}}+(k-k_3+k)_\lambda{g_\gamma^\rho}+(-k-k_3)_\gamma{g^\rho_\lambda}]
\nonumber\\
& &\times \frac{1}{(k+k_2)^2k^2(k-k_3)^2}\nonumber\\
&=&(-ig)^3(-\frac{1}{2}C_1f^{abc})\int_0^1dx_1\int_0^{x_1}dx_2{\intdk}\frac{\Gamma(3)}{\Gamma(1)^3}
\frac{1}{[(1-x_1)(k+k_2)^2+(x_1-x_2)k^2+x_2(k-k_3)^2]^3} \nonumber\\
& &\times
[(k_1-k-k_2)^\gamma{g_{\mu\alpha}}+(k+k_2-k_3+k)_\mu{g_\alpha^\gamma}
+(k_3-k-k_1)_\alpha{g^\gamma_\mu}] \nonumber\\
& &\times [(k_2-k)^\alpha{g_{\nu\rho}}+(k+k+k_2)_\nu{g_\rho^\alpha}+(-k-k_2-k_2)_\rho{g^\alpha_\nu}] \nonumber\\
& &\times [(k_3-k+k_3)^\rho{g_{\lambda\gamma}}+(k-k_3+k)_\lambda{g_\gamma^\rho}+(-k-k_3)_\gamma{g^\rho_\lambda}]\nonumber\\
&=&-iC_1g^3f^{abc}\int_0^1dx_1\int_0^{x_1}dx_2{\intdk}\frac{1}{\{[k+(1-x_1)k_2-x_2k_3]^2
+x_1(1-x_1)k_2^2+x_2(1-x_2)k_3^2\}^3} \nonumber\\
& &\times
[(k_1-k-k_2)^\gamma{g_{\mu\alpha}}+(k+k_2-k_3+k)_\mu{g_\alpha^\gamma}
+(k_3-k-k_1)_\alpha{g^\gamma_\mu}] \nonumber\\
& &\times [(k_2-k)^\alpha{g_{\nu\rho}}+(k+k+k_2)_\nu{g_\rho^\alpha}+(-k-k_2-k_2)_\rho{g^\alpha_\nu}]\times\nonumber\\
& &[(k_3-k+k_3)^\rho{g_{\lambda\gamma}}+(k-k_3+k)_\lambda{g_\gamma^\rho}+(-k-k_3)_\gamma{g^\rho_\lambda}]\nonumber\\
&=&-iC_1g^3f^{abc}\int_0^1dx_1\int_0^{x_1}dx_2{\intdk}\frac{1}{(k^2-M_{6c}^2)^3} \nonumber\\
& &\times [(-k+(-1-x_1)k_2+(-1-x_2)k_3)^\gamma{g_{\mu\alpha}}+(2k+(-1+2x_1)k_2+(-1+2x_2)k_3)_\mu{g_\alpha^\gamma}+\nonumber\\
& &(-k+(2-x_1)k_2+(2-x_2)k_3)_\alpha{g^\gamma_\mu}] \nonumber\\
& &\times [(-k+(2-x_1)k_2-x_2k_3)^\alpha{g_{\nu\rho}}+(2k+(-1+2x_1)k_2+2x_2k_3)_\nu{g_\rho^\alpha}+\nonumber\\
& &(-k+(-1-x_1)k_2-x_2k_3)_\rho{g^\alpha_\nu}] \nonumber\\
& &\times [(-k+(1-x_1)k_2+(2-x_2)k_3)^\rho{g_{\lambda\gamma}}+(2k+(-2+2x_1)k_2+(-1+2x_2)k_3)_\lambda{g_\gamma^\rho}+\nonumber\\
& &(-k+(1-x_1)k_2+(-1-x_2)k_3)_\gamma{g^\rho_\lambda}]\nonumber\\
\end{eqnarray}
with $M_{6c}^2=-x_1(1-x_1)k_2^2-x_2(1-x_2)k_3^2$. The divergent part
reads:
\begin{eqnarray}
L(6c)_{\mu\nu\lambda;div}^{abc}&=&-iC_1g^3f^{abc}\int_0^1dx_1\int_0^{x_1}dx_2{\intdk}
\frac{1}{(k^2-M_{6c}^2)^3}\times\nonumber\\
&
&\{[-k^\alpha{g_{\nu\rho}}+2k_\nu{g_\rho^\alpha}-k_\rho{g^\alpha_\nu}][-k^\rho{g_{\lambda\gamma}}
+2k_\lambda{g_\gamma^\rho}-k_\gamma{g^\rho_\lambda}]\times
[((-1-x_1)k_2+(-1-x_2)k_3)^\gamma{g_{\mu\alpha}}\nonumber
\\
& & +((-1+2x_1)k_2+(-1+2x_2)k_3)_\mu{g_\alpha^\gamma}
+((2-x_1)k_2+(2-x_2)k_3)_\alpha{g^\gamma_\mu}]+\nonumber\\
& &[-k^\gamma{g_{\mu\alpha}}+2k_\mu{g_\alpha^\gamma}-k_\alpha{g^\gamma_\mu}][-k^\rho{g_{\lambda\gamma}}
+2k_\lambda{g_\gamma^\rho}-k_\gamma{g^\rho_\lambda}]\times\nonumber\\
& &[((2-x_1)k_2-x_2k_3)^\alpha{g_{\nu\rho}}+((-1+2x_1)k_2+2x_2k_3)_\nu{g_\rho^\alpha}
+((-1-x_1)k_2-x_2k_3)_\rho{g^\alpha_\nu}]+\nonumber\\
&
&[-k^\gamma{g_{\mu\alpha}}+2k_\mu{g_\alpha^\gamma}-k_\alpha{g^\gamma_\mu}][-k^\alpha{g_{\nu\rho}}
+2k_\nu{g_\rho^\alpha}-k_\rho{g^\alpha_\nu}]\times
[((1-x_1)k_2+(2-x_2)k_3)^\rho{g_{\lambda\gamma}} \nonumber
\\
& & +((-2+2x_1)k_2+(-1+2x_2)k_3)_\lambda{g_\gamma^\rho}
+((1-x_1)k_2+(-1-x_2)k_3)_\gamma{g^\rho_\lambda}]\}\nonumber\\
&=&-iC_1g^3f^{abc}\int_0^1dx_1\int_0^{x_1}dx_2{\intdk}\frac{1}{(k^2-M_{6c}^2)^3}\times\nonumber\\
& &
\{(k^\alpha{k_\gamma}g_{\lambda\nu}-2k^\alpha{k_\lambda}g_{\nu\gamma}-k_\gamma{k_\lambda}
g^\alpha_\nu-k^\alpha{k_\nu}g_{\lambda\gamma}-2k_\nu{k_\gamma}g^\alpha_\lambda+4k_\nu{k_\lambda}
g^\alpha_\gamma+k^2g^\alpha_\nu{g}_{\gamma\lambda})\times [((-1-x_1)k_2 \nonumber \\
& &
+(-1-x_2)k_3)^\gamma{g_{\mu\alpha}}+((-1+2x_1)k_2+(-1+2x_2)k_3)_\mu{g_\alpha^\gamma}
+((2-x_1)k_2+(2-x_2)k_3)_\alpha{g^\gamma_\mu}]+\nonumber\\
& &(-2k_\alpha{k_\lambda}g^\rho_\mu-k_\alpha{k_\mu}g^\rho_\lambda+4k_\mu{k_\lambda}
g^\rho_\alpha+k^\rho{k_\alpha}g_{\mu\lambda}-k^\rho{k_\lambda}g_{\mu\alpha}-2k^\rho{k_\mu}
g_{\alpha\lambda}+k^2g^\rho_\lambda{g}_{\alpha\mu})\times\nonumber\\
& &[((2-x_1)k_2-x_2k_3)^\alpha{g_{\nu\rho}}+((-1+2x_1)k_2+2x_2k_3)_\nu{g_\rho^\alpha}
+((-1-x_1)k_2-x_2k_3)_\rho{g^\alpha_\nu}]+\nonumber\\
&
&(-k^\gamma{k_\mu}g_{\nu\rho}+k^2g^\gamma_\mu{g}_{\nu\rho}-2k^\gamma{k_\nu}g{\mu\rho}
+4k_\mu{k_\nu}g^\gamma_\rho+k^\gamma{k_\rho}g_{\mu\nu}-2k_\mu{k_\rho}g^\gamma_\nu-k_\nu{k_\rho}
g^\gamma_\mu)\times [((1-x_1)k_2 \nonumber \\
& & +(2-x_2)k_3)^\rho{g_{\lambda\gamma}}+((-2+2x_1)k_2
+(-1+2x_2)k_3)_\lambda{g_\gamma^\rho}+((1-x_1)k_2+(-1-x_2)k_3)_\gamma{g^\rho_\lambda}]\}
\end{eqnarray}
After adopting the relation
$I_{0\mu\nu}^R=\frac{1}{4}g_{\mu\nu}I_0^R$, we arrive at the
following result:
\begin{eqnarray}
L(6c)_{\mu\nu\lambda;div}^{abcR}&& =-iC_1g^3f^{abc}\int_0^1dx_1\int_0^{x_1}dx_2I_0^R(M_{6c}^2)\times\nonumber\\
&
&\{(\frac{1}{4}g^\alpha\gamma{g}_{\lambda\nu}-\frac{2}{4}g^\alpha_\lambda{g}_{\nu\gamma}
-\frac{1}{4}g_{\gamma\lambda}g^\alpha_\nu-\frac{1}{4}g^\alpha_\nu{g}_{\lambda\gamma}
-\frac{2}{4}g_{\nu\gamma}g^\alpha_\lambda+\frac{4}{4}g_{\nu\lambda}g^\alpha_\gamma
+g^\alpha_\nu{g}_{\gamma\lambda})\times [((-1-x_1)k_2 \nonumber
\\
& &
+(-1-x_2)k_3)^\gamma{g_{\mu\alpha}}+((-1+2x_1)k_2+(-1+2x_2)k_3)_\mu{g_\alpha^\gamma}
+((2-x_1)k_2+(2-x_2)k_3)_\alpha{g^\gamma_\mu}]+\nonumber\\
& &(-\frac{2}{4}g_{\alpha\lambda}g^\rho_\mu-\frac{1}{4}g_{\alpha\mu}g^\rho_\lambda
+\frac{4}{4}g_{\mu\lambda}g^\rho_\alpha+\frac{1}{4}g^\rho_\alpha{g}_{\mu\lambda}
-\frac{1}{4}g^\rho_\lambda{g}_{\mu\alpha}-\frac{2}{4}g^\rho_\mu{g}_{\alpha\lambda}
+g^\rho_\lambda{g}_{\alpha\mu})\times\nonumber\\
& &[((2-x_1)k_2-x_2k_3)^\alpha{g_{\nu\rho}}+((-1+2x_1)k_2+2x_2k_3)_\nu{g_\rho^\alpha}
+((-1-x_1)k_2-x_2k_3)_\rho{g^\alpha_\nu}]+\nonumber\\
& &(-\frac{1}{4}g^\gamma_\mu{g}_{\nu\rho}+g^\gamma_\mu{g}_{\nu\rho}
-\frac{2}{4}g^\gamma_\nu{g}{\mu\rho}+\frac{4}{4}g_{\mu\nu}g^\gamma_\rho
+\frac{1}{4}g^\gamma_\rho{g}_{\mu\nu}-\frac{2}{4}g_{\mu\rho}g^\gamma_\nu
-\frac{1}{4}g_{\nu\rho}g^\gamma_\mu)\times [((1-x_1)k_2 \nonumber
\\
& & +(2-x_2)k_3)^\rho{g_{\lambda\gamma}}+((-2+2x_1)k_2
+(-1+2x_2)k_3)_\lambda{g_\gamma^\rho}+((1-x_1)k_2+(-1-x_2)k_3)_\gamma{g^\rho_\lambda}]\}
\end{eqnarray}

Similarly, the fourth diagram (Fig.6d) is given:
\begin{center}
\includegraphics[scale=1]{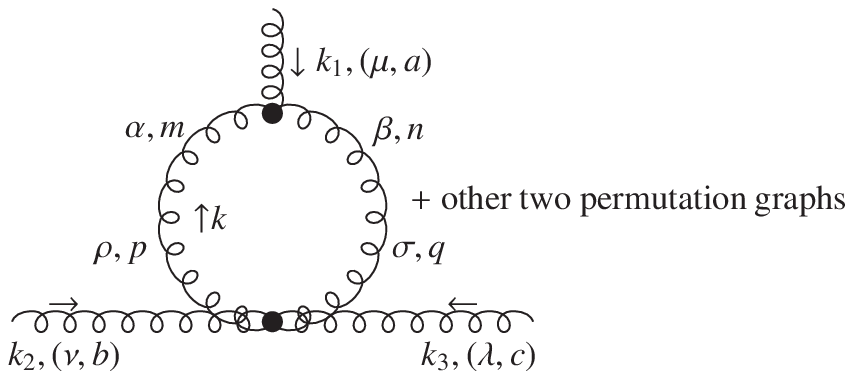}
\end{center}
\begin{center}{\sl Fig.6d.}\end{center}

\begin{eqnarray}
L(6d)_{\mu\nu\lambda}^{abc}&=&\frac{1}{2!}{\intdk}gf^{amn}[(k_1-k)_\beta{g}_{\mu\alpha}
+(k+k+k_1)_\mu{g}_{\alpha\beta}+(-k-k_1-k_1)_\alpha{g}_{\beta\mu}]
\frac{-i\delta^{mp}g^{\alpha\rho}}{k^2}\frac{-i\delta^{nq}g^{\beta\sigma}}{(k+k_1)^2}\nonumber\\
& &\times
(-ig^2)[f^{epq}f^{ecb}(g_{\rho\lambda}g_{\sigma\nu}-g_{\nu\rho}g_{\lambda\sigma})
+f^{epc}f^{ebq}(g_{\nu\rho}g_{\lambda\sigma}-g_{\rho\sigma}g_{\nu\lambda})
+f^{epb}f^{eqc}(g_{\rho\sigma}g_{\nu\lambda}-g_{\rho\lambda}g_{\nu\sigma})]\nonumber\\
& &+permutation\ graphs\nonumber\\
&=&\frac{1}{2}ig^3f^{apq}[f^{epq}f^{ecb}(g_{\rho\lambda}g_{\sigma\nu}
-g_{\nu\rho}g_{\lambda\sigma})+f^{epc}f^{ebq}(g_{\nu\rho}g_{\lambda\sigma}
-g_{\rho\sigma}g_{\nu\lambda})+f^{epb}f^{eqc}(g_{\rho\sigma}g_{\nu\lambda}
-g_{\rho\lambda}g_{\nu\sigma})]\nonumber\\
& &\times
{\intdk}[(k_1-k)^\sigma{g}^\rho_\mu+(k+k+k_1)_\mu{g}^{\rho\sigma}
+(-k-k_1-k_1)^\rho{g}^\sigma_\mu]\frac{1}{k^2(k+k_1)^2}\nonumber\\
& &+permutation\ graphs\nonumber\\
&=&\frac{i}{2}g^3C_1f^{abc}[-(g_{\rho\lambda}g_{\sigma\nu}-g_{\nu\rho}g_{\lambda\sigma})
+\frac{1}{2}(g_{\nu\rho}g_{\lambda\sigma}-g_{\rho\sigma}g_{\nu\lambda})
+\frac{1}{2}(g_{\rho\sigma}g_{\nu\lambda}-g_{\rho\lambda}g_{\nu\sigma})]\int_0^1dx_1\times\nonumber\\
& &{\intdk}\frac{\Gamma(2)}{\Gamma(1)^2}\frac{1}{[(1-x_1)k^2+x_1(k+k_1)^2]^2}[(k_1-k)^\sigma{g}^\rho_\mu
+(k+k+k_1)_\mu{g}^{\rho\sigma}+(-k-k_1-k_1)^\rho{g}^\sigma_\mu]\nonumber\\
& &+permutation\ graphs\nonumber\\
&=&\frac{i}{2}g^3C_1f^{abc}\frac{3}{2}(g_{\nu\rho}g_{\lambda\sigma}-g_{\rho\lambda}g_{\sigma\nu})
\int_0^1dx_1{\intdk}\frac{1}{[(k+x_1k_1)^2+x_1(1-x_1)k_1^2]^2}\times\nonumber\\
& &[(k_1-k)^\sigma{g}^\rho_\mu+(k+k+k_1)_\mu{g}^{\rho\sigma}+(-k-k_1-k_1)^\rho{g}^\sigma_\mu]
+permutation\ graphs\nonumber\\
&=&\frac{3i}{4}g^3C_1f^{abc}(g_{\nu\rho}g_{\lambda\sigma}-g_{\rho\lambda}g_{\sigma\nu})
\int_0^1dx_1{\intdk}\frac{1}{(k^2-M_{6d}^2)^2}
[(-k+(1+x_1)k_1)^\sigma{g}^\rho_\mu \nonumber \\
& & +(2k+(1-2x_1)k_1)_\mu{g}^{\rho\sigma}
+(-k+(-2+x_1)k_1)^\rho{g}^\sigma_\mu]+permutation\ graphs
\end{eqnarray}
with $M_{6d}^2=-x_1(1-x_1)k_1^2$. The finally result is simply given
by
\begin{eqnarray}
L(6d)_{\mu\nu\lambda;div}^{abcR}&=&\frac{3i}{4}g^3C_1f^{abc}(g_{\nu\rho}g_{\lambda\sigma}
-g_{\rho\lambda}g_{\sigma\nu})\int_0^1dx_1I_0(M_{6d}^2)[(1+x_1)k_1^\sigma{g}^\rho_\mu
+(1-2x_1)k_{1\mu}g^{\rho\sigma}+(-2+x_1)k_1^\rho{g}^\sigma_\mu]\nonumber\\
& &+permutation\ graphs
\end{eqnarray}

\subsection{four-gluon vertex renormalization}

There are five diagrams which can contribute to four-gluon vertex
renormalizaion. The first one (Fig.7a) is given by:
\begin{center}
\includegraphics[scale=1]{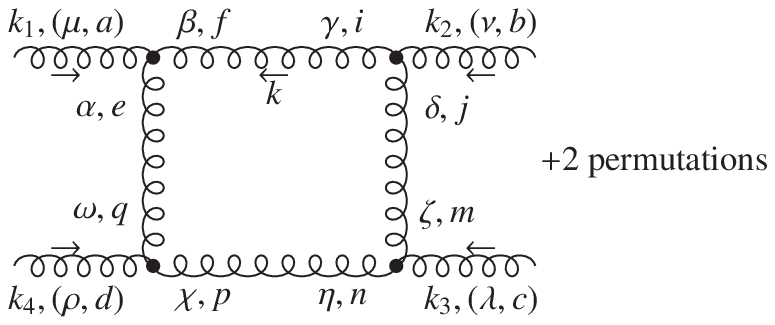}
\end{center}
\begin{center}{\sl Fig.7a.}\end{center}

\begin{eqnarray}
L(7a)_{\mu\nu\lambda\rho}^{abcd}&=&{\intdk}gf^{aef}[(k_1+k+k_1)_\beta{g}_{\mu\alpha}
+(-k-k_1-k)_\mu{g}_{\alpha\beta}+(k-k_1)_\alpha{g}_{\beta\mu}]\frac{-i\delta^{fi}g^{\beta\gamma}}{k^2}\times\nonumber\\
& &gf^{bij}[(k_2+k)_\delta{g}_{\nu\gamma}+(-k-k+k_2)_\nu{g}_{\gamma\delta}
+(k-k_2-k_2)_\gamma{g}_{\delta\nu}]\frac{-i\delta^{jm}g^{\delta\zeta}}{(k-k_2)^2}\times\nonumber\\
& &gf^{cmn}[(k_3-k_2+k)_\eta{g}_{\lambda\zeta}+(k_2-k-k+k_2+k_3)_\lambda{g}_{\zeta\eta}
+(k-k_2-k_3-k_3)_\zeta{g}_{\eta\lambda}]\frac{-i\delta^{pn}g^{\eta\chi}}{(k-k_2-k_3)^2}\times\nonumber\\
& &
f^{dpq}[(k_4+k+k_1+k_4)_\omega{g}_{\rho\chi}+(-k-k_1-k_4-k-k_1)_\rho{g}_{\chi\omega}
+(k+k_1-k_4)_\chi{g}_{\omega\rho}]\frac{-i\delta^{qe}g^{\omega\alpha}}{(k+k_1)^2}\nonumber\\
& &+2\ permutations
\end{eqnarray}
its divergent part reads:

\begin{eqnarray}
L(7a)_{\mu\nu\lambda\rho;div}^{abcd}&=&{\intdk}gf^{aef}(k_\beta{g}_{\mu\alpha}-2k_\mu{g}_{\alpha\beta}
+k_\alpha{g}_{\beta\mu})\frac{-i\delta^{fi}g^{\beta\gamma}}{k^2}gf^{bij}(k_\delta{g}_{\nu\gamma}
-2k_\nu{g}_{\gamma\delta}+k_\gamma{g}_{\delta\nu})\frac{-i\delta^{jm}g^{\delta\zeta}}{(k-k_2)^2}\times\nonumber\\
& &gf^{cmn}(k_\eta{g}_{\lambda\zeta}-2k_\lambda{g}_{\zeta\eta}+k_\zeta{g}_{\eta\lambda})
\frac{-i\delta^{pn}g^{\eta\chi}}{(k-k_2-k_3)^2}f^{dpq}(k_\omega{g}_{\rho\chi}-2k_\rho{g}_{\chi\omega}
+k_\chi{g}_{\omega\rho})\frac{-i\delta^{qe}g^{\omega\alpha}}{(k+k_1)^2}\nonumber\\
& &+2\ permutations\nonumber\\
&=&(-ig)^4f^{aef}f^{bfj}f^{cjn}f^{dne}{\intdk}\times\nonumber\\
&
&\frac{(k^\gamma{g}_{\mu\alpha}-2k_\mu{g}_\alpha^\gamma+k_\alpha{g}^\gamma_\mu)(k^\zeta{g}_{\nu\gamma}
-2k_\nu{g}_\gamma^\zeta+k_\gamma{g}^\zeta_\nu)
(k^\chi{g}_{\lambda\zeta}-2k_\lambda{g}_\zeta^\chi+k_\zeta{g}^\chi_\lambda)(k^\alpha{g}_{\rho\chi}
-2k_\rho{g}_\chi^\alpha+k_\chi{g}^\alpha_\rho)}{k^2(k-k_2)^2(k-k_2-k_3)^2(k+k_1)^2}\nonumber\\
& &+2\ permutations\nonumber\\
&=&g^4f^{aef}f^{bfj}f^{cjn}f^{dne}\int_0^1dx_1\int_0^{x_1}dx_2\int_0^{x_2}dx_3{\intdk}
\frac{\Gamma(4)}{\Gamma(1)^4}\times\nonumber\\
&
&\frac{(k^\gamma{g}_{\mu\alpha}-2k_\mu{g}_\alpha^\gamma+k_\alpha{g}^\gamma_\mu)(k^\zeta{g}_{\nu\gamma}
-2k_\nu{g}_\gamma^\zeta+k_\gamma{g}^\zeta_\nu)
(k^\chi{g}_{\lambda\zeta}-2k_\lambda{g}_\zeta^\chi+k_\zeta{g}^\chi_\lambda)(k^\alpha{g}_{\rho\chi}
-2k_\rho{g}_\chi^\alpha+k_\chi{g}^\alpha_\rho)}
{[(1-x_1)k^2+(x_1-x_2)(k-k_2)^2+(x_2-x_3)(k-k_2-k_3)^2+x_3(k+k_1)^2]^4}\nonumber\\
& &+2\ permutations\nonumber\\
&=&6g^4f^{aef}f^{bfj}f^{cjn}f^{dne}\int_0^1dx_1\int_0^{x_1}dx_2\int_0^{x_2}dx_3{\intdk}\times\nonumber\\
&
&\frac{(k^\gamma{g}_{\mu\alpha}-2k_\mu{g}_\alpha^\gamma+k_\alpha{g}^\gamma_\mu)(k^\zeta{g}_{\nu\gamma}
-2k_\nu{g}_\gamma^\zeta+k_\gamma{g}^\zeta_\nu)
(k^\chi{g}_{\lambda\zeta}-2k_\lambda{g}_\zeta^\chi+k_\zeta{g}^\chi_\lambda)(k^\alpha{g}_{\rho\chi}
-2k_\rho{g}_\chi^\alpha+k_\chi{g}^\alpha_\rho)}
{\{[k-(x_1-x_2)k_2-(x_2-x_3)(k_2+k_3)+x_3k_1]^2-M_{7a}^2\}^4}\nonumber\\
& &+2\ permutations\nonumber\\
&\sim&6g^4f^{aef}f^{bfj}f^{cjn}f^{dne}\int_0^1dx_1\int_0^{x_1}dx_2\int_0^{x_2}dx_3{\intdk}
\frac{1}{(k^2-M_{7a}^2)^4}\times\nonumber\\
&
&(k^\gamma{g}_{\mu\alpha}-2k_\mu{g}_\alpha^\gamma+k_\alpha{g}^\gamma_\mu)(k^\zeta{g}_{\nu\gamma}
-2k_\nu{g}_\gamma^\zeta+k_\gamma{g}^\zeta_\nu)
(k^\chi{g}_{\lambda\zeta}-2k_\lambda{g}_\zeta^\chi+k_\zeta{g}^\chi_\lambda)(k^\alpha{g}_{\rho\chi}
-2k_\rho{g}_\chi^\alpha+k_\chi{g}^\alpha_\rho)\nonumber\\
& &+2\ permutations\nonumber\\
&=&6g^4f^{aef}f^{bfj}f^{cjn}f^{dne}\int_0^1dx_1\int_0^{x_1}dx_2\int_0^{x_2}dx_3{\intdk}
\frac{1}{(k^2-M_{7a}^2)^4}\times\nonumber\\
& &(g_{\mu\nu}g_{\lambda\rho}k^4+g_{\mu\rho}g_{\nu\lambda}k^4+3g_{\mu\rho}k_\nu{k}_\lambda{k}^2
+3g_{\lambda\rho}k_{\mu}k_{\nu}k^2+3g_{\mu\nu}k_{\rho}k_{\lambda}k^2+3g_{\nu\lambda}k_{\mu}k_{\rho}k^2
+34k_{\mu}k_{\nu}k_{\lambda}k_{\rho})\nonumber\\
& &+2\ permutations\nonumber\\
&\sim&6g^4f^{aef}f^{bfj}f^{cjn}f^{dne}\int_0^1dx_1\int_0^{x_1}dx_2\int_0^{x_2}dx_3[(g_{\mu\nu}g_{\lambda\rho}
+g_{\mu\rho}g_{\nu\lambda})I_{0}(M_{7a}^2)+\nonumber\\
& &3g_{\mu\rho}I_{0\nu\lambda}(M_{7a}^2)+3g_{\lambda\rho}I_{0\mu\nu}(M_{7a}^2)
+3g_{\mu\nu}I_{0\rho\lambda}(M_{7a}^2)+3g_{\nu\lambda}I_{0\mu\rho}(M_{7a}^2)
+34I_{0\mu\nu\lambda\rho}(M_{7a}^2)]\nonumber\\
& &+2\ permutations
\end{eqnarray}
with
$$M_{7a}^2=-(x_1-x_2)(1-x_1+x_2)k_2^2-(x_2-x_3)(1-x_2+x_3)(k_2+k_3)^2-x_3(1-x_3)k_1^2.$$
and we have used in the last step the identities:
$$k^2=(k^2-M^2)+M^2\ \ and\ \ k^4=(k^2-M^2)^2+2M^2(k^2-M^2)+M^4.$$
Taking the relations
\begin{eqnarray}
I_{0\mu\nu}^R=\frac{1}{4}g_{\mu\nu}I_0^R \qquad
I_{0\mu\nu\rho\sigma}^R=\frac{1}{24}(g_{\mu\nu}g_{\rho\sigma}+g_{\mu\rho}g_{\nu\sigma}+g_{\mu\sigma}g_{\nu\rho})I_0^R.
\end{eqnarray}
the divergent part can be expressed in term of $I_0^R$:
\begin{eqnarray}
L(7a)_{\mu\nu\lambda\rho;div}^{abcdR}&=&6g^4f^{aef}f^{bfj}f^{cjn}f^{dne}\int_0^1dx_1\int_0^{x_1}dx_2
\int_0^{x_2}dx_3[\frac{5}{2}(g_{\mu\nu}g_{\lambda\rho}+g_{\mu\rho}g_{\nu\lambda})+\nonumber\\
&
&\frac{34}{24}(g_{\mu\nu}g_{\lambda\rho}+g_{\mu\lambda}g_{\nu\rho}+g_{\mu\rho}g_{\nu\lambda})]I_{0}^R(M_{7a}^2)+2\
permutations
\end{eqnarray}

The second diagram (Fig.7b) is given by:
\begin{center}
\includegraphics[scale=1]{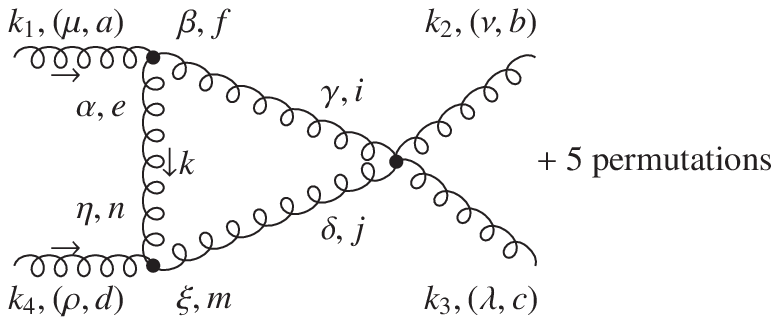}
\end{center}
\begin{center}{\sl Fig.7b.}\end{center}
\begin{eqnarray}
L(7b)_{\mu\nu\lambda\rho}^{abcd}&=&{\intdk}gf^{aef}[(k_1+k)_{\beta}g_{\mu\alpha}
+(-k-k+k_1)_{\mu}g_{\alpha\beta}+(k-k_1-k_1)_{\alpha}g_{\beta\mu}]\frac{-i\delta^{en}g^{\alpha\eta}}{k^2}\nonumber\\
& &gf^{dmn}[(k_4+k+k_4)_{\eta}g_{\rho\chi}+(-k-k_4-k)_{\rho}g_{\chi\eta}+(k-k_4)_{\chi}g_{\eta\rho}]
\frac{-i\delta^{mj}g^{\chi\delta}}{(k+k_4)^2}\nonumber\\
& &(-i)g^2[f^{lib}f^{lcj}(g_{\gamma\lambda}g_{\nu\delta}-g_{\gamma\delta}g_{\nu\lambda})
+f^{lic}f^{ljb}(g_{\gamma\delta}g_{\lambda\nu}-g_{\gamma\nu}g_{\lambda\delta})+\nonumber\\
& &f^{lij}f^{lbc}(g_{\gamma\nu}g_{\delta\lambda}-g_{\gamma\lambda}g_{\delta\nu})]
\frac{-i\delta^{if}g^{\gamma\beta}}{(k-k_1)^2}+5\ permutations\nonumber\\
&=&g^4{\intdk}f^{aef}[(k_1+k)_{\beta}g_{\mu\alpha}+(-k-k+k_1)_{\mu}g_{\alpha\beta}
+(k-k_1-k_1)_{\alpha}g_{\beta\mu}]f^{dme}[(k_4+k+k_4)^{\alpha}g_{\rho\chi}+\nonumber\\
&
&(-k-k_4-k)_{\rho}g_{\chi}^{\alpha}+(k-k_4)_{\chi}g^{\alpha}_{\rho}][f^{lfb}f^{lcm}
(g^{\beta}_{\lambda}g_{\nu}^{\chi}-g^{\beta\chi}g_{\nu\lambda})+
f^{lfc}f^{lmb}(g^{\beta\chi}g_{\lambda\nu}-g^{\beta}_{\nu}g_{\lambda}^{\chi})+\nonumber\\
&
&f^{lfm}f^{lbc}(g^{\beta}_{\nu}g^{\chi}_{\lambda}-g^{\beta}_{\lambda}g^{\chi}_{\nu})]\frac{1}{k^2(k+k_4)^2(k-k_1)^2}+5\
permutations
\end{eqnarray}
its divergent part reads:
\begin{eqnarray}
L(7b)_{\mu\nu\lambda\rho;div}^{abcd}&=&g^4{\intdk}f^{aef}f^{dme}
[f^{lfb}f^{lcm}(g^{\beta}_{\lambda}g_{\nu}^{\chi}-g^{\beta\chi}g_{\nu\lambda})
+f^{lfc}f^{lmb}(g^{\beta\chi}g_{\lambda\nu}-g^{\beta}_{\nu}g_{\lambda}^{\chi})
+f^{lfm}f^{lbc}(g^{\beta}_{\nu}g^{\chi}_{\lambda}-g^{\beta}_{\lambda}g^{\chi}_{\nu})]\nonumber\\
& &[k_{\beta}g_{\mu\alpha}-2k_{\mu}g_{\alpha\beta}+k_{\alpha}g_{\beta\mu}]
[k^{\alpha}g_{\rho\chi}-2k_{\rho}g^{\alpha}_{\chi}+k_{\chi}g^{\alpha}_{\rho}]
\frac{1}{k^2(k+k_4)^2(k-k_1)^2}+5\ permutations\nonumber\\
&=&g^4{\intdk}f^{aef}f^{dme}
[f^{lfb}f^{lcm}(g^{\beta}_{\lambda}g_{\nu}^{\chi}-g^{\beta\chi}g_{\nu\lambda})
+f^{lfc}f^{lmb}(g^{\beta\chi}g_{\lambda\nu}-g^{\beta}_{\nu}g_{\lambda}^{\chi})
+f^{lfm}f^{lbc}(g^{\beta}_{\nu}g^{\chi}_{\lambda}-g^{\beta}_{\lambda}g^{\chi}_{\nu})]\nonumber\\
& &\int_0^1\int_0^{x_1}dx_1dx_2\frac{\Gamma(3)}{\Gamma(1)\Gamma(1)\Gamma(1)}
\frac{[k_{\beta}g_{\mu\alpha}-2k_{\mu}g_{\alpha\beta}+k_{\alpha}g_{\beta\mu}]
[k^{\alpha}g_{\rho\chi}-2k_{\rho}g^{\alpha}_{\chi}+k_{\chi}g^{\alpha}_{\rho}]}{[(1-x_1)k^2
+(x_1-x_2)(k+k_4)^2+x_2(k-k_1)^2]^3}+5\ permutations\nonumber\\
&=&2g^4{\intdk}f^{aef}f^{dme}
[f^{lfb}f^{lcm}(g^{\beta}_{\lambda}g_{\nu}^{\chi}-g^{\beta\chi}g_{\nu\lambda})+f^{lfc}f^{lmb}
(g^{\beta\chi}g_{\lambda\nu}-g^{\beta}_{\nu}g_{\lambda}^{\chi})+f^{lfm}f^{lbc}
(g^{\beta}_{\nu}g^{\chi}_{\lambda}-g^{\beta}_{\lambda}g^{\chi}_{\nu})]\nonumber\\
&
&\int_0^1\int_0^{x_1}dx_1dx_2\frac{k^2g_{\beta\mu}g_{\rho\chi}-k_{\beta}k_{\mu}g_{\rho\chi}
-2k_{\beta}k_{\rho}g_{\mu\chi}+4k_{\mu}k_{\rho}g_{\beta\chi}+k_{\beta}k_{\chi}g_{\mu\rho}
-2k_{\mu}k_{\chi}g_{\beta\rho}-k_{\rho}k_{\chi}g_{\mu\beta}}{(k^2-M_{7b}^2)^3}+\nonumber\\
& &+5\ permutations
\end{eqnarray}
with $M_{7b}=(x_1-x_2)(x_1-x_2-1)k_4^2+x_2(x_2-1)k_1^2$. Taking the
relation $I_{0\mu\nu}^R=\frac{1}{4}g_{\mu\nu}I_0^R$, we finally
yield:
\begin{eqnarray}
L(7b)_{\mu\nu\lambda\rho;div}^{abcdR}&=&2g^4f^{aef}f^{dme}
[f^{lfb}f^{lcm}(g^{\beta}_{\lambda}g_{\nu}^{\chi}-g^{\beta\chi}g_{\nu\lambda})
+f^{lfc}f^{lmb}(g^{\beta\chi}g_{\lambda\nu}-g^{\beta}_{\nu}g_{\lambda}^{\chi})
+f^{lfm}f^{lbc}(g^{\beta}_{\nu}g^{\chi}_{\lambda}-g^{\beta}_{\lambda}g^{\chi}_{\nu})]\nonumber\\
&
&\int_0^1\int_0^{x_1}dx_1dx_2(g_{\beta\mu}g_{\rho\chi}-\frac{1}{4}g_{\beta\mu}g_{\rho\chi}
-\frac{2}{4}g_{\beta\rho}g_{\mu\chi}+\frac{4}{4}g_{\mu\rho}g_{\beta\chi}+\frac{1}{4}g_{\beta\chi}g_{\mu\rho}
-\frac{2}{4}g_{\mu\chi}g_{\beta\rho}-\frac{1}{4}g_{\rho\chi}g_{\mu\beta})I_0(M_{7b})\nonumber\\
& &+5\ permutations
\end{eqnarray}

The third diagram (Fig.7c) is evaluated to be:
\begin{center}
\includegraphics[scale=1]{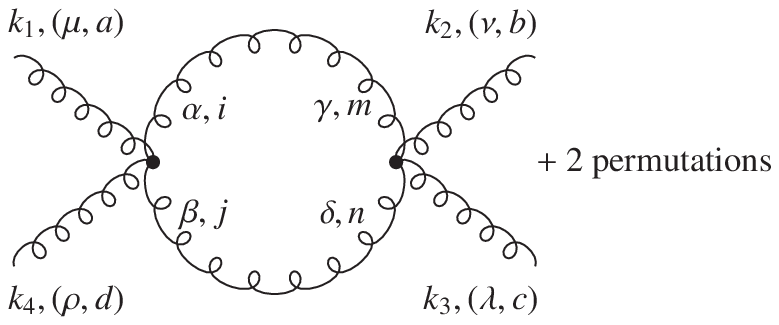}
\end{center}
\begin{center}{\sl Fig.7c.}\end{center}
\begin{eqnarray}
L(7c)_{\mu\nu\lambda\rho}^{abcd}&=&\frac{1}{2}{\intdk}(-i)g^2[f^{eai}f^{ejd}(g_{\mu\beta}g_{\alpha\rho}
-g_{\mu\rho}g_{\alpha\beta})+f^{eaj}f^{edi}(g_{\mu\rho}g_{\beta\alpha}-g_{\mu\alpha}g_{\beta\rho})
+f^{ead}f^{eij}(g_{\mu\alpha}g_{\rho\beta}-g_{\mu\beta}g_{\rho\alpha})]\nonumber\\
& &\times
(-i)g^2[f^{fmb}f^{fcn}(g_{\gamma\lambda}g_{\nu\delta}-g_{\gamma\delta}g_{\nu\lambda})
+f^{fmc}f^{fnb}(g_{\gamma\delta}g_{\lambda\nu}-g_{\gamma\nu}g_{\lambda\delta})+f^{fmn}f^{fbc}
(g_{\gamma\nu}g_{\delta\lambda}-g_{\gamma\lambda}g_{\delta\nu})] \nonumber\\
& &\times
\frac{-i\delta^{im}g^{\alpha\gamma}}{k^2}\frac{-i\delta^{jn}g^{\beta\delta}}{(k+k_1+k_4)^2}
+2\ permutations\nonumber\\
&=&\frac{1}{2}g^4{\intdk}[f^{eai}f^{ejd}(g_{\mu\beta}g_{\alpha\rho}-g_{\mu\rho}g_{\alpha\beta})
+f^{eaj}f^{edi}(g_{\mu\rho}g_{\beta\alpha}-g_{\mu\alpha}g_{\beta\rho})+f^{ead}f^{eij}
(g_{\mu\alpha}g_{\rho\beta}-g_{\mu\beta}g_{\rho\alpha})]\nonumber\\
& &\times
[f^{fib}f^{fcj}(g^{\alpha}_{\lambda}g_{\nu}^{\beta}-g^{\alpha\beta}g_{\nu\lambda})
+f^{fic}f^{fjb}(g^{\alpha\beta}g_{\lambda\nu}-g^{\alpha}_{\nu}g_{\lambda}^{\beta})+f^{fij}f^{fbc}
(g^{\alpha}_{\nu}g^{\beta}_{\lambda}-g^{\alpha}_{\lambda}g^{\beta}_{\nu})]\frac{1}{k^2(k+k_1+k_4)^2}\nonumber\\
& &+2\ permutations\nonumber\\
&=&\frac{1}{2}g^4[f^{eai}f^{ejd}(g_{\mu\beta}g_{\alpha\rho}-g_{\mu\rho}g_{\alpha\beta})
+f^{eaj}f^{edi}(g_{\mu\rho}g_{\beta\alpha}-g_{\mu\alpha}g_{\beta\rho})+f^{ead}f^{eij}
(g_{\mu\alpha}g_{\rho\beta}-g_{\mu\beta}g_{\rho\alpha})] \nonumber\\
& &\times
[f^{fib}f^{fcj}(g^{\alpha}_{\lambda}g_{\nu}^{\beta}-g^{\alpha\beta}g_{\nu\lambda})
+f^{fic}f^{fjb}(g^{\alpha\beta}g_{\lambda\nu}-g^{\alpha}_{\nu}g_{\lambda}^{\beta})
+f^{fij}f^{fbc}(g^{\alpha}_{\nu}g^{\beta}_{\lambda}-g^{\alpha}_{\lambda}g^{\beta}_{\nu})]\nonumber\\
&
&\int_0^1dx_1{\intdk}\frac{\Gamma[2]}{\Gamma[1]^2}\frac{1}{(k^2-M_{7c}^2)^2}+2\
permutations
\end{eqnarray}
with $M_{7c}=x_1(x_1-1)(k_1+k_4)^2$. The divergent part is:
\begin{eqnarray}
L(7c)_{\mu\nu\lambda\rho;div}^{abcdR}&=&\frac{1}{2}g^4[f^{eai}f^{ejd}(g_{\mu\beta}g_{\alpha\rho}
-g_{\mu\rho}g_{\alpha\beta})+f^{eaj}f^{edi}(g_{\mu\rho}g_{\beta\alpha}-g_{\mu\alpha}g_{\beta\rho})
+f^{ead}f^{eij}(g_{\mu\alpha}g_{\rho\beta}-g_{\mu\beta}g_{\rho\alpha})]\times\nonumber\\
& &[f^{fib}f^{fcj}(g^{\alpha}_{\lambda}g_{\nu}^{\beta}-g^{\alpha\beta}g_{\nu\lambda})
+f^{fic}f^{fjb}(g^{\alpha\beta}g_{\lambda\nu}-g^{\alpha}_{\nu}g_{\lambda}^{\beta})
+f^{fij}f^{fbc}(g^{\alpha}_{\nu}g^{\beta}_{\lambda}-g^{\alpha}_{\lambda}g^{\beta}_{\nu})]\nonumber\\
& &\int_0^1dx_1I_0(M_{7c})+2\ permutations
\end{eqnarray}

The fourth diagram (Fig.7d) is given by:
\begin{center}
\includegraphics[scale=1]{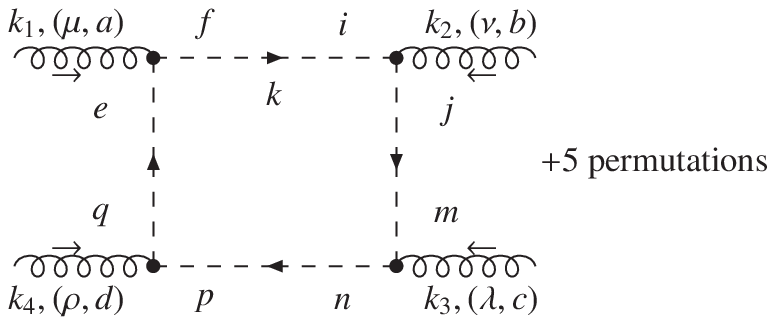}
\end{center}
\begin{center}{\sl Fig.7d.}\end{center}
\begin{eqnarray}
L(7d)_{\mu\nu\lambda\rho}^{abcd}&=&-{\intdk}(-g)f^{afe}k_{\mu}\frac{i\delta^{fi}}{k^2}(-g)f^{bji}(k+k_2)_{\nu}
\frac{i\delta^{jm}}{(k+k_2)^2}(-g)f^{cnm}(k+k_2+k_3)_{\lambda}\nonumber \\
& & \frac{i\delta^{np}}{(k+k_2+k_3)^2}\times (-g)f^{dqp}(k-k_1)_{\rho}\frac{i\delta^{qe}}{(k-k_1)^2}\nonumber\\
&=&-g^4f^{aie}f^{bmi}f^{cpm}f^{dep}{\intdk}\frac{k_{\mu}(k+k_2)_{\nu}(k+k_2+k_3)_{\lambda}
(k-k_1)_{\rho}}{k^2(k+k_2)^2(k+k_2+k_3)^2(k-k_1)^2}\nonumber\\
\end{eqnarray}
so the divergent part is:
\begin{eqnarray}
L(7d)_{\mu\nu\lambda\rho;div}^{abcd}&=&-g^4f^{aie}f^{bmi}f^{cpm}f^{dep}{\intdk}
\frac{k_{\mu}k_{\nu}k_{\lambda}k_{\rho}}{k^2(k+k_2)^2(k+k_2+k_3)^2(k-k_1)^2}+ 5 permutations\nonumber\\
&=&-g^4f^{aie}f^{bmi}f^{cpm}f^{dep}\int_0^1dx_1\int_0^{x_1}dx_2\int_0^{x_2}dx_3{\intdk}
\frac{\Gamma[4]}{\Gamma[1]\Gamma[1]\Gamma[1]\Gamma[1]}
\frac{k_{\mu}k_{\nu}k_{\lambda}k_{\rho}}{(k^2-M_{7d}^2)^4}\nonumber\\
& &+5 permutations\nonumber\\
&=&-6g^4f^{aie}f^{bmi}f^{cpm}f^{dep}\int_0^1dx_1\int_0^{x_1}dx_2\int_0^{x_2}dx_3I_{0\mu\nu\lambda\rho}(M_{7d}^2)+5\
permutations
\end{eqnarray}
where
$M_{7d}=(x_1-x_2)(x_1-x_2-1)k_2^2+(x_2-x_3)(x_2-x_3-1)(k_2+k_3)^2+x_3(x_3-1)k_1^2$.
Using the relation
\begin{eqnarray}
I_{0\mu\nu\rho\sigma}^R=\frac{1}{24}(g_{\mu\nu}g_{\rho\sigma}+g_{\mu\rho}g_{\nu\sigma}+g_{\mu\sigma}g_{\nu\rho})I_0^R.
\end{eqnarray}
the divergent part becomes:
\begin{eqnarray}
L(7d)_{\mu\nu\lambda\rho;div}^{abcdR}&=&-\frac{1}{4}g^4f^{aie}f^{bmi}f^{cpm}f^{dep}(g_{\mu\nu}g_{\lambda\rho}
+g_{\mu\lambda}g_{\nu\rho}+g_{\mu\rho}g_{\nu\lambda})
\int_0^1dx_1\int_0^{x_1}dx_2\int_0^{x_2}dx_3I_{0}^R(M_{7d}^2)\nonumber\\
& &+5\ permutations
\end{eqnarray}

The last diagram (Fig.7e) is evaluated to be
\begin{center}
\includegraphics[scale=1]{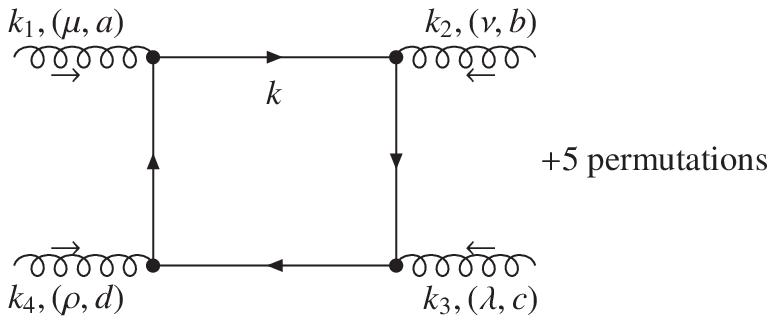}
\end{center}
\begin{center}{\sl Fig.7e.}\end{center}
\begin{eqnarray}
L(7e)_{\mu\nu\lambda\rho}^{abcd}&=&-Tr{\intdk}ig\gamma_{\mu}T^{a}\frac{i}{k\sla-k\sla_4-m}ig\gamma_{\rho}
T^{d}\frac{i}{k\sla+k\sla_2+k\sla_3-m}ig\gamma_{\lambda}T^{c}\frac{i}{k\sla+k\sla_2}-mig\gamma_{\nu}
T^{b}\frac{i}{k\sla-m}\nonumber\\
& &+5\ permutations\nonumber\\
&=&-g^4Tr(T^{a}T^{d}T^{c}T^{b}){\intdk}\frac{tr[\gamma_{\mu}(k\sla-k\sla_4+m)\gamma_{\rho}
(k\sla+k\sla_2+k\sla_3+m)\gamma_{\lambda}(k\sla+k\sla_2+m)\gamma_{\nu}(k\sla+m)]}{[(k-k_4)^2-m^2]
[(k+k_2+k_3)^2-m^2][(k+k_2)^2-m^2][k^2-m^2]}\nonumber\\
& &+5\ permutations
\end{eqnarray}
the divergent part reads:
\begin{eqnarray}
L(7e)_{\mu\nu\lambda\rho;div}^{abcd}&=&-g^4Tr(T^{a}T^{d}T^{c}T^{b}){\intdk}
\frac{tr(\gamma_{\mu}k\sla\gamma_{\rho}k\sla\gamma_{\lambda}k\sla\gamma_{\nu}k\sla)}{[(k-k_4)^2-m^2]
[(k+k_2+k_3)^2-m^2][(k+k_2)^2-m^2][k^2-m^2]}\nonumber\\
& &+5\ permutations\nonumber\\
&\sim&-g^4Tr(T^{a}T^{d}T^{c}T^{b})\int_0^1dx_1\int_0^{x_1}dx_2\int_0^{x_2}dx_3{\intdk}
\frac{\Gamma[4]}{\Gamma[1]\Gamma[1]\Gamma[1]\Gamma[1]}\frac{tr(\gamma_{\mu}k\sla\gamma_{\rho}k\sla
\gamma_{\lambda}k\sla\gamma_{\nu}k\sla)}{(k^2-M_{7e}^2)^4}\nonumber\\
& &+5\ permutations\nonumber\\
&=&-24g^4Tr(T^{a}T^{d}T^{c}T^{b})\int_0^1dx_1\int_0^{x_1}dx_2\int_0^{x_2}dx_3{\intdk}
\frac{1}{(k^2-M_{7e}^2)^4}\nonumber\\
& &\times
(g_{\mu\nu}g_{\lambda\rho}k^4-g_{\mu\lambda}g_{\nu\rho}k^4+g_{\mu\rho}g_{\nu\lambda}k^4
-2g_{\mu\nu}k_{\lambda}k_{\rho}k^2-2g_{\nu\lambda}k_{\rho}k_{\mu}k^2-2g_{\lambda\rho}k_{\mu}k_{\nu}k^2
\nonumber \\
& &
-2g_{\rho\mu}k_{\nu}k_{\lambda}k^2+8k_{\mu}k_{\nu}k_{\lambda}k_{\rho})
+5 permutations
\end{eqnarray}
where
$M_{7e}=m^2+(x_1-x_2)(x_1-x_2-1)k_4^2+(x_2-x_3)(x_2-x_3-1)(k_2+k_3)^2+x_3(x_3-1)k_2^2$.
Using the relation $I_{0\mu\nu}^R=\frac{1}{4}g_{\mu\nu}I_0^R$, we
finally obtain:
\begin{eqnarray}
L(7e)_{\mu\nu\lambda\rho;div}^{abcd R}&=&-8g^4Tr(T^{a}T^{d}T^{c}T^{b})(g_{\mu\nu}g_{\lambda\rho}
-2g_{\mu\lambda}g_{\nu\rho}+g_{\mu\rho}g_{\nu\lambda})\int_0^1dx_1\int_0^{x_1}dx_2
\int_0^{x_2}dx_3I_0^R(M_{7e})+\nonumber\\
& & 5\ permutations
\end{eqnarray}


\end{document}